\documentclass[journal]{IEEEtran}
\usepackage{color}
\usepackage{verbatim}
\usepackage{amsfonts}
\usepackage{amssymb}
\usepackage{stfloats}
\usepackage{chngpage}
\usepackage{cite}
\usepackage{graphicx}
\usepackage{psfrag}
\usepackage{subfigure}
\usepackage{amsmath}
\usepackage{array}
\usepackage{epstopdf}
\usepackage{authblk}
\usepackage{graphicx} 
\usepackage{amsthm} 
\usepackage{lipsum}
\usepackage{verbatim} 
\usepackage{authblk}
\usepackage{mathtools}
\usepackage{hyperref}
\usepackage{cuted}
\usepackage[lined,boxed,ruled]{algorithm2e}
\usepackage{algpseudocode}
\usepackage{framed} 
\usepackage{subfigure}
\usepackage{soul}
\usepackage{bm}
\usepackage{setspace}
\usepackage{multirow}
\usepackage{makecell}

\def\phi{\varphi}

\def\({\left(}
\def\){\right)}

\def\b0{{\mathbf{0}}}

\title{Integrated Sensing and Edge AI: Realizing Intelligent Perception in 6G}

\author{Zhiyan Liu, \emph{Graduate Student Member}, \emph{IEEE}, Xu Chen, Hai Wu, Zhanwei Wang, \emph{Graduate Student Member}, \emph{IEEE}, Xianhao Chen, \emph{Member}, \emph{IEEE}, Dusit Niyato, \emph{Fellow}, \emph{IEEE}, and Kaibin Huang, \emph{Fellow}, \emph{IEEE}

\thanks{Manuscript received 12 January 2025; revised 29 May 2025; accepted 19 July 2025. The work described in this paper was supported in part by the Research Grants Council of the Hong Kong Special Administrative Region, China under a fellowship award (HKU RFS2122-7S04), NSFC/RGC CRS (CRS\_HKU702/24), the Areas of Excellence scheme grant (AoE/E-601/22-R), Collaborative Research Fund (C1009-22G), and the Grants 17212423 \& 17304925, and in part by the Shenzhen-Hong Kong-Macau Technology Research Programme (Type C) (SGDX20230821091559018). The work of Dusit Niyato is supported by the Singapore Ministry of Education (MOE) Tier 1 (RG87/22 and RG24/24), the NTU Centre for Computational Technologies in Finance (NTU-CCTF), and the RIE2025 Industry Alignment Fund - Industry Collaboration Projects (IAF-ICP) (Award I2301E0026), administered by A*STAR, as well as supported by Alibaba Group and NTU Singapore through Alibaba-NTU Global e-Sustainability CorpLab (ANGEL). The work of Xianhao Chen is supported in part by the Research Grants Council of Hong Kong under Grant 27213824 and CRS HKU702/24, and in part by HKU IDS Research Seed Fund under Grant IDS-RSF2023-0012.

Zhiyan Liu, Xu Chen, Hai Wu, Zhanwei Wang, Xianhao Chen, and Kaibin Huang are with the Department of Electrical and Electronic Engineering, The University of Hong Kong, Hong Kong (e-mails: zyliu@eee.hku.hk; chenxu@eee.hku.hk; wuhai@eee.hku.hk; zhanweiw@eee.hku.hk; xchen@eee.hku.hk; huangkb@eee.hku.hk). Dusit Niyato is with the College of Computing
and Data Science, Nanyang Technological University, Singapore (e-mail: dniyato@ntu.edu.sg). The corresponding author is Kaibin Huang.}
    }

\makeatletter
\newcommand{\removelatexerror}{\let\@latex@error\@gobble}
\makeatother

\IEEEoverridecommandlockouts

\begin{document}

\maketitle
\begin{abstract}
Sensing and edge \emph{artificial intelligence} (AI) are envisioned as two essential and interconnected functions in \emph{sixth-generation} (6G) mobile networks. On the one hand, sensing-empowered applications rely on powerful AI models to extract features and understand semantics from ubiquitous wireless sensors. On the other hand, the massive amount of sensory data serves as the fuel to continuously refine edge AI models. This deep integration of sensing and edge AI has given rise to a new task-oriented paradigm known as \emph{integrated sensing and edge AI} (ISEA), which features a holistic design approach to communication, AI computation, and sensing for optimal sensing-task performance. In this article, we present a comprehensive survey for ISEA. We first provide technical preliminaries for sensing, edge AI, and new communication paradigms in ISEA. Then, we study several use cases of ISEA to demonstrate its practical relevance and introduce current standardization and industrial progress. Next, the design principles, metrics, tradeoffs, and architectures of ISEA are established, followed by a thorough overview of ISEA techniques, including digital air interface, over-the-air computation, and advanced signal processing. Its interplay with various 6G advancements, e.g., new physical-layer and networking techniques, are presented. Finally, we present future research opportunities in ISEA, including the integration of foundation models, convergence of ISEA and \emph{integrated sensing and communications} (ISAC), ultra-low-latency ISEA, and practicality issues.

\end{abstract}

\section{Introduction}
The {sixth-generation} (6G) mobile networks mark a remarkable evolution in the digital era. As anticipated by ITU-R in June 2023, the 6G landscape is shaped by a range of usage scenarios, including hyper-reliable and low-latency communications (HRLLC), immersive communications, massive connectivity, ubiquitous connectivity, integrated sensing and communications (ISAC), and integrated artificial intelligence (AI) and communications (IAAC)\cite{ITUR2023}. With these usage scenarios, 6G will support not only real-time and intensive data transmissions but also numerous intelligence applications requiring real-time intelligent operations, such as remote surgery, autonomous driving, holographic telepresence, and digital twin\cite{Saad_6G, Huawei2022}.

As reflected by ITU usage scenarios, two prominent trends in 6G are edge AI and sensing. On the one hand, one of the key trends of 6G is the full integration of edge AI into mobile networks. 6G networks can support AI provisioning with the mobile edge computing (MEC) paradigm, delivering diverse AI applications to resource-limited mobile devices with low latency, reduced bandwidth, and better privacy. This leads to a new field called edge intelligence or edge AI\cite{Letaief2022JSAC,GreenEdgeAI,GX2020CM}. On the other hand, edge intelligence must be fueled with massive amounts of sensory data\cite{JCSC_2021,RN244}. Aligned with this trend, 6G is expected to support multi-modal sensing through not only ISAC but also the collaborative perception of massive 6G edge devices, including connected autonomous vehicles (CAVs), unmanned aerial vehicles (UAVs), surveillance cameras, Internet of Things (IoT) sensors\cite{ISCC_Survey_DZ}, etc. These two trends, i.e., edge AI and sensing, are interdependent, because edge AI relies on massive sensory data for model training and inference, while sensing depends on advanced AI algorithms running at the network edge to maximize data value with reasonable communication and computing overhead.

Although the coexistence and interdependence of communications, AI, and sensing is a natural trend in 6G, their cooperation and integrated design for real-timeness and high reliability (e.g., 30 ms inference latency and near 100\% accuracy for autonomous driving\cite{Challenges-Level4/5,zw2025AIoutage}) is faced with several critical challenges, including communication and computation capacity, operational complexity, security and privacy, and sustainability. For instance, it is essential for current networking architectures and air interface techniques to evolve to accommodate the demanding requirements in latency and reliability of transmission and analytics of sensory data. These challenges necessitate innovative approaches for resource management, integrated frameworks, signal processing techniques, networking architectural design, etc., all unified under a new set of metrics and principles oriented towards optimized performance of sensing applications. This gives rise to a new task-oriented paradigm, \emph{integrated sensing and edge AI} (ISEA), which is the theme of this paper. In what follows, we first provide a historical perspective of ISEA, presenting the milestones of communications, sensing, and AI to elaborate on the grand convergence of these three domains, and demonstrate how 6G, as a unified communication-computing platform, can support integrated sensing and AI in a harmonious manner. After that, we present the major contributions of this paper with comparisons to other works and the paper's organization.

\subsection{Milestones of Integrated Sensing and Edge AI}
\subsubsection{Milestones of Communications}
The development of modern communication technologies dates back to the 19th century, when Samuel Morse created the first single-wire telegraph system using the Morse code, followed by the invention of the telephone by Alexander Graham Bell in 1876. In terms of wireless connectivity, the major breakthrough was made by Guglielmo Marconi, who prototyped the first long-distance radio wave-based wireless telegraphy system and conducted the world's first cross-ocean radio message transmission in 1902. The famous article ``A Mathematical Theory of Communication'' by Claude Shannon published in 1948 opened the field of information theory and has guided the modern communication system design until today, which defines the communication problem as ``reproducing at one point either exactly or approximately a message selected at another point'', while the semantic aspects of the said messages are regarded as ``irrelevant to the engineering problem''\cite{shannon}. Therein, the Shannon-Hartley theorem characterizes the maximum rate of errorless information transmission over a noisy communication channel, underpinning the rate-centric paradigms of communication system design ever since. 

Building upon Shannon's theory, the development of mobile cellular networks from its first to fifth generation has revolutionized society and lifestyles. The first generation of mobile networks (1G), launched in 1979, were analog systems dedicated to voice communication between mobile phones and fixed-line telephones, adopting frequency modulation and frequency division multiple access (FDMA). 
The second generation (2G) was commercially launched in Finland in 1991. A fully digital network, 2G is revolutionary in many aspects, such as encryption of radio signals, higher spectrum efficiency, and the commencement of data service provision to mobile phones albeit at a low rate of around 40 Kbps, which makes it suitable for pure text messages. 
Significantly improving the data rate to the wireless broadband level, the third generation (3G) was first commercially launched in 2001 and reached its peak in the early 2010s, satisfying applications such as mobile Internet access and video streaming. Such an improvement is attributed to several technical advances, such as code division multiple access (CDMA) and high-speed packet access (HSPA). 

From the fourth-generation (4G) onwards, mobile networks started supporting pervasive ``sensing'' functionalities, marked by techniques such as Narrowband Internet of Things (NB-IoT), Long-Term Evolution Machine Type Communication (LTE-M), and LTE Vehicle-to-Everything (LTE-V2X) communications to support machine-type and IoT techniques. The 4G networks were first commercially deployed in 2009, pushing the downlink data rate beyond 100 Mbps. Long-term evolution (LTE) is the major standard for 4G, featuring orthogonal frequency-division multiple access (OFDMA) and all-IP mobile networks. In 3GPP Release 13 (LTE Advanced Pro), frozen in June 2016, both NB-IoT and LTE-M were developed by 3GPP to support massive IoT devices and sensors in low-power wide-area networks (LPWANs), transiting mobile networks to a network of sensors. In the 3GPP Release 14 LTE standard in 2017, cellular V2X was first introduced to support communications between vehicles and road infrastructure, where vehicles are equipped with powerful multi-modal sensors to perceive environments.

The fifth-generation (5G), beginning its deployment in 2019, evolves into an integrated platform for communications, computing, and sensing, which provides enhancements for the aforementioned services in 4G. It features three main application areas, i.e., Enhanced Mobile Broadband (eMBB), Massive Machine-type Communications (mMTC), and Ultra-reliable and Low Latency Communications (URLLC). Therein, eMBB specifies higher data rates up to 1 Gbps for indoor, URLLC targets mission-critical applications that require ultra-high reliability, e.g., 99.9999\%, and low end-to-end (E2E) latency, e.g., 50ms and mMTC supports scenarios with dense IoT sensor access. Moreover, 5G New Radio (NR) V2X supports more advanced sensing applications for connected and autonomous vehicles, including extended sensor sharing that enables the exchange of raw or processed data gathered among vehicles, roadside units, devices of pedestrian, and V2X application servers~\cite{bagheri20215g}. As evident, the mission of 5G extends beyond supporting data exchange from massive low-end IoT sensors. It also supports intensive data transmissions from high-end sensors, such as cameras and Light Detection and Ranging (LiDAR) on vehicles.

Following the pace of evolution, i.e., 10 years for each generation, 6G is projected to be launched around 2030. The academia is actively pushing forward several potential technologies envisioned to play vital roles in 6G. HRLLC, immersive communications, and massive connectivity are extended from 5G usage scenarios, namely enhanced eMBB, mMTC, and URLLC, taking into account more stringent performance requirements in 6G applications. For instance, HRLLC will extend the performance limits of 5G URLLC, aiming to enable real-time (sub-millisecond level) applications in diverse sectors such as remote surgery and autonomous driving~\cite{NE2020_6G}. Immersive communications will support advanced applications like holographic telepresence, providing lifelike, three-dimensional communication experiences to revolutionize remote work, education, and social interactions~\cite{Saad_6G,ZhangZ_6G_VTM}. Meanwhile, massive communications will connect billions of devices seamlessly, fostering the dramatic growth of the IoT. 
On the other hand, as alluded to earlier, ISAC and IAAC are two new usage scenarios compared with 5G. ISAC with multi-functional waveforms, which can be traced back to 1960s\cite{FLiu2018TSPdual}, has recently gained emerging attention owing to massive Multiple-Input Multiple-Output (MIMO) and higher frequency bands. Another important shift is the sink of AI provisioning to edge nodes, resulting in the integrated design of wireless communications and AI. Joint efforts from academia and standardization organizations share the common objective of moving from the rate-centric paradigm to task-oriented application provisioning underpinned by integrated AI, communications, and sensing.

\subsubsection{Milestones of Sensing}
Modern sensing emerged with the development of electrical and electronic techniques which enabled machines to capture, record, and transmit sensory information. The 20th century saw the first use of radio waves in sensing, such as radio detection and ranging (Radar) developed by Robert Watson-Watt, which has since played a central role in remote sensing applications. Radar techniques were used to process data gathered from the propagation and reflection of radio waves, revealing the distance, speed, and direction of objects. 

Since the late 20th century, advancements in electronics and communication techniques have led to an era where sophisticated sensing systems were established. For instance, evolving with satellite technology, the Global Positioning System (GPS), was launched in 1978 to enable precise location sensing worldwide. Later than this, the development of other global navigation systems began: GLONASS (1982), Beidou (2000), and Galileo (2011). On the other hand, the invention of transistors at Bell Labs in 1947 marked the beginning of the miniaturization and integration of sensing devices. The emergence of microelectromechanical systems (MEMS) in the 1980s further allowed for the miniaturization of mechanical and electro-mechanical elements integrated into chips, which are now widely used in various sensors including accelerometers, gyroscopes, and pressure sensors. Concurrently, advanced optical sensors and fiber optics were developed for efficient telecommunications and precise measurements.

The early 2000s witnessed the emergence of sensing devices, termed ``smart devices'', that are endowed with the capability of real-time data processing by integrating lightweight processors and chips with sensors. This allows for access to not only raw sensory data but also a timely interpretation of the meaning underlying them. For example, wearable devices including fitness trackers and smartwatches can monitor and record health metrics and report immediately when human abnormalities are detected. 

From the 2010s to the present, IoT sensing gained extensive attention, where smart environments from homes to cities were enabled via the vast networks of interconnected diverse sensors including cameras, motion sensors, depth sensors, accelerometers, etc. The capture multi-modal sensory data gathered at IoT sensors are shared and analyzed (globally or in a distributed manner) to track trends and gather insights about everything from efficiency and energy use in factories to traffic jams to personal conditions. Furthermore, powerful AI algorithms emerged to support complicated sensory data processing and interpretation, leading to more autonomous and intelligent systems\cite{AI_Sensor_Rev}. Moreover, interest in quantum sensing and its deployment grew significantly. Quantum sensors can leverage entanglement, single photons, and squeezed states to perform extremely precise measurements (e.g., nanometer-level resolution~\cite{Quantum_enhance_sensing}), beating limits in traditional sensing technology.

Looking ahead to the future trend, we envision AI-driven autonomous sensory devices with hyper-connection. For one thing, the sensors are expected to be integrated with biological systems, in which the realized tactile sensing will be the key component for future digital twins and real-time e-healthcare including health monitoring and diagnostics. For another, the miniaturization of sensors will continue. Extremely small sensors can find wide applications in medical diagnostics, environmental monitoring, and industrial processes. More importantly, enhanced connectivity in 5G and 6G will facilitate the deployment and integration of advanced sensing technologies in real-time applications.

\subsubsection{Milestones of AI}
The history of AI has witnessed remarkable milestones and significant advancements since its inception. Tracing back to 1950, Alan Turing introduced the concept of machine intelligence and proposed the Turing Test as a criterion for evaluating a machine's ability to exhibit human-like intelligence. 
In 1956, the Dartmouth Conference marked the birth of AI as a formal research field.
Throughout the 1960s, early AI programs, such as Arthur Samuel's checkers-playing program and John McCarthy's Lisp programming language, showcased basic problem-solving and symbolic processing capabilities. 
However, Marvin Minsky and Seymour Papert's book ``Perceptrons'' (1969) emphasized the limitations of early neural networks, resulting in a temporary decline in AI research funding. 
The 1980s saw machine learning emerge as a subfield of AI, with algorithms such as decision trees, reinforcement learning, and support vector machines enabling AI systems to learn from data and improve their performance over time. In the 1990s, researchers developed probabilistic and statistical methods, including Bayesian networks and hidden Markov models, allowing AI systems to handle uncertainty and make inferences from incomplete or noisy data. A significant milestone was achieved in 1997 when IBM's Deep Blue defeated world chess champion Garry Kasparov, showcasing the capabilities of AI in complex decision-making tasks. 

The rise of big data and the development of deep learning (DL) techniques in the 2000s led to significant improvements in areas like computer vision and natural language processing. AlexNet, a deep convolutional neural network, won the ImageNet Large Scale Visual Recognition Challenge in 2012, marking a turning point in the adoption of DL for image recognition tasks. This achievement spurred the development of more advanced architectures, such as VGG, Inception, and ResNet. In 2016, DeepMind's AlphaGo, an AI system that combines DL and Monte Carlo Tree Search, defeated the world champion in the ancient game of Go, demonstrating advances in AI's ability to tackle complex problems. The advent of Transformer models, introduced by Vaswani et al. in 2017, has brought significant advancements in natural language processing (NLP) tasks. Transformers utilize self-attention mechanisms to capture long-range dependencies and parallelize computation, making them highly effective in processing large-scale language data. The success of DL can be largely attributed to the availability of large datasets, powerful computational resources such as graphics processing units (GPUs), and improvements in neural network architectures and optimization techniques. In particular, the exploding computational capability enables exhaustive pre-training over a vast amount of data on the Internet, nurturing cutting-edge foundation models like Generative Pre-trained Transformer 3 (GPT-3) by OpenAI. ChatGPT, a sibling model to GPT-3, exemplifies the success of foundation models in generating human-like text. By leveraging the pre-training and fine-tuning approach, ChatGPT can understand context, answer questions, provide recommendations, and even engage in creative tasks like writing and storytelling. 

Despite the booming of AI, challenges remain in terms of computational efficiency, robustness, and ethical considerations, which are essential aspects for the future development of AI. As researchers and practitioners continue to push the boundaries of AI, especially pushing AI towards the network edge to support ubiquitous and real-time intelligent applications in the era of 6G, where edge AI is a core research direction.

\subsection{Evolving towards Integrated Sensing and Edge AI}
As the technologies of communication, sensing, and AI evolve, an emerging trend is the convergence of the three technologies due to their tight coupling in 6G killer applications, with the examples given shortly in Section \ref{section-case-studies}. Specifically, sensing and AI become deeply coupled, as state-of-the-art sensing techniques are largely built on AI-based backbones for sensing analytics, and conversely, AI model training relies on the large bulk of data generated by sensors to distill intelligence. However, this training-and-inference process is not achieved by a single all-in-one node but through the cooperation of sensors, terminal user devices, and computation nodes distributed in the network, rendering communication as an indispensable component and key factor in the E2E task performance.

    Traditional designs of communication systems, however, treat sensing and AI as independent applications that operate on the Application Layer of the network, while the underlying communication infrastructure, e.g., previous generations of mobile networks, is designed in a rate-oriented manner agnostic to the information semantics. Such paradigms are inherently sub-optimal in terms of task performance and inadequate for satisfying the Quality of Service (QoS) requirements as the data volume, AI model complexity, and number of nodes scale dramatically. Instead, 6G is envisioned to be AI-native with deep communication-computation (C$^2$) integration and task-oriented optimization for the highest efficiency in supporting AI-based applications.  {The air-interface technologies for AI-empowered tasks are thus designed with different principles from previous ones aimed at errorless bit transmission, focusing instead on efficient reproduction of sensory information in a manner effective for the downstream sensing task. Examples include joint source-channel coding (JSCC) as well as data-aware access control and resource allocation. For multi-access scenarios, an outstanding example is {over-the-air computation} (AirComp), which exploits the waveform superposition property of wireless channels to realize computation of aggregation functions by simultaneous access of sensors. These new communication paradigms are expected to be a key system component seamlessly coupled with and optimized for sensing tasks.} In addition, the network itself shall feature sensing as an inherent capability, e.g., network-as-a-sensor, through multifunctional waveforms and native support for a massive number of networked sensors. Thus, we argue that the integrated design of communication, sensing and AI, which involves cross-layer synergy for optimized task performance, is an inevitable trend of technical advancements. On the other hand, it has become a consensus that the support of low-latency and high-reliability AI applications would require the AI computation power to be distributed in edge networks instead of relying solely on cloud data centers. Hence, in this paper, we incorporate sensing, AI, and communications featuring edge networks into a unified framework, ISEA, oriented towards E2E task performance by integrated design and optimization of the three components.

\begin{figure*}
    \centering
    \includegraphics[width=1.5\columnwidth]{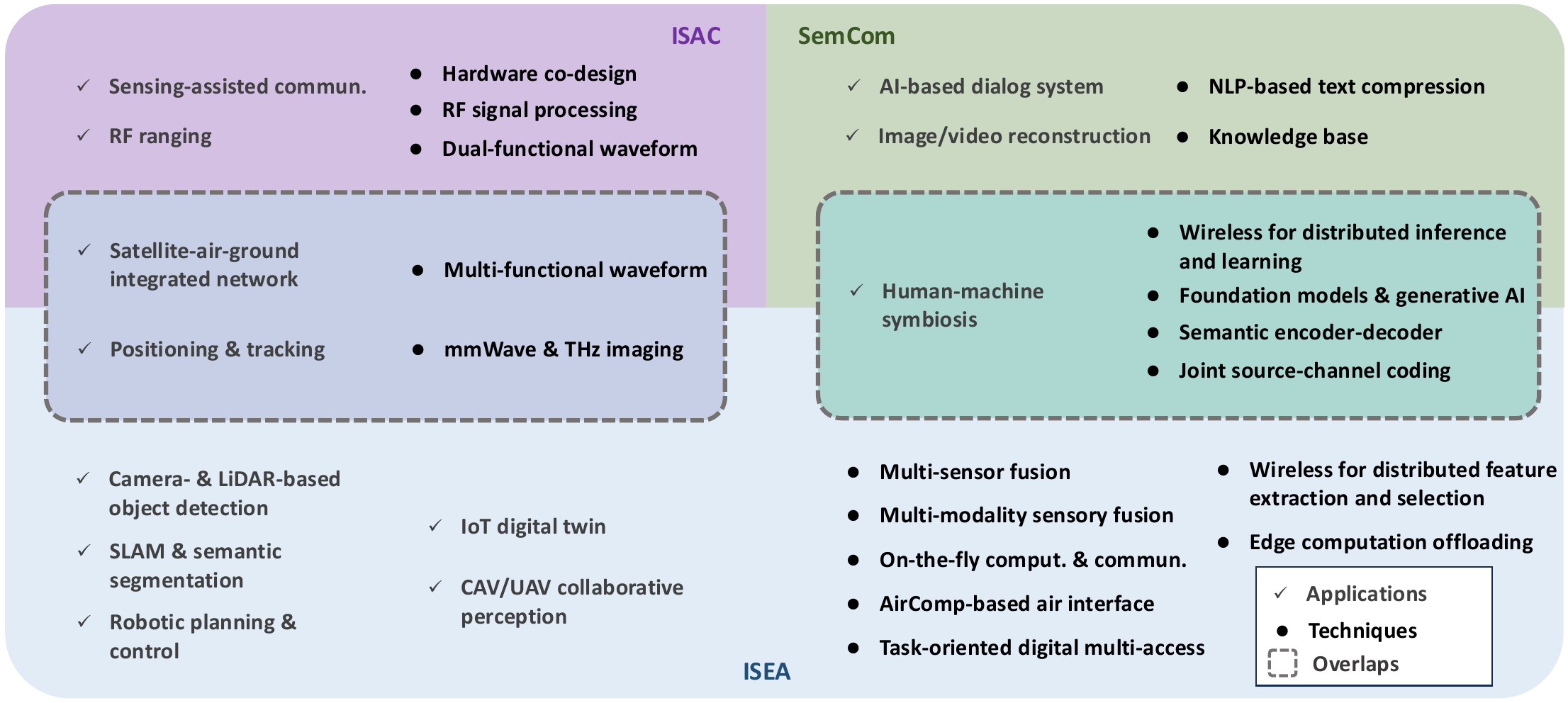}
    \caption{Differences and overlappings in terms of the scopes of ISEA compared with ISAC and SemCom. Compared with ISAC, which focuses on the integration and mutual assistance of RF sensing and communication, ISEA features multiple sensing modalities and joint communication-computation design for optimized performance of AI-based sensing tasks, serving applications that require rich environmental semantics. Compared with SemCom, ISEA targets specific sensing tasks and involves the joint design of a broader range of operations and components. }
    \label{fig: diff_scope}
\end{figure*}

\subsection{Comparisons with Relevant Works and Our Contribution}

{ISEA is a novel paradigm that integrates the acquisition, storage and transmission of sensing data with AI at the mobile edge to achieve low latency, high reliability and native intelligence for various sensing-assisted applications. ISEA is connected to two other emerging concepts in 6G, ISAC and {semantic communications} (SemCom), but is still fundamentally different due to its new application scenarios, design goals and approaches, as elaborated below.

\begin{itemize}
\item \textbf{Comparisons with ISAC:} ISAC jointly designs RF sensing and communication, improving resource utilization through their coexistence and mutual assistance within the same hardware, spectrum and waveform\cite{Liu2022JSAC,JRCCLS_Survey,ISAC_survey2025}. {Both ISEA and ISAC view sensing as a built-in function of the mobile network, not an independent module built on separate devices that interface with the network's application layer. Also, both paradigms exploit radio-frequency (RF) sensing potentially with multi-functional waveforms in millimeter wave (mmWave) or terahertz (THz) frequency bands.
Positioning, ranging, tracking, and satellite-air-ground network are considered as important application scenarios by both ISEA and ISAC, as depicted in Fig.~\ref{fig: diff_scope}. }

Despite these similarities, ISEA and ISAC still exhibit fundamental differences. Sensing in ISAC almost solely refers to RF sensing, and the main motivation of ISAC studies is the integration of RF sensing and wireless communication systems due to their similar hardware and signaling. In ISEA, however, sensing is a broader concept not limited to RF sensing but includes multiple modalities such as Red-Green-Blue (RGB) cameras, LiDARs, and event cameras. This leads to several unique characteristics of ISEA. First, the rich but latent semantics embedded in these sensing modalities, e.g., images and videos, render the AI model indispensable for understanding sensing data and making decisions in ISEA, as compared to ISAC, which mainly focuses on RF signal processing. Moreover, ISEA requires a dedicated communication design for sensing tasks due to the high data dimensionality and the constant data exchanges among network nodes for computation offloading and cooperation. On the other hand, in ISAC, communication is generally not for exchanging sensory data but for normal data traffic coexisting with sensing. This results in distinctions between ISEA and ISAC in terms of key supporting techniques, as illustrated in Fig.~\ref{fig: diff_scope}. In terms of application scenarios, ISEA supports higher-level sensing tasks such as semantic segmentation and robotic planning and control due to multi-modality and AI capabilities, while ISAC mainly targets positioning, ranging, and physical layer applications, i.e., sensing-assisted communications.

\item \textbf{Comparisons with SemCom:}
SemCom goes beyond the traditional paradigm of building reliable bit pipes, aiming instead at the efficient transmission of information semantics via AI-based encoder-decoders and shared knowledge bases\cite{SemComSurv1,2021What}. The greatest common point between ISEA and SemCom is the native integration of AI models into the communication system to understand, compress and reconstruct semantics from multi-modal data for efficient communication beyond Shannon's limit.
{
As illustrated in Fig. \ref{fig: diff_scope},
both paradigms are empowered by a common set of enabling techniques, including wireless architectures for distributed inference and learning, semantic encoder-decoder frameworks, joint source-channel coding, and the adoption of foundation models and generative AI for enhanced semantic representation and generalization. These techniques collectively facilitate a shift from symbol-level fidelity to meaning-level accuracy, providing the platform for diverse human-machine interaction applications.
}

However, the goals of ISEA and SemCom are fundamentally different. SemCom aims at best semantic reconstruction over a transmitter-receiver pair under given semantic similarity metrics, e.g., sentence similarity\cite{DLSemCom1} for texts and DL-based similarity for images. In contrast, ISEA focuses on the effectiveness aspect of communication, i.e., the maximization of E2E sensing task performance, e.g., object detection precision and robotic object retrieval accuracy. In addition, while SemCom focuses on one or multiple communication links, ISEA requires the orchestration of a broader coverage of components and operations as involved by the sensing task, e.g., sensing signal processing, data storage, model fine-tuning with sensory data, etc. Further, the different scopes in terms of applications and techniques are illustrated in Fig.~\ref{fig: diff_scope}.

\end{itemize}

Overall, the proposed ISEA framework fundamentally differs from ISAC's resource-sharing paradigm (optimizing communication-sensing trade-offs) and SemCom's semantic compression (balancing bandwidth against contextual distortion). By jointly optimizing AI, communication, and sensing subsystems under an end-to-end performance criterion, ISEA transforms traditional component-level trade-offs - such as data delivery precision-latency compromise or model splitting's computation-communication balance - into tunable parameters for global optimization\cite{Model_Splitting_RRM_TWC2024}. This system-level approach enables superior adaptability where conventional paradigms must make hard resource partitioning decisions.

\begin{table*}[!ht]
\centering
\caption{Comparisons with Related Survey Papers}
\scriptsize
\begin{center}
\setlength{\tabcolsep}{0.8mm}{%
{
\begin{tabular}{|>{\centering\arraybackslash}m{0.08\textwidth}|>{\centering\arraybackslash}m{0.13\textwidth}|>{\centering\arraybackslash}m{0.14\textwidth}|>{\centering\arraybackslash}m{0.16\textwidth}|>{\centering\arraybackslash}m{0.14\textwidth}|>{\centering\arraybackslash}m{0.20\textwidth}|}
\hline
\textbf{Ref.} & \textbf{Sensing Modalities} & \textbf{Performance Metrics} & \textbf{Communication Techniques} & \textbf{Design Principles} & \textbf{Research Opportunities} \\ \hline
\cite{ISCC_Survey_GX} & RF signals, LiDAR, RGB cameras, CSI-based sensing & Task accuracy, latency, energy consumption & AirComp, JSCC, semantic communication & Task-oriented ISCC, semantic-aware optimization & Semantic-aware FL, edge coordination, and wireless–AI co-design \\ \hline
\cite{GreenEdgeAI} & Wearables, RF sensors, RGB cameras, LiDAR, ambient sensors & Energy efficiency, inference latency, accuracy & Gradient compression, quantized transmission, adaptive scheduling & Energy-first design, model–hardware co-optimization & Generative edge AI, neuromorphic computing, adaptive resource scheduling \\ \hline
\cite{IMMSC_Survey} & Radar, LiDAR, RGB cameras, GPS, IMU, RF signals & Beam prediction, channel estimation, sensing accuracy & Sensor-aided beamforming, communication-aware sensing & Cross-modal coordination, ANN-enabled fusion, data-centric design & Unified SoM framework, multi-sensor fusion, dynamic dataset integration \\ \hline
\cite{JRCCLS_Survey} & Radar, LiDAR, RF signals, ultrasonic sensors, RGB cameras & Sensing accuracy, coverage, E2E latency & Passive radar communication, cognitive radio, full-duplex, MIMO & Functional reuse, scalable JSAC, spectrum efficiency & RIS-enhanced JSAC, backscatter sensing, integrated localization \\ \hline
\cite{ISCC_Survey_DZ} & Radar, LiDAR, RGB cameras, mobile crowdsensing & Trade-offs among sensing, communication, and computation & AirComp, ISAC, triple-functional waveforms & Integrated waveform design, cross-layer orchestration & ISCC in digital twins, SAGIN, and computing networks \\ \hline
\cite{EdgeAIIoT_Survey} & LiDAR, radar, RGB cameras, wearables, industrial and environmental sensors & Latency, reliability, scalability, energy consumption & Bandwidth-aware compression, lightweight IoT protocols & Edge–cloud synergy, privacy-aware collaboration & Cross-layer scheduling, infrastructure-free AIoT, SAGIN-based fusion \\ \hline
\cite{CollabSense_Survey} & RGB cameras, RF signals, WSNs, mobile sensors & Coverage ratio, QoM, activation efficiency & Sensor scheduling, duty cycling, cluster-based communication & Coverage-aware coordination, role-based activation & Probabilistic models, adaptive full-view coverage, sensor selection \\ \hline
\textbf{Ours (ISEA)} & Fused multi-modal data, RGB camera, LiDAR, mmWave and UWB radar, event cameras & Task-oriented metrics: perception accuracy, E2E inference latency & Modality-aware JSCC, AirComp, short packet transmission & Task-performance co-optimization across sensing, comm, computation & Integration with foundation models, low-latency ISEA, convergence of ISEA and ISAC \\ \hline
\end{tabular}%
}
}
\end{center}
\label{table:comparison}
\end{table*}

Several surveys and reviews are related to the topic of ISEA \cite{ISCC_Survey_DZ,ISCC_Survey_GX,IMMSC_Survey,GreenEdgeAI, CollabSense_Survey, JRCCLS_Survey, EdgeAIIoT_Survey}.
Specifically, \cite{EdgeAIIoT_Survey} introduces the integration of AI, IoT, and edge computing, surveys its potential applications, and enables techniques for AI inference in IoT with end-edge-cloud orchestration as well as decentralized edge learning. In \cite{GreenEdgeAI}, the authors review energy-efficient design techniques for edge AI operations, i.e., data acquisition, training and inference, to fit AI capabilities into resource-limited wireless edge networks. However, these works \cite{EdgeAIIoT_Survey,GreenEdgeAI} focus on the general edge AI where sensing is one of its applications without a thorough discussion on sensing-oriented techniques. 
From the sensing perspective, \cite{CollabSense_Survey} investigates the design issues for collaborative sensing including sensing models, deployment and scheduling techniques, and metrics. However, the main focus is solutions for sensing coverage maximization, not covering wireless link issues or downstream inference. \cite{JRCCLS_Survey} discusses the advances in joint radar sensing and communications with applications to localization and sensing applications in IoT, which, however, does not involve other sensing modalities or AI-empowered data processing. The integration between RF sensing, multi-modal sensing and communication modules, named {Synesthesia of Machines} (SoM), is introduced in \cite{IMMSC_Survey}, which reviews current progress on the mapping between sensory data and channel data as well as mutual assistance of sensing and communications. Nevertheless, its main focus is the relationship and interaction between RF and other sensing modalities in the coexistence of communicational functions, not a generalized framework for efficient sensory data processing and exchange as in ISEA. 
The integration of sensing, communication and computation (ISCC), as reviewed in \cite{ISCC_Survey_GX,ISCC_Survey_DZ}, is an emerging framework exploiting coupling of the three operations for a wide range of performance metrics, e.g., 6G communication rates and E2E task latency. Compared with \cite{ISCC_Survey_GX,ISCC_Survey_DZ}, our paper focuses on the perspective on the native integration of AI models into sensing in edge networks oriented towards the performance of AI-empowered sensing tasks.} Our main contributions are summarized as follows.

\begin{itemize}
\item We first provide technical preliminaries on  ISEA from perspectives of sensing, edge AI and new communication techniques, and introduce the use cases and industrial and standardization progress of ISEA. Then, we give clear definitions of ISEA in terms of principles, metrics, tradeoffs, and architectures, highlighting the key differences between ISEA and conventional edge AI systems.

\item We provide the first comprehensive survey on how 6G wireless communications can support ISEA, including digital air interface for ISEA, AirComp-based air interface for ISEA, and advanced signal processing for ISEA. We focus on the adaption of a wide range of techniques, e.g., radio resource management (RRM), access control, and JSCC, to AI-empowered sensing tasks for optimized E2E performance. We further review how ISEA can be boosted by other trending advancements in 6G.

\item We point out several future research directions for ISEA, including the interplay between ISEA and foundation models, the convergence of ISEA and ISAC, and ultra-low-latency ISEA.
\end{itemize}

{
To highlight the key distinctions of ISEA from related surveys, Table~\ref{table:comparison} provides a comprehensive comparison in terms of sensing modalities, performance metrics, communication techniques, design principles, and research opportunities.
}

\subsection{Paper Organization}

\begin{figure*}
    \centering
    \includegraphics[width=1.6\columnwidth]{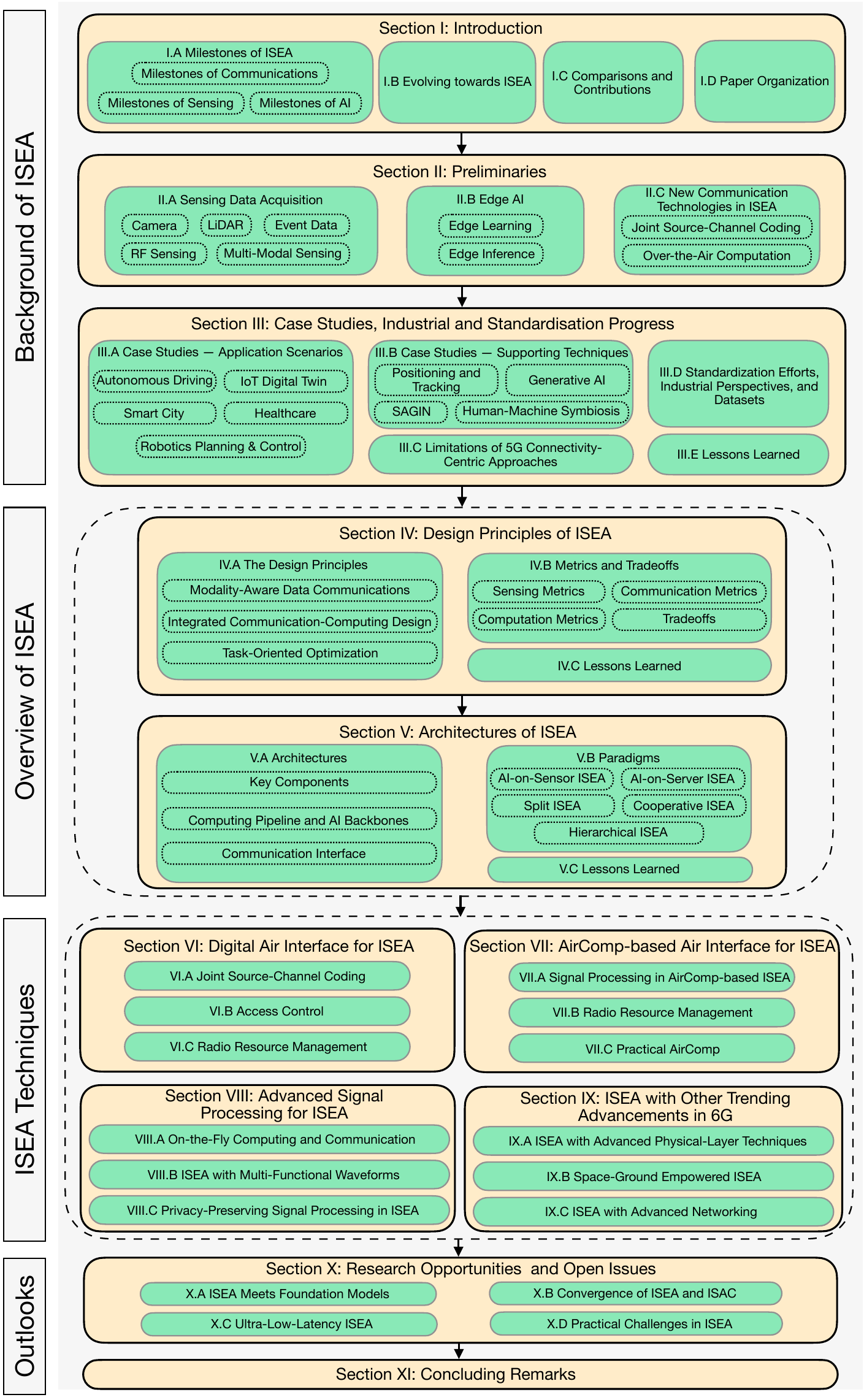}
    \caption{The outline of this survey. We first provide background information on the motivation, technical preliminaries and usage scenarios of ISEA, followed by an overview defining the design principles and architecture of ISEA. A comprehensive review of ISEA techniques is then presented, followed by future outlooks. }
    \label{fig: outline}
\end{figure*}

The remainder of this paper is organized as follows. We first introduce useful preliminary knowledge for ISEA involving sensing data acquisition, edge AI and new communication technologies, respectively in Section~\ref{subsection-preliminaries-A}, Section~\ref{subsection-preliminaries-B} and Section~\ref{subsection: preliminaries_C}. Then, in Section~\ref{section-case-studies}, we provide case studies for ISEA ranging from its application scenarios to supporting techniques that fit into the ISEA paradigm, investigate the standardization efforts for and industrial perspectives on ISEA, and introduce available datasets. The key design principles of ISEA are defined in Section~\ref{subsection-principles-A}, followed by the metrics and tradeoffs in ISEA in Section~\ref{subsection-principles-B}. The general architectures of ISEA are introduced in Section~\ref{subsection-architectures-A}, and several specific ISEA paradigms are sketched in Section~\ref{subsection-architectures-B}. ISEA techniques based on digital air interface are presented in Section~\ref{section-dai}, while those based on AirComp air interfaces are presented in Section~\ref{section-aircomp}. The incorporation of advanced signal processing techniques into ISEA is introduced in Section~\ref{section-advsp}. We discuss opportunities to boost ISEA using other trending 6G advancements and review relevant works in Section~\ref{section-6gadv}. Outlooks for research opportunities and open issues are provided in Section~\ref{section:outlook}. Finally, we conclude this paper in Section~\ref{section-conclusions}. For the readers' convenience, we have included Fig.~\ref{fig: outline} to illustrate the outline of this paper. 

\section{Preliminaries}\label{section-preliminaries}
In this section, we provide technical preliminaries underpinning the establishment of ISEA. We first present several prevalent modalities of sensing data acquisition, detailing their principles, data types, application scenarios, etc. Then, we introduce edge AI as the enabler for low-latency and high-reliability sensing tasks empowered by AI models in two aspects, i.e., edge learning and edge inference.

\subsection{Sensing Data Acquisition}\label{subsection-preliminaries-A}
\subsubsection{Camera}
Camera-based visual data is arguably the most prevalent sensing modality with a wide presence in various devices, e.g., automated vehicles, UAVs, and robots. 
A monocular RGB camera system generally comprises of a lens and an image sensor converting light intensity at the focal plane in different color channels into digital signals. Although usually associated with a low cost, a monocular camera can constantly stream high-resolution image data carrying texture and color information. Such information is critical for semantic-relevant sensing, including traffic-sign detection and person re-identification\cite{RN234}. However, the lack of depth information is an inherent drawback of monocular cameras. To resolve this issue, RGB-Depth (RGB-D) camera systems have been developed to endow the visual data with depth information, which can be realized by multiple techniques with different application scenarios\cite{RN165}. A classical technique is binocular camera system with two cameras simultaneously capturing two-dimensional (2D) images from different view angles, which is relatively robust to ambient illumination but incurs high computational complexity and calibration efforts\cite{aqel2016review}.  As another technique adopted by commercial products, e.g., Kinect V1 and Apple Face ID, infrared structured light with predefined patterns can be projected to the sensing target, which is reflected and detected by an infrared sensor for three-dimensional (3D) reconstruction\cite{geng2011structured}. Structured light-based sensing provides high-resolution depth estimates and can be operated in dark environments while being mostly limited to indoor scenarios due to the short perception range and strong ambient illumination interference in outdoor environments. Another key advantage of camera-based sensing is the extremely rich library of AI models operating on image or video data for a wide variety of downstream tasks.
\subsubsection{Light Detection and Ranging}
LiDAR is another prevalent sensing modality that exploits the reflection of emitted lights on targets for 3D ranging and imaging. Due to the short wavelengths of optical waves, LiDAR enjoys a high spatial resolution, which, along with its robustness to ambient lights, makes it a favorable choice for mission-critical or resolution-demanding applications such as autonomous driving and drone-based environment perception despite its relatively high cost. For example, Google's autonomous driving vehicle, Waymo, features a 360-degree LiDAR as the core of its sensor suite. DJI's latest drone model, Matrice 300 RTK, is also equipped with a LiDAR module for applications such as terrain modeling and accident scene mapping. The basic architecture of a LiDAR involves a transmitter emitting light pulses, usually lasers, towards the target and a receiver measuring the round-trip-delay of the reflected light, also known as time of flight, to determine the distance to the target\cite{behroozpour2017lidar}. Resolution and maximum sensing range are two important performance metrics. The typical form of LiDAR output is point cloud, which is a set of points with 3D coordinates and additional information such as reflection intensity.  State-of-the-art LiDARs with high spatial resolution and wide sensing range generate point-cloud data at a data rate up to several Gbps\cite{autolidar}, posing significant challenges to signal processing and data transmission among multiple agents. 
Traditionally, point cloud data are processed by manual feature engineering with expert knowledge, such as ground filtering to differentiate between ground and non-ground points and subsequent clustering of the latter into different objects\cite{li2020lidar}. However, with the emergence of AI, it has become a consensus that DL-based backbones will be fundamental in understanding LiDAR data for complex downstream tasks such as semantic segmentation and object detection. Celebrated DL backbones including VoxelNet\cite{zhou2018voxelnet} and PointPillars\cite{lang2019pointpillars}, take raw point cloud data as input and output spatial features consisting of one feature vector for each region in the regularly spaced grid of the sensing range, known as a voxel. The features can then be fed into a downstream inference head for particular tasks, such as region proposal networks for object detection. An important property of the spatial features is its sparsity in the voxel dimension, as a large portion of the voxels is empty, i.e., without data points, which results in all-zero feature vectors. 

\begin{figure}
    \centering
    \includegraphics[width=0.99\linewidth]{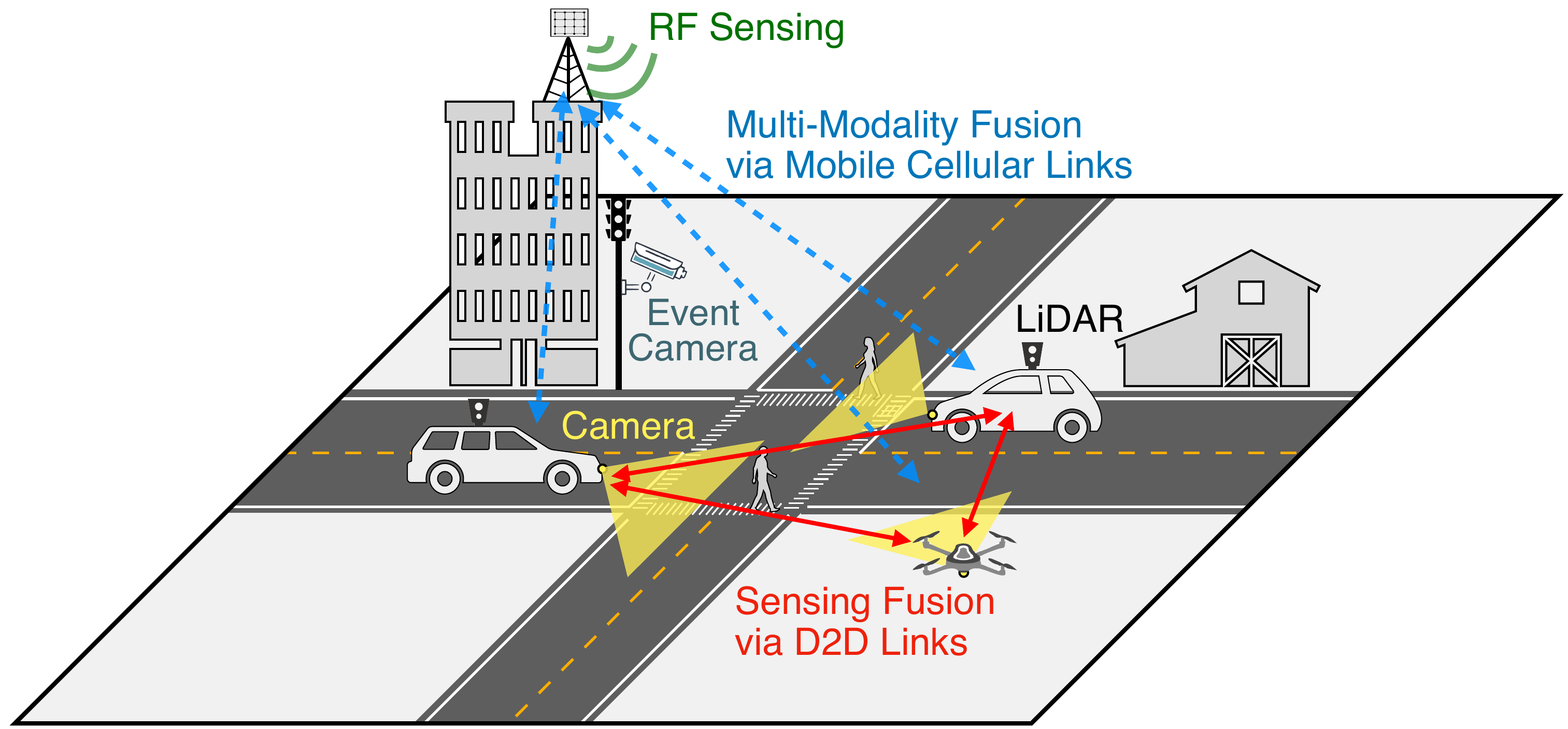}
    \caption{In an urban traffic scenario, multiple sensing modalities are installed, including RGB cameras, LiDARs, event cameras and radars. Multi-modal sensory fusion is achieved via device-to-device (D2D) and mobile cellular links such that each participant has a holistic view of all vehicles and pedestrians.}
    \label{fig: sensing_modalities}
\end{figure}

\subsubsection{Radio-Frequency Sensing}
RF sensing refers to collecting information about physical objects using RF waves. In RF sensing, RF signals are first emitted by a transmitter, then reflected by objects like vehicles, buildings, and people, and eventually captured by a receiver using RADAR technology~\cite{ISEA_ISAC2,behroozpour2017lidar}. The RF signals commonly used in RF sensing are composed of frequency-modulated continuous wave (FMCW), wireless fidelity (Wi-Fi), and ultra-wideband (UWB) signals. Specifically, FMCW radar is shown to be efficient and accurate in estimating the range, velocity, and angle information~\cite{RN142}. UWB Doppler radars are powerful in capturing vital signs of the human body, such as heart rate and respiration signals~\cite{UWB_tracking}. According to transmitter/receiver positions, RF sensing can be categorized into 1) passive sensing—the transmitter and receiver are located at different positions 2) and active sensing—the transmitter is the same as the receiver. Compared to light-based sensing, RF sensing can realize stable and low-cost identification and classification in non-line-of-sight (non-LoS) scenarios~\cite{ISAC_ISEA5}. At the same time, the broadcasting property of radio signal transmission enables RF-sensing to simultaneously monitor multiple objects and record changes in a vast region.
To this end, raw sensory data need to be processed to generate informative knowledge about objects. In this class, low-level features such as speeds, range, and angles can be efficiently extracted by using analytical methods including principal component analysis (PCA) and ambiguity analysis~\cite{Bekkerman2006}. Currently, extensive efforts are also made to leverage advanced DL methods to facilitate complicated RF-sensing applications. Exemplified by DL-based human activity identification, range-Doppler maps, micro-Doppler spectrograms (mDS), and spectrogram envelopes can be first extracted from raw RF data and are then fed into a bidirectional long short-term memory (LSTM) recurrent neural network to make classification decision~\cite{RN133,Micro_Doppler}. 

In 5G and beyond, RF-sensing and radar technologies are expected to be fully integrated within communication systems, termed ISAC. ISAC is deemed useful for improving utilization efficiency and reducing implementation costs, being adopted as a key technology in 6G~\cite{RN113}. It introduces new research opportunities for RF-sensing from three aspects: realizing communication by pure radar signals, executing sensing via communication signals, and dual-functional waveform design. The coexistence between radar and communication systems spreads over the spectrum from 1 GHz to 300 GHz and diverse architectures including MIMO radar, Orthogonal Frequency-Division Multiplexing (OFDM) systems, and mmWave communications. 

\subsubsection{Event Data}
Different from traditional frame-based cameras, which capture images by continuously sampling the scene at a fixed rate and storing the entire frame as a grid of pixels, event-based cameras capture and report the changes in light intensity at the pixel level in real-time.  
Specifically, the pixels in event-based cameras operate independently and asynchronously. 
Each pixel continuously monitors the light intensity changes at its location and generates an event when the intensity crosses a certain threshold. 
The event consists of the originating pixel's coordinates, timestamp, and polarity, indicating the trend of intensity, i.e., increasing or decreasing.
These components form a stream of event data that represents the visual information captured, which can be transmitted, processed, and analyzed to extract valuable insights or perform specific computer vision tasks, e.g., object tracking and optical flow estimation.
It is emphasized that event data is highly sparse compared with the aforementioned frame-based video data since the event-based camera produces data at a much lower rate.
The sparse nature contributes to its advantages of low power consumption, high temporal resolution, and high dynamic range, which benefits its processing in the considered next-generation wireless communication system as well \cite{Event_Camera_2021}.
The sparsity nature of event data can be utilized to compress data size for low-cost communication.

\subsubsection{Multi-Modal Sensing}
Multi-modal sensing is motivated by the fact that the above three sensing modalities have their advantages and inherent limitations, respectively. For example, LiDAR and RF sensing both provide precise 3D spatial information while lacking the rich semantic information (e.g., color and texture) provided by camera images. Exploiting complementary sensing features in multiple modalities is thus expected to boost sensing performance. Sensors with multiple modalities can co-exist on a single device, such as autonomous vehicles equipped with both LiDAR and cameras, where multi-modal fusion can be executed on board without incurring extra communication costs. In a more general scenario, sensors with different modalities are distributed to multiple devices in the network. For example, in an edge robotics system, the robots are equipped with LiDAR and a camera, while the BS is equipped with ISAC hardware that provides the RF-sensing modality. In this case, the communication cost is a critical aspect in designing the fusion strategy. 

\begin{figure}
    \centering
    \includegraphics[width=0.99\linewidth]{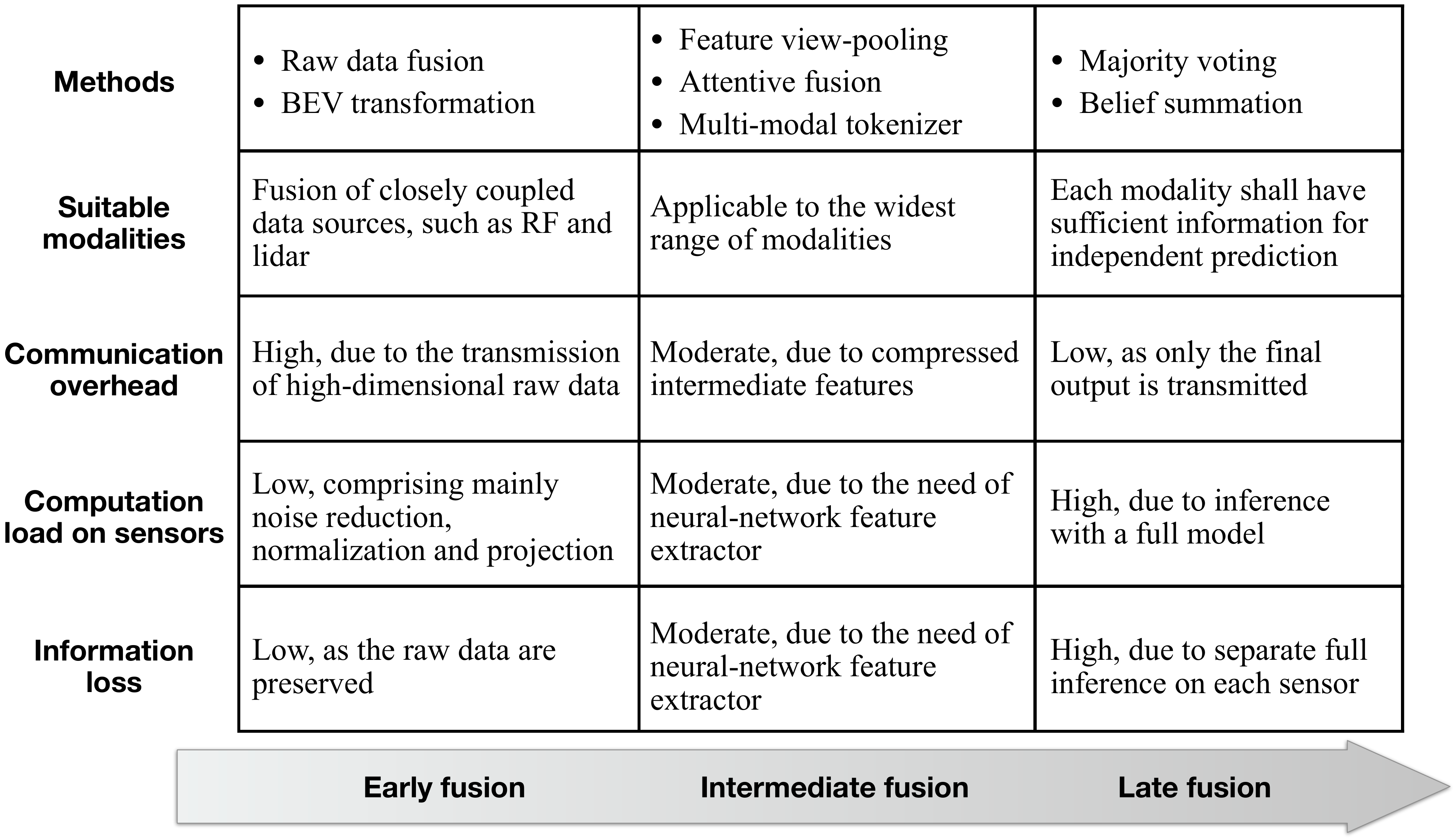}
    \caption{Different data fusion strategies with tradeoffs in versatility, communication overhead, computation workload, and information loss\cite{Fusion_survey1,Fusion_survey2,Fusion_survey3}.}
    \label{fig: fusion_methods}
\end{figure}

A rich literature exists on sensing models that exploit multiple sensing modalities to complete a downstream task. The fusion of LiDAR and camera image is arguably the most prevalent fusion strategy applied for various tasks including object detection, tracking, and semantic segmentation \cite{RN233,RN236,RN232,RN166}. 
The authors of\cite{wang2022multi} complement LiDAR with RF sensing to improve the detection performance. Depending on the model architecture design, the fusion methods can happen at different stages, e.g., early fusion at the raw-data level, intermediate fusion at the feature level, and late fusion at the decision level. Generally, modality fusion at higher levels results in a smaller communication volume for information exchange between sensors, while the information loss may increase due to the reduced data dimension. Detailed descriptions of these fusion methods and tradeoffs are shown in Fig.~\ref{fig: fusion_methods}.

\subsection{Edge AI}\label{subsection-preliminaries-B}
\subsubsection{Edge Learning}
Extensive studies have demonstrated the effectiveness of machine-learning algorithms in distilling knowledge from raw sensory data to support downstream sensing tasks like classification, semantic segmentation, and 3D stereo. In practice, however, the enormous sensory datasets are collected by edge IoT devices such as cameras, radars, and diverse sensors. Centralizing these local data at a cloud center for AI training and data analysis is undesirable due to the resultant privacy constraints and communication bottlenecks~\cite{RN184}. To tackle this issue, leveraging learning algorithms at the network edge to distill intelligence in a distributed manner to avoid direct transmission of raw data has emerged as a promising approach~\cite{RN184,chen2024Eng}. In this respect, federated learning (FL) has gained significant traction~\cite{CMZ2021JSAC,RN147,RN365}. FL builds upon the iterative stochastic-gradient descent (SGD) algorithm and involves three main operations: 1) local sensors train a local model using their data via SGD, 2) local sensors upload these intermediate results (e.g., gradients and network weights) over wireless links, and 3) a central server aggregates local results to update a global model and broadcasts the latest updates to local sensors. These operations iterate until convergence. FL methods have been well investigated in wireless systems by utilizing diverse air interface techniques, such as AirComp and broadband transmission~\cite{RN365}. However, empowering sensing with edge learning methods encounters several challenges that may not be adequately addressed by conventional algorithms. First of all, the presence of data and model heterogeneity across different sensors presents both opportunities and challenges for integrating sensing and edge AI. Consider a multi-view sensing scenario. Even when observing the same target, the variations in sensor spatial locations and hardware settings (e.g., RGB and depth cameras) naturally introduce non-identically distributed (Non-IID) properties and distinct representations to on-device data. Although the data inconsistency enables the trained models to acquire information from different domains by multi-view or multi-modal fusion to improve sensing performance, it also requires the design of edge learning algorithms considering more complicated operations including feature pooling, attention-based fusion, and spatial alignment~\cite{airfusion}.

\subsubsection{Edge Inference}
Whilst edge learning aims at distilling intelligence from data collected by edge devices, edge inference focuses on the efficient provision of well-trained AI models as services to heterogeneous edge devices under possibly stringent latency and reliability requirements. In contrast to the conventional cloud inference which places the intensive computation load of AI model inference at the central cloud, edge inference deploys computation resources on edge nodes in proximity to the device requesting service, thus promising lower round-trip latency and avoiding traffic jams in the backbone network. These advantages render edge inference as a key enabling technology for sensing with latency-demanding and mission-critical natures. Depending on the deployment of the model computation load, edge inference can be categorized into on-device inference, on-server inference, split inference, and distributed inference. On-device and on-server inference places model computation solely on end devices or servers, aiming at scenarios with extremely scarce communication resources or low-end on-device computation capabilities, respectively. Split inference is arguably the most prevalent approach where the AI model is split into a low-complexity device sub-model and a server sub-model due to its privacy-preserving nature and high flexibility. The device uses its sub-model to extract features from the raw data, which are uploaded to the server for further inference with the server sub-model. Therein, the feature uploading phase is often confronted with a communication bottleneck since the extracted features may have a higher dimension than raw data, especially when handling bursting arrivals of inference requests from multiple users. A set of techniques have been developed for split inference to boost its efficiency under QoS requirements, including communication and computation resource allocation\cite{RN155,Zhou2020IoTJ}, feature compression/pruning\cite{Niu2019Infocom,Deniz2020SPAWC}, JSCC \cite{Zhang2020CM}, progressive transmission\cite{LQ2021arxiv}, early exiting\cite{dualTSP,Chen2020TWC}, batch processing\cite{Co-inference_Shi,xu2025batching}, and model partitioning point selection\cite{Chen2021IOTJ}. Different from previous paradigms involving a device-server pair, distributed inference involves the distribution of computation load onto multiple edge devices to complete a single inference request \cite{RN156,edgeflow}, which usually requires the partition of an AI model into multiple independent execution units for distributed computing.
{
\subsection{New Communication Technologies in ISEA}\label{subsection: preliminaries_C}
The traditional design paradigm is based on the separation of source and channel coding, where the former handles data compression and the latter aims at errorless bit transmission over the air interface. However, this paradigm becomes sub-optimal as ISEA shifts the focus of communication from achieving high rates to attaining high performance for particular tasks with a low E2E latency, necessitating the incorporation of new communication technologies. Next, we introduce two representative communication technologies, which, in a general sense, target point-to-point and multiple access scenarios, respectively.

\subsubsection{Joint Source-Channel Coding}
JSCC aims to push the limits of separated source-channel coding, overcoming its drawbacks, e.g., cliff effect, in various 6G scenarios, including the ultra-low-latency regime, AI training/inference with E2E performance requirements, and knowledge base-enabled communications\cite{DenisJSCCSurvey}. Despite its recent revival, JSCC is a long-running research topic, but classical techniques primarily focused on statistical source distributions and joint tuning and decoding of separately designed source and channel codes. The emerging DL-empowered JSCC comprises an encoder that realizes the mapping from source to channel inputs and a decoder that takes in the channel inputs for information reconstruction for downstream tasks, while both are parameterized by neural networks \cite{DenizJSCC1}. Such a design enables E2E training of the JSCC framework towards maximizing a desired target over training dataset, where channels are considered as a frozen intermediate layer as opposed to the trainable encoder-decoder pair. Examples of E2E optimization targets include peak signal-to-noise ratio (PSNR) for images and multi-scale structural similarity index measures for videos. Advanced DL-based JSCC techniques involve adaptive rate control, finite-constellation implementations, etc. In ISEA systems, the design goal of JSCC shall go beyond the level of data/semantic reconstruction towards the effectiveness level, aiming at optimized downstream task performance. For example, the E2E communication of point-cloud features shall be optimized such that the downstream object detection model yields the highest precision instead of the feature reconstruction error.

\subsubsection{Over-the-Air Computation}
Over-the-air computation (AirComp)  leverages the waveform superposition property of wireless multi-access channels to realize the desired aggregation function of distributed data by simultaneous transmissions\cite{GX2021WCM}. Compared to orthogonal access where the communication latency grows with the number of devices, AirComp's latency scales by $\mathcal{O}(1)$ w.r.t. the number of devices due to simultaneous access, rendering it a promising solution for ISEA applications that often require low-latency data aggregation from many devices. The general architecture of AirComp is illustrated in Fig.~7~(a).
The typical procedure of AirComp can be elaborated by considering the scenario of value averaging where both devices and the server are equipped with a single antenna. First, aiming for analog wireless transmissions, each device normalizes its data into a zero-mean distribution, and then modulates the normalized values onto the magnitude of transmission symbols. Then, {zero-forcing} transmitter is applied at each device based on its transmit {channel state information} (CSI) to pre-compensate channel fading ~\cite{GX2019IOTJ}, which is also known as \emph{magnitude alignment}. Subsequently, the devices transmit simultaneously with symbol-level synchronization by e.g., time-advance techniques in 4G/5G standards such that the signal magnitudes sum up in the air. The server then receives and scales the aggregated signal to recover the averaged data. 

Beyond realizing data averaging, AirComp supports the computation of a family of aggregation functions, known as \emph{nomographic functions}, by adding due data pre-processing and post-processing at the device and server respectively. These functions include, for example, multiplication, geometric mean, and vector norms. Moreover, AirComp can be implemented in multi-antenna systems, e.g., single-input-multiple-output (SIMO)~\cite{Aggregation_gain_XChen} and MIMO~\cite{GX2019IOTJ}), further exploiting channel diversity to multiplex aggregations or reduce errors. }

\section{Case Studies, Industrial and Standardization Progress}\label{section-case-studies}
In this section, we conduct use case studies for ISEA, first introducing 6G application scenarios enabled by ISEA infrastructure and then several key supporting techniques for these scenarios as functions of ISEA. Next, we summarize the standardization efforts, industrial perspectives and datasets regarding ISEA development. 
\subsection{Case Studies -- Application Scenarios}
\subsubsection{Autonomous Driving}
Undoubtedly, the evolving capabilities of sensing and AI will play an indispensable role in next-level autonomous driving, which requires the vehicle to perceive the environment, analyze sensory data, and make decisions in a real-time, intelligent, and reliable manner. On the one hand, each vehicle itself is equipped with moderate-scale AI models to fuse and understand multi-modal sensory data, supporting subsequent decision-making. This technique is known as single-vehicle intelligence. On the other hand, vehicle-to-vehicle (V2V) and vehicle-to-infrastructure (V2I) cooperation is envisioned to solve several critical challenges including object occlusion, extreme weather, contesting, etc. A typical scenario of V2V collaboration involves multiple vehicles communicating over wireless links to exchange sensory information in a shared feature space, e.g., bird's-eye view space, to detect occluded vehicles. V2I generally involves a roadside server assisting vehicles with localization, perception, and coordinated decision-making. ISEA in cooperative autonomous driving aims to achieve integrated sensing and networking of intelligent vehicles and servers under stringent latency, reliability, and E2E accuracy demands by orchestrating AI computation and sensory data communications.

\subsubsection{Robotics Planning and Control}
State-of-the-art robotics relies heavily on AI models to understand human languages, perceive the environment, make decisions, and interact with the environment and user\cite{RN509}. Specifically, natural-language instruction is transferred by language models into semantic representations of the physical world. Meanwhile, sensory data captured by various sensors such as cameras, RGB-D sensors, and range sensors are analyzed by sensing models to create a high-dimensional feature space that encodes the objects and environments perceived by a robot. By projecting the instruction semantics into the sensory space, the robot can associate certain perceived patterns with the instruction and control its actuators to accomplish the task. The heavy computational load of large language and sensing models renders pure on-robot model deployment infeasible under battery life constraints. Thus, robotic task accomplishment frequently involves sending sensory data and input prompts to edge servers in proximity to robots for low-latency remote execution. Beyond the single-robot workflow, another emerging paradigm is multi-robot collaboration via edge servers for localization and perception, where each robot sends sensory data generated from LiDAR, cameras, inertial measurement unit (IMU), etc., to the edge server for fusion into holistic perception results. 

\subsubsection{Internet-of-Things Digital Twin}
Digital twin involves creating a software replica of a real-world entity for testing, monitoring, and behavior prediction. Some typical scenarios include virtual sensors for industrial production, digital patients for health monitoring, and digital city. The digital replica will be created from real-time data collected by IoT sensors. Meanwhile, the critical role of AI models has become a consensus in various aspects of digital twins, e.g., virtualization models for mirroring sensory observation to virtual representation and analytics models for calculating physical motion and predicting future behavior\cite{RN511}. However, digital twin is challenged by the computation-intensive nature of such models and low latency required by real-time data analytics, for which ISEA can be a promising solution by offloading sensory data to the edge cloud for real-time processing.


\subsubsection{Smart City}
Enabled by massive IoT sensors and AI-based intelligent systems, smart city is envisioned as the solution to many urban challenges\cite{RN204}. For example, the concept of ``City Brain''\cite{citybrain} is reshaping city management through streamlined sensing, data analytics, and decision making. With the development of inter-domain data sharing and multi-modal data fusion, City Brain is evolving from local intelligence to a general intelligent engine for use cases ranging from traffic control to emergency response to city planning. It relies on edge AI for real-time intelligent analytics of sensory data collected by millions of sensors deployed in smart cities, such as visual cameras, thermal cameras, wearable devices, etc., and subsequent decision-making. With C$^2$ integration and cross-layer synergy, ISEA provides the infrastructure for efficient selection, transmission, and analytics of a massive amount of sensory data in fast response time as typically required by smart city applications. For example, assisted by on-sensor AI models, the transmission of sensory data to the edge server is triggered with importance awareness such that emergency events are reliably detected with minimal network traffic. 

\subsubsection{Healthcare}
Smart healthcare is empowered by data from a variety of sensors ranging from wearable devices to patient monitors to surveillance cameras, which are widely deployed in hospitals, homes, and elderly communities\cite{RN188}. The collected data, typical in multiple modalities, are analyzed via AI models to monitor health conditions, recognize human gestures, respond to emergencies, and combat epidemic outbreaks\cite{RN195}. The privacy sensitiveness of medical data, extensive bandwidth costs, and stringent latency constraints require the AI model computation to be deployed on edge servers instead of remote cloud centers. ISEA provides a versatile framework for aggregating and analyzing multi-modal data from healthcare sensors with highly heterogeneous communication and computation capabilities. 

\subsection{Case Studies -- Supporting Techniques}
\subsubsection{Positioning and Tracking}
The fields of positioning and object tracking feature a rich set of technologies that collectively enable precise localization and continuous monitoring of objects with high accuracy, high reliability, and low latency. Advanced positioning systems include satellite-based systems like GPS and BeiDou for global navigation~\cite{RN214} and terrestrial technologies like radio frequency identification (RFID), Bluetooth Low Energy (BLE), and (UWB) for indoor positioning and tracking~\cite{UWB_tracking,RFID}. AI-based positioning and tracking are increasingly employed in applications with complex environments, multi-modal data (e.g., the fusion of visual information and LiDAR~\cite{RN213,RN221,RN216,RN219}) and high-precision requirements like UAV swarms and autonomous driving~\cite{RN208,RN210}. ISEA is deemed a promising framework contributing to real-time intelligent positioning and tracking by offering low-latency access to edge AI resources and support for cooperation between devices.

\subsubsection{Generative AI}
Generative AI is expected to be native in 6G mobile networks for automating a broad range of tasks in mobile applications including human semantic communications, personal assistants, auto-pilot, and robotic control. The ability of generative AI to comprehend and respond to nuanced inputs makes it a powerful tool for ISEA, supporting sensory data processing and automating ISEA operations. First, with its general intelligence capabilities, generative models can support versatile sensory data understanding and decision-making, adapting to new tasks in few-shot and zero-shot cases. Second, generative AI holds the potential to realize automated ISEA operations, e.g., dynamic scheduling and sensor access control adaptive to the system status. For example, in a mobile network, multiple sensors need to access a server for cooperative field perception and multi-modal fusion. Generative AI, aided with chain-of-thought (CoT) prompting and multi-modal adaptors, can analyze network status and patterns in multi-sensor observations to provide predictive modeling and real-time actuation for content-driven resource allocation and energy management among sensor devices. In addition, the development of generative AI is expected to benefit from access to a huge number of sensing datasets. Further, personalized sensory data allows for task-oriented fine-tuning of a general large language model, making it more efficient in specific application scenarios. The process of data provisioning and training shall be supported by ISEA with techniques such as device-edge cooperative fine-tuning.

\subsubsection{Satellite-Air-Ground Integrated Network (SAGIN)}
SAGIN features the integration of satellite systems, aerial networks, and terrestrial communications, targeting large coverage and strong resilience~\cite{RN357}. The majority of research efforts focus on the cooperative execution between heterogeneous systems in SAGIN, which includes both joint network design and protocol optimization~\cite{RN367}. On the other hand, lifting AI above the ground is becoming attractive for supporting applications requiring expansive field perception and intelligent task execution, such as intelligent transportation systems, military missions, and disaster rescue~\cite{RN385}. This leads to a natural fusion between ISEA and SAGIN. In this case, due to the distinct levels of coverage (and thus the available sensing range) offered by different components in SAGIN, the architecture of ISEA is reshaped into a hierarchical one that consists of three levels: 1) edge devices/servers (e.g., sensors, smart devices, and edge servers) serve as the edge for ground systems, 2) UAVs form the edge for air-ground systems, and 3) satellites act as the edge for space systems. Simultaneously, the components of  AI models will be employed at different levels of the network edge to execute their corresponding functionalities (e.g. compression, feature extraction, fusion, and prediction) according to the task requirements and resource availability. Despite the flexibility achieved by SAGIN resilience, heterogeneity exists in both sensory datasets, computation capability, and data rate, which poses several challenges to the materialization of ISEA in SAGIN.

\subsubsection{Human-Machine Symbiosis}
Advanced sensor technologies, such as IoT devices, wearable sensors, and smart environments, enable a seamless flow of data between humans and machines. These techniques, coupled with AI-based data analytics, offer personalized and responsive interactions for cooperative interaction and close coupling between humans and machines~\cite{2021What}, known as human-machine symbiosis. Several practical examples illustrate how AI-sensing integrated techniques are intricately woven into various facets of our lives, enriching human experiences and decision-making. For instance, leveraging sensory data (e.g., based on motion and temperature sensors, thermostats, cameras, or wireless sensing) within smart homes enables automation based on the recognition of human behavior (e.g., voice and gestures), adjusting the environment according to occupants' preferences and optimizing energy efficiency and comfort~\cite{2021What}. ISEA, propelled by and designed for low-latency intelligent sensory data analytics, can serve as a powerful platform for collectively implementing sensing techniques and AI to offer real-time human-machine symbiosis services.

\subsection{Limitations of 5G Connectivity-Centric Approaches}
{

The conventional 5G New Radio (NR) architecture, while optimized for high-throughput communication, exhibits fundamental inefficiencies when applied to integrated sensing, AI computation, and communication. These limitations stem from three key aspects:

\begin{itemize}
    \item Functional isolation: 5G NR employs distinct waveform designs for communication (e.g., OFDM) and sensing (e.g., sparse reference signals), leading to: 1) Resource inefficiency: Redundant pilot transmissions for channel estimation (communication) and radar-like sensing result in extra overhead in joint positioning and tracking scenarios; 2) Latency bottlenecks: AI-driven applications (e.g., generative AI) require real-time sensor-data fusion, yet 5G’s sequential processing (sense → transmit → compute) introduces prohibitive delays for time-sensitive inference tasks.
    \item Rigid resource allocation: 5G’s slot-based scheduling lacks the agility to dynamically optimize: 1) Time-frequency resources for ultra-low-latency tracking (e.g., µs-level updates in drone swarms); 2) Computing-aware transmission for distributed AI workloads (e.g., split inference across devices and edge servers).
    \item Semantic-agnostic operation: Current protocols lack native support for: 1) Task-oriented QoS metrics: Traditional bit-error-rate (BER) optimizations do not align with perceptual quality in generative AI or sensing accuracy in SLAM applications; 2) Cross-domain tradeoffs: Independent optimization of communication capacity and sensing resolution (Cramér-Rao bound) leads to suboptimal performance in integrated networks.

\end{itemize}

}
\subsection{Standardization Efforts, Industrial Perspective, and Datasets}
\subsubsection{Standardization Efforts}
Substantial efforts are made by international standardization organizations like ITU, IEEE, and 3GPP, to promote the development of ISEA-involved specifications. To be specific, the ITU’s Telecommunication sector (ITU-R) recommended two new usage scenarios in June 2023 for 6G, namely IAAC and ISAC, and initially offered a roadmap for technology/standard development in its report ITU-R M.2516. Other ITU sectors like ITU’s Telecommunication standardization sector (ITU-T) have also developed a set of recommendations including ITU-T Y.4207, Y.4213, Y.4216, Y.4420, and X.1363 to describe and initialize requirements, capabilities, and frameworks of IoT sensing within different scenarios. At the same time, IEEE has several existing sensing-related standard activities like Sensor Performance and Quality and Smart Manufacturing and Smart Factories, where IEEE 2700 and P2806 propose a framework for sensor performance specification terminology and digital representation of physical objects, respectively. IEEE also emphasizes specific endeavors related to sensor interfaces for cyber-physical systems within its IEEE 2888 standards family. These initiatives encompass defining sensor interfaces for both the cyber and physical realms, establishing standards for actuator interfaces, and coordinating digital synchronization between the cyber and physical domains. In the technical reports released by 3GPP, such as 3GPP TR 23.700-80 and 3GPP TR 22.874, ISEA is introduced as a vital component associated with specific performance requirements for different 5G assistance applications, including FL, robotic control, and automotive networks. Moreover, there exist other important standardization activities and events that will provide key functional components and technical guidance for ISEA. For example, the Internet Engineering Task Force (IETF) is looking at a data format to represent sensor measurements based on its packet transmission protocol RFC 7428. A project initialized by the International Electrotechnical Commission (IEC), called PWI TR JTC1-SC41, is developing several components of the correspondence measure in digital twins such as similarity, resolution, latency, level of detail, which can be considered as performance metrics in ISEA.

\subsubsection{Industrial Perspective}
The telecommunications industry widely recognizes sensing and edge AI as key functions of 6G networks and invests consistent efforts for their materialization. For edge AI, Qualcomm envisions connected edge intelligence with sensor fusion and edge analytics to enable ultra-low latency and privacy-preserving services\cite{Qualcomm2022}. As concrete progress, it has released several models of edge AI stations, named Edge AI Boxes, with dedicated chips integrating wireless communication modules and AI accelerators. These products target various use cases, such as surveillance cameras, smart manufacturing, smart cities, and autonomous driving, which require real-time analytics of massive sensing data. Huawei has also rolled out Atlas 500 AI Edge Station designed for edge applications. The leading semiconductor companies are increasing their focus on edge AI and sensing chipsets, including the NVIDIA Jetson series and the Intel Movidius vision processing units, to realize AI computing on edge servers/devices for intelligent sensors, like smart cameras. In the automotive industry, it has become a consensus that sensors distributed on multiple vehicles and infrastructure units should be connected and cooperate for improved situation awareness (see, e.g., the standards J3216 of SAE International). Leading companies in the auto-driving industry, e.g., Volkswagen, Huawei, and BYD, are interested in cooperative sensing enabling techniques ranging from sensor fusion to V2V and V2I data communications~\cite{HuaweiIntelligentNetwork}. AI models powered by multi-source data and real-time edge computing are expected to enhance autonomous driving capabilities significantly.

\subsubsection{Datasets}
As AI-empowered sensing draws growing attention, abundant datasets are available for benchmarking the E2E performance of ISEA protocols and control algorithms on various tasks with different modalities. For object detection tasks, KITTI dataset \cite{RN237} provides real-world visual, LiDAR, and GPS/IMU data recorded by sensors installed on a test vehicle while providing object annotations. For cooperative sensing between vehicles, OPV2V\cite{RN170} features sensing data simultaneously collected from multiple vehicles while providing similar sensing modalities as KITTI, and is thus suitable for testing the impact of feature fusion, sensor selection and multi-access on detection accuracy. For multi-sensor object recognition tasks, the ModelNet10 \cite{modelnet_paper} and ShapeNetCore \cite{shapenet2015} datasets provide accurate 3D models for object samples divided into 10 and 12 classes, respectively. For each object sample, multiple 2D images from various angles can be rendered to be used as captured images by sensors. Beyond traditional sensing modalities, several event camera-based datasets are available, e.g., MVSEV\cite{mvsev} and M3ED\cite{m3ed}, both providing real-world event camera data captured by multiple devices ranging from ground vehicles to UAVs. The types of ground-truth data include reference poses, depth image, point clouds, semantic segmentations, etc., enabling benchmarking a variety of cooperative sensing tasks such as video reconstruction, activity detection, and semantic segmentation. While the abovementioned datasets focus on sensory data and do not contain wireless channel measurements, most provide sufficient information of the propagation environment, e.g., scatterers and propagation distance, to model channels between user devices and edge servers.

{
\subsection{Lessons Learned}
\begin{itemize}
    \item A wide variety of emerging 6G applications and techniques requires intelligent perception for environment awareness, semantic understanding, and rapid response. Their real-time and mission-critical nature results in unprecedented QoS requirements, including millisecond-level communication-and-computation latency, ultra-high reliability in E2E task performance, and scalability to thousands of cooperating devices\cite{3gpp.22.874}. 
    \item These requirements cannot be satisfied with existing 5G networks, which focuses on creating high-throughput bit pipes but are suboptimal in E2E task metrics.
    \item ISEA, on the contrary, provides a versatile framework for intelligent perception provisioning in edge networks with E2E optimality. Standardization organization and industry leaders have recognized the potential of ISEA and initiated projects and prototypes that pave the way for real-world implementations of ISEA\cite{Qualcomm2022,HuaweiIntelligentNetwork}. Furthermore, a variety of datasets \cite{RN237,RN170,modelnet_paper,shapenet2015,mvsev,m3ed} have been made available to simulate multi-modal sensing data acquisition, mobility, and wireless propagation in real-world environments, facilitating benchmarking of ISEA performance. These insights indicate that ISEA represents a promising research direction in the context of 6G technologies.
\end{itemize}
}

\section{Design Principles of ISEA}\label{section-principles}
The existing communication system, which is intrinsically modular and rate-oriented, is insufficient for meeting the demanding QoS requirements of emerging AI-empowered sensing applications in terms of, e.g., accuracy, reliability, and latency, calling for the integration of sensing and edge AI. In this section, we first sketch the design principles of ISEA with a detailed breakdown into three key aspects. We further present the task-oriented metrics from sensing, communications, and computing perspectives, followed by a discussion on the fundamental tradeoffs in ISEA. 
\subsection{The Design Principles}\label{subsection-principles-A}
ISEA aims to integrate sensing, AI, and communications to achieve optimized task performance. The general principle of ISEA design is \emph{to process and transmit task-relevant sensory information for various applications with improved accuracy, latency, and energy efficiency}. It aligns with not only task-oriented communications of 6G but also the design principle of E2E perception applications. In the sequel, we break down the general principle into several key aspects that distinguish ISEA from conventional rate-oriented mobile communication systems and other emerging paradigms such as semantic communications and ISAC.

\subsubsection{Modality-Aware Data Communications}
The conventional digital mobile networks designed under the rate-maximization design principle aim at transmitting each data bit without awareness of data content. However, the ultimate goal of ISEA is to efficiently convey sensory information, which, as a consequence, shall be designed in light of the modality, embedded semantics, and inherent characteristics of sensory data. Specifically, ISEA is envisioned to convey the critical semantics in sensory data for higher communication efficiency, which can be achieved by techniques such as projection onto a global Bird's-eye-view (BEV) coordinate with lower dimensions, shared knowledge base between transmitter and receiver (e.g., known information of the sensing environment), and AI-based cross-layer coding design. For example, for an event detection task with millions of urban sensors, ISEA protocols shall be tailored to exploit its time-domain data traffic to balance efficiency and reliability, where sensing data is typically sporadic but can surge if an event is detected. Also, certain sensing features, e.g., voxel features, show structured sparsity in dimensions with real-world correspondence, which can be adjusted adaptively to reduce the communication cost by orders of magnitude. 
    
\subsubsection{Integrated Communication-Computing Design} Conventional mobile networks generally adopt a communication-computing separated design paradigm, being agnostic of the downstream AI models. On the contrary, ISEA adopts an integrated communication-computing approach by designing transmission protocols and optimizing resource allocation for AI-empowered sensing models, evolving from MEC that assumes generic computation models. An important theme of ISEA is thus designing communications to accommodate the need for sensing AI model computation. For example, the general-purpose backbone model for sensory data analytics for various downstream tasks, usually computationally intensive, can be accelerated by parallel processing of a batch of sensory data from multiple devices, requiring dedicated design of caching protocols and resource allocation. Also, certain characteristics of sensing AI model computation can be leveraged, and some components can be integrated with communication modules to boost the system efficiency, e.g., exploiting the inherent robustness of AI models for data compression, unequal feature protection, joint design of transceiver and feature coding, and over-the-air sensory data fusion. 
    
\subsubsection{Task-Oriented Optimization} The ultimate goal of ISEA design and optimization is to maximize the sensing task performance, as elaborated in the following subsection. Achieving this requires E2E system optimization crossing the physical and application layers under sensing QoS constraints. While conventional metrics of communication systems, e.g., latency, can still serve as useful tools, their relationships with the downstream task performance shall be re-established in the sensing context. A notable example is accelerating model training/inference in the presence of channel noise to maximize the detection accuracy, e.g., classification accuracy or intersection over Unions (IoUs), on collected sensory data. 

\subsection{Metrics and Tradeoffs}\label{subsection-principles-B}
The development of ISEA relies on cross-disciplinary architecture designs that involve the concepts of sensing, communication, and computing. Consequently, the ISEA performance can be comprehensively characterized from these three aspects with the metrics introduced below.
\subsubsection{Sensing Metrics}
Hierarchical performance evaluation can be utilized in ISEA, which involves task-level accuracy, estimation error, range and resolution, and timeliness, as elaborated below. 
\begin{itemize}
    \item \textbf{Accuracy/uncertainty:} Downstream sensing tasks largely encompass target classification, detection, and semantic segmentation. The primary performance metric in this domain is accuracy, which quantifies the probability of correct classification or the probability of error in hypothesis testing. In ISEA, accuracy consistently serves as the key objective function for optimizing E2E algorithms and framework designs. Moreover, a strong duality exists between accuracy and inference uncertainty, which is defined as the entropy of posteriors of classification classes given observations \cite{TIT_Entropy_Err,LQ2021arxiv}. The good analytical properties of inference uncertainty make it an alternative for sensing performance evaluation in both theoretical analysis and engineering practice.
    \item \textbf{Estimation error:} Parameter estimation and scene reconstruction play an important role in supporting sensing applications. The resulting performance in these scenarios can be evaluated by quantifying the gap between the estimated results and their ground truth into the metrics of {mean square error} (MSE), ambiguity functions, dissimilarity (e.g., Hausdorff distance), and mean intersection over union (mIoU). These metrics have been widely used in sensing waveform design (e.g., the dual-functional waveform design in ISEA), data fusion strategy, and point cloud processing. It has also been shown that estimation error can be connected to accuracy according to the concepts of classification margin and perturbation theory. Additionally, estimation error can be described through information metrics for sensing tasks with generative modeling, such as the generative point cloud learning in multi-robot field perception.
    \item \textbf{Coverage and Resolution:} The sensing coverage is defined as the union of reliable sensing range by each connected sensing node. It is jointly determined by network connectivity, capabilities and allocated resources of sensors, and occlusions. Resolution, on the other hand, encompasses both angle and ranging resolution: angle resolution refers to the minimum angle interval between two resolvable objects while ranging resolution denotes the minimum distance between them. For RF sensing, it has been shown that angle and range resolution are inversely proportional to the effective aperture of the sensing array and the leveraged bandwidth, respectively. 
    \item \textbf{Timeliness:} Continuous detection and tracking are crucial for enabling time-sensitive applications such as autonomous driving and healthcare. It requires timely observation and understanding of dynamic environments. Timeliness quantifies the performance of these sensing tasks in terms of the quality of the environment sensing and analysis being done at a favorable time slot. Particular timeliness metrics include {age of information} (AoI), rate of false alarms, and track latency, etc.
\end{itemize}
\subsubsection{Communication Metrics}
In the context of ISEA systems, several communication metrics defined in traditional wireless networks are expected to be adopted and extended for performance evaluation. These metrics include:
\begin{itemize}
    \item \textbf{Throughput:} As one of the most important metrics in traditional communication systems, throughput reflects the efficiency of bit stream transmission. Broadly, it relates to the data rate metric during system optimization, which quantifies the number of bits reliably transmitted over a unit bandwidth per second.  The well-known Shannon equation has explicitly revealed the relationship between data rate and relevant communication parameters including bandwidth, signal-to-noise-ratio (SNR), multi-antenna precoding, and beamforming.  In ISEA, communication throughput extends beyond merely achieving reliable bit transmission. It encompasses the transmission efficiency of semantics associated with specific sensing tasks. For instance, in multi-view sensing, AirComp-based feature pooling exemplifies how local features can be fused using noisy transmission to significantly enhance transmission efficiency while maintaining comparable sensing accuracy to scenarios with reliable transmission. Furthermore, ISEA emphasizes the effectiveness of sensing execution, prompting throughput measurements to adopt an importance-aware approach. This approach ensures that the transmission of critical semantic information is prioritized, aligning with the overarching objectives of the sensing task at hand.
    
    \item \textbf{Latency:} Communication latency in ISEA mainly comprises the time usage of the exchange and aggregation of raw sensory data and intermediate results (e.g., features and model parameters), although the control signaling latency also becomes non-negligible for ultra-low-latency applications. Limited communication resources and hostile wireless channels often exacerbate this challenge, particularly in scenarios with extensive access demands. To meet the stringent timeliness constraints inherent in ISEA, optimizing communication latency necessitates a holistic approach that integrates the design of air interface techniques and computation strategies. 
    
    \item \textbf{Energy efficiency:} Energy consumption emerges as a critical domain in ISEA design, as it directly influences both the system's lifetime and implementation costs. The overall energy usage in an ISEA system consists of the power consumption for sensing, computation, and communications. Energy efficiency can be used to evaluate energy consumption, which is defined as the ratio of the attained performance gain to the total consumed energy. To ensure a high level of energy efficiency, power consumption needs to be minimized via power control strategies and resource management subject to resource constraints and sensing operations.
    
    \item \textbf{Mobility:} Mobility management is another important domain concerned by ISEA due to the natural movements of both sensors and targets. Unlike traditional takeover strategies employed in cellular networks, which primarily revolve around device mobility and base station (BS) coverage, mobility management in ISEA necessitates a more nuanced approach. Specifically, ISEA mobility management must consider not only the movement of sensors but also the association between sensors and targets. This association is guided by the observation capabilities of sensors, including fields of view and equipment types. By incorporating these factors into mobility management strategies, ISEA can optimize sensor-target associations to maximize sensing effectiveness and ensure accurate data collection in dynamic environments.
\end{itemize}

\subsubsection{Computation Metrics}
The execution of ISEA entails a collection of computing operations, ranging from pre-processing raw sensory data to feature extraction to data fusion. The efficiency of these computations is characterized by the algorithms' time and space complexity, as introduced below.
\begin{itemize}
    \item \textbf{Time complexity:}  Time complexity, which can be quantified by the required floating-point operations per second (FLOPS), denotes the running time of an algorithm within given computing resources, such as CPU frequency. Integrated with communication latency into overall ISEA timelines, time complexity and its associated computing latency serve as key metrics for system optimization, influencing decisions regarding task offloading and resource management. 

    \item \textbf{Space complexity:} Conversely, space complexity focuses on the implementation of learning algorithms, particularly being concerned if high-dimensional data reading and writing occur. For example, employing a complex neural network on a lightweight device may be undesirable due to the frequent high-dimensional data operations it requires and the resulting memory requirements. To address storage constraints, it is essential to design low-complexity algorithms that strike a balance between performance and efficiency, ensuring optimal system operation while conserving resources.
\end{itemize}

\subsubsection{Tradeoffs}
Although there may be consensus on optimizing certain metrics in ISEA systems, such as improving ISEA timeliness by minimizing both communication and computation latency, the tradeoffs between the aforementioned metrics often manifest in an impossible trinity: sensing performance, communication efficiency, and computation efficiency. Specifically, achieving high energy efficiency and low latency while simultaneously maintaining optimal sensing accuracy is often infeasible. This inherent trade-off arises because of the limited system resources. For example, optimizing communication efficiency and latency may involve aggressive data compression to remote servers, potentially compromising sensing accuracy. In contrast, prioritizing high sensing accuracy may necessitate more extensive computations and data transmissions, leading to increased energy consumption and resource requirements. As a result, designers of ISEA systems must carefully balance these competing objectives based on the application's specific requirements and available resources. This may involve employing novel signal processing techniques, adaptive algorithms, and resource allocation strategies that dynamically optimize system performance within the constraints of the impossible trinity.

\subsection{Lessons Learned}
\begin{itemize}
    \item  Different from the rate-maximization principle in conventional mobile networks, ISEA aims at optimized E2E task performance, e.g., sensing accuracy, E2E latency, and energy efficiency. This involves the coordinated integration of sensing, AI, and communication to efficiently process and transmit task-relevant sensory data.
    \item Specifically, communication links are designed to be aware of the semantics and characteristics of the conveyed sensing data. By leveraging the interaction between these factors and the time-varying network connectivity, ISEA implements modality-aware communication schemes that outperforms modality-agnostic approaches in efficiency. Second, ISEA adopts an integrated communication and computing design to optimize resource allocation and develop transmission protocols tailored for AI-empowered sensing models. Finally, all ISEA designs and optimizations target the E2E sensing task performance, allowing for the utilization of challenges in rate-oriented communication systems, such as channel noise and interference, to benefit downstream tasks. 
    \item Following the task-oriented principle, the ISEA performance is evaluated with a broad set of metrics that go beyond the scope of existing communication systems. Notably, sensing metrics play a crucial role in ISEA design, encompassing accuracy/uncertainty, estimation error, coverage, resolution, and timeliness. Communication metrics, such as throughput, latency, energy efficiency, and mobility, are also integral to ISEA but are adapted to align more closely with downstream tasks. For instance, throughput in ISEA extends beyond mere transmission efficiency to encompass rates of sensing semantic transmission\cite{JSCC_Imp2} and task completion\cite{RN155}. Additionally, computation metrics like time complexity and space complexity must be factored in for end-to-end system functionality. While achieving high performance across all metrics may not always be achievable, ISEA designers must strike a balance among these metrics within resource constraints to meet the needs of specific tasks.
\end{itemize}

\section{Architectures of ISEA}\label{section-architectures}

\begin{figure}
    \centering
    \includegraphics[width=0.95\linewidth]{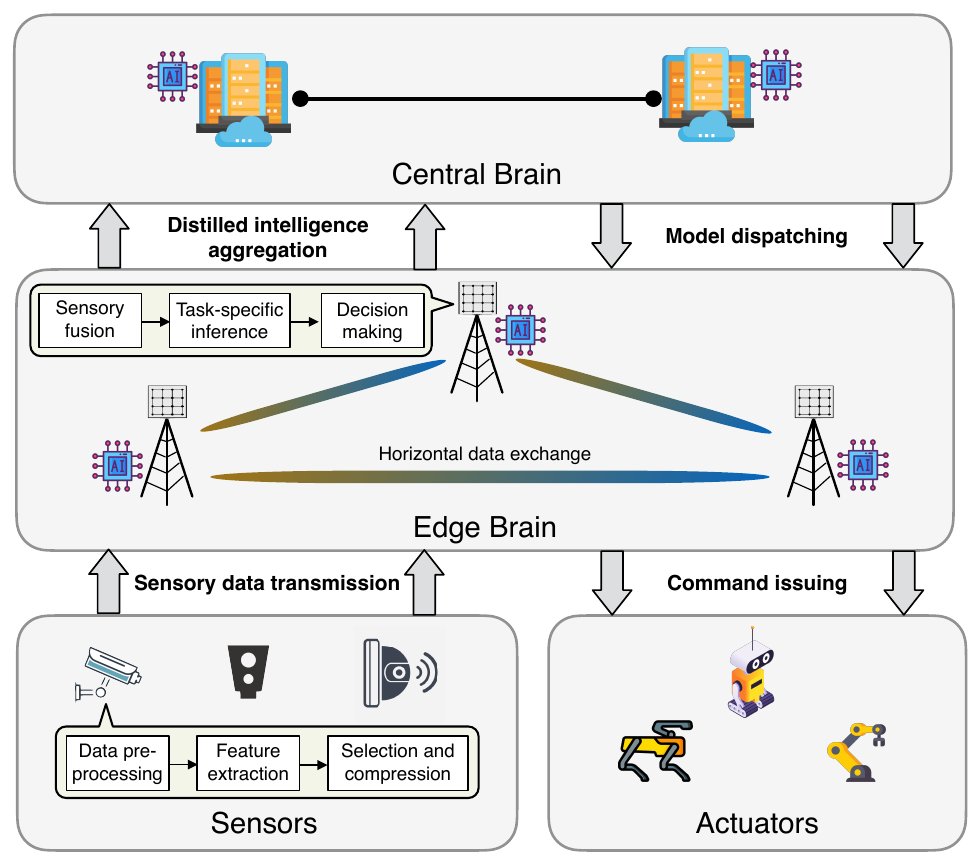}
    \caption{ISEA consists of four main components, namely, sensors, edge brain, central brain, and actuators. The execution of latency-sensitive sensing tasks involves sensors transmitting sensory data to the edge brain for semantic understanding, inference, and command issuing to actuators. Data exchanges between the edge brain and the central brain are mainly longer-term model updates.}
    \label{fig: isea_gen_archi}
\end{figure}

ISEA is a comprehensive framework orchestrating key components and operations involved in the sensing task execution for task performance. Meanwhile, the ISEA architecture design shall possess high versatility such that it can be specificized for a wide range of applications with highly heterogeneous scales and complexity, as well as on various platforms with different communication capabilities, network topologies, and computing power. In what follows, we first describe the general architecture of ISEA, covering its key components, computation operations, and communication interfaces, which is illustrated in Fig.~\ref{fig: isea_gen_archi}. Then, we specify five ISEA paradigms with different function placements and topologies.
\subsection{General Architecture}\label{subsection-architectures-A}
\subsubsection{Key Components}
In this subsection, we introduce key components of the ISEA architecture, namely, sensors, edge brain, central brain, and actuators. 

\begin{itemize}
    \item \textbf{Sensors:} Sensors convert the information of the physical world, such as light intensity and depth information, into digital signals. The sensory data stream can be either sporadic or at a fixed rate. While conventionally being independent modules with dedicated hardware, sensors can be software-defined, sharing hardware and interfaces with other functions, of which the most notable examples include virtual sensors in industrial IoT and ISAC-empowered RF sensing using communication signals. Sensors are often equipped with application-specific integrated circuits or processors for basic front-end signal processing such as denoising, filtering, and encoding. By now, advances in microprocessors allow sensors to be equipped with higher-level computation capabilities, even including lightweight AI computations. Examples include GPUs and Field Programmable Gate Arrays (FPGAs) equipped on UAVs, and digital signal processors (DSPs) tailored for AI processing on image sensors\cite{SonyAISensor}. This opens up the possibility of moving higher-level functions, such as feature extraction, compression, and coarse pattern recognition to sensors to reduce the network traffic and computation load at edge or cloud servers. 
    \item \textbf{Edge brain:} Edge brain refers to the collection of intelligent edge nodes in proximity to sensors, capable of data reception, caching, aggregation, and AI-empowered analytics of sensory data in a concerted manner. While still subject to limited power and computing budgets, these edge nodes are expected to carry moderate AI computing performance (e.g., $\sim 10$ TFLOPS in half precision) and storage capacity for sensing inference or training on state-of-the-art AI models. In the edge brain, neural network layers can be distributed among interconnected edge servers for collaborative computing based on distributed machine learning approaches such as split machine learning~\cite{lin2024split, Lin-TMC-EPSFL,
lin2024H-splitFL}, thereby reducing E2E latency. When edge nodes collaborate, smashed data (activations at neural network layers) can be extracted from sensory data, moving across the edge networks.
    Also, the sensory data from various distributed sensors should be appropriately fused at the edge brain to achieve global environmental awareness. In this way, distributed edge nodes aggregate asynchronous-arriving sensory data for multi-view multi-modal fusion, gathering ever-changing environmental data for on-the-fly fine-tuning, and maintaining a compact library of context-aware AI models for inference, thereby enhancing detection accuracy and making informed decisions. Not only network equipment such as BS, routers, or dedicated edge AI stations can be part of the edge brain. Depending on the dynamic wireless link and battery states, user devices, e.g., mobile phones and vehicles, can also be scheduled for collaborative computing as part of the edge brain.
    \item \textbf{Central brain:} Located at cloud data centers, the central brain is equipped with massive computation capability for handling computation-intensive tasks, such as training large language models and executing global decision-making. Nevertheless, the long geographic distance to users and potential traffic congestion within the backbone network make it unsuitable for inference tasks with demanding latency requirements. In ISEA, the central brain mainly performs the following functions. First, the central cloud is capable of global aggregation of distilled information from the edge brain to accomplish computation-intensive yet time-insensitive tasks, such as urban planning. Second, it serves as the highest-level parameter server in hierarchical FL, consistently aggregating trained model parameters from geographically distributed edge nodes for global model updates and distribution. At last, the central brain stores a large number of AI models for diverse sensing tasks of different sizes and complexities. This enables in-time downloading of suitable AI models onto the edge brain. 
    \item \textbf{Actuators:} Actuators receive real-time commands from the edge brain and interact with the real world accordingly. For example, a robotic arm receives movement inputs from a robot planning server and grasps the object referred to by human instructions; a traffic light has its pattern controlled by an intelligent controller analyzing real-time traffic of surrounding blocks. 
\end{itemize}

\subsubsection{Computation Operations and AI Backbones}
{While ISEA is envisioned to support various sensing tasks using different AI models, they share similarities in the computing stage. A typical pipeline of computation operations in the ISEA system includes data pre-processing, feature extraction, feature selection and compression, sensory fusion, and task-specific inference. }
\begin{itemize}
    \item \textbf{Data pre-processing:} This stage involves pre-processing of the raw sensory data into suitable forms for input to the AI model or wireless transmission, e.g., de-noising, filtering, and normalization. Usually, this stage does not involve neural-network operations, which hence can be implemented with on-sensor circuits.
    \item \textbf{Feature extraction:} In this stage, feature extractors are applied to the pre-processed sensory data to generate semantic representation in the feature space, which can be fed into a task-specific \emph{head} for downstream inference in the following computation stages. The feature extractors are typically generic backbone models trained for specific data types but not restricted to specific downstream tasks. In other words, the extracted features can be reused for different tasks by simply inputting the corresponding downstream model. For example, ConvNeXt and vision transformers (ViTs) for 2D images have been proven to be efficient backbones for multiple computer vision tasks ranging from object detection to classification to segmentation, delivering satisfactory performance without backbone fine-tuning. For point cloud data, PointPillar\cite{lang2019pointpillars} and VoxelNet\cite{zhou2018voxelnet} are prevalent backbones for object detection and semantic segmentation. The advantages of generic feature extractors include more efficient storage usage and less communication cost due to feature reuse for multiple tasks. 
    \item \textbf{Feature selection and compression:} The selection and compression of features to reduce communication cost is necessary for settings where features should be exchanged between ISEA components, e.g., when extracted features are transmitted to edge nodes for task-specific inference. Generic feature compression techniques include encoder-decoder and information bottleneck to encode high-dimensional features into a compact but informative representation for transmission or feature pruning to remove redundant feature dimensions. Aiming at a specific goal, e.g., finding a missing person, local gateway models can determine whether the features need to be transmitted by evaluating the correlation between the features and task-relevant information, such as a semantic query.
    \item \textbf{Sensory fusion:} Sensory fusion is a key stage for sensing with multiple modalities, multiple views, or both. In ISEA, the fusion occurs at the edge brain which collects and aggregates sensory information from multiple sensors. Due to the high dimensionality of raw data and privacy concerns, the fusion of abstract features uploaded by sensors is necessary. Concatenation is arguably the most straightforward approach without information loss, but meanwhile results in a high feature dimension, slowing down training and inference. For multi-view sensing, a more efficient approach, called view-pooling, is widely adopted for both image and point cloud features which incurs element-wise averaging or max-out across all feature maps. The fusion method is embedded in the model architecture. For features whose dimensions correspond to spatial locations, e.g., voxel features obtained from the point cloud, the individual observation needs to be projected onto a global coordinate defined by the edge brain, known as spatial alignment, via coordinate transformation or learned models such as BEVFusion.  
    \item \textbf{Task-specific inference:} This stage applies a task-specific head onto the generic, fused feature to obtain the desired sensing result. Examples include convolutional neural network (CNN)-based image classifiers for object recognition and regional proposal networks for detection tasks. The computation overhead of such models can vary, from simple linear classifiers to sophisticated designs. 
\end{itemize}
{Depending on the heterogeneous computing capabilities and communication links, these computation operations can be deployed onto different components in ISEA, resulting in various paradigms as detailed in Section~\ref{subsection-architectures-B}. }

\subsubsection{Communication Interface}
ISEA relies heavily on the seamless delivery of sensory information among the distributed network entities, e.g., sensors/actuators, edge brains, and cloud brains. An ISEA system typically comprises two primary modes of data transmission:
\begin{itemize}
    \item \textbf{Vertical data transmission:} In this mode, ISEA participants deliver sensory data, extracted features, or computing results to another level to facilitate hierarchical data analytics. For instance, cooperative sensing normally involves multi-access uplink transmission for data aggregation from distributed devices, after which the global decision or perception results can be disseminated to sensors via downlink broadcasting.

    \item \textbf{Horizontal data exchange:} Participants at the same level can share their perception data nearby, thereby reducing transmission latency. For instance, in a V2V system, vehicles can broadcast information such as sensory data, speed, headings, and brake status for rapid danger identification and warning triggers.
\end{itemize}

The communication interface within ISEA can be established through wired and wireless channels. Wired communication, carrying data over wired links such as fiber optics, offers stable and efficient data transfer over long distances. Conversely, wireless communication utilizes radio waves, providing flexible and ubiquitous connectivity for short-distance communications. However, the dynamic nature of channel conditions and resource constraints (e.g., bandwidth, energy, and time) in wireless transmission necessitates ongoing task-specific designs, contributing to its rapid evolution.

Conventional wireless communication techniques applicable to ISEA can be broadly categorized into digital and analog schemes, each adhering to diverse design principles, transmitted data formats, and applicable scenarios. The digital air interface is primarily built on the theoretical foundations of Shannon's theory, where sensory data and intermediate results are converted into bits via sequential execution of quantization, source (entropy) coding, and channel coding, and the bit streams are then modulated onto constellations according to well-designed modulation schemes for radio transmission. Compared to its digital counterpart, analog transmission employs uncoded linearly modulated ISEA data as the transmission symbols. It retains the structure of ISEA data during transmission at the cost of channel distortion. Therefore, basic computation operations can be integrated with wireless transmissions, empowering efficient sensory data delivery such as AirComp-based aggregation. 

\subsection{Paradigms}\label{subsection-architectures-B}
\begin{figure*}[t]
    \centering
    \includegraphics[width=0.85\textwidth]{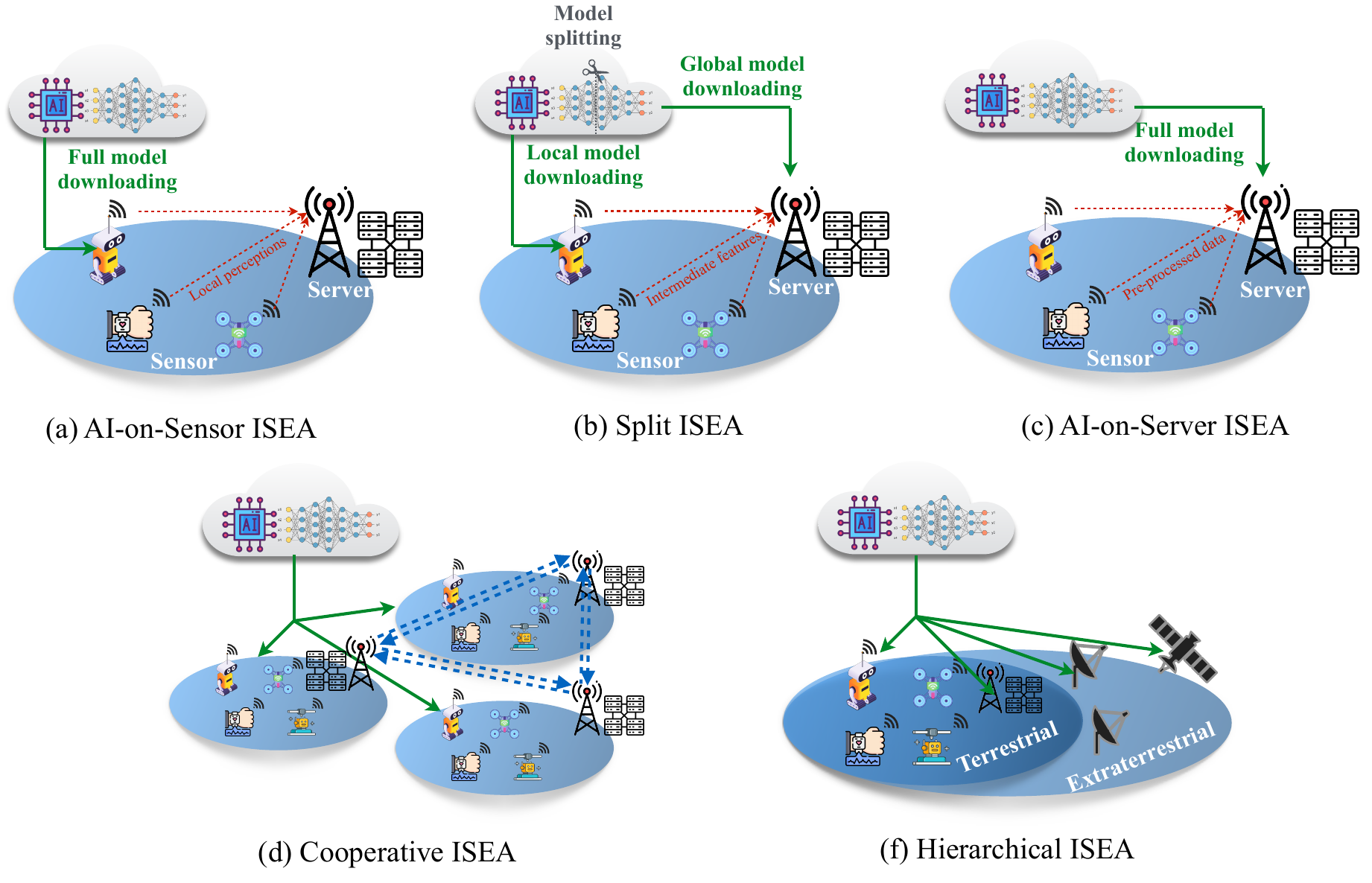}
    \caption{ISEA can be instantiated into different paradigms depending on the placement of AI models and computation powers, network topologies, and communication capabilities.}
    \label{fig:architecture}
\end{figure*}

Next, we introduce five paradigms of ISEA as illustrated in Fig.~\ref{fig:architecture} with different configurations of computation workload and communication links and discuss their application scenarios. 
\subsubsection{AI-on-Sensor ISEA}
In this paradigm, an AI model is completely deployed on sensors, which perceive and analyze sensory information and make decisions on their own. This paradigm is suitable for the most latency-critical scenarios as it does not incur data transmission latency. However, the size and computational complexity of on-sensor AI models are strictly limited by the storage and computation capability of sensors. While primarily relying on sensors' computation power, there can still be interactions between sensors and edge nodes. The first situation is achieving proposal-level fusion, i.e., receiving and aggregating local inference results, such as categorical probability and detection boxes, via fusion schemes such as majority vote. The communication latency is significantly lower than the feature-level fusion due to the low data volume of inference results. Second, sensors can participate in FL by transmitting their local updates to an edge brain, or the edge brain can distribute update-to-date AI models to heterogeneous sensors.
\subsubsection{AI-on-Server ISEA}
In this paradigm, the computation is completely offloaded to edge servers with rich computation resources. The uploaded sensory data can either be raw data or pre-processed data that only requires minimal computation capabilities at sensors. AI-on-Server ISEA usually refers to the case where a large number of low-cost sensors cooperatively sense environments, monitoring temperature, humidity and sound intensity. In this case, only the aggregated data conveying critical underlying information requires computing, while data obtained by each sensor alone carries little information. Meanwhile, on-sensor feature extraction is unnecessary and/or infeasible due to the locally sensed data's simplicity and low dimensionality (e.g., only one numerical value per frame for air quality sensing). While the data rate requirement for each sensor is low, a communication bottleneck can still occur due to multiple access by a potentially massive number of sensors and extremely stringent RF power constraints at sensors.
\subsubsection{Split ISEA}
Split ISEA primarily relies on split inference~\cite{RN21,Co-inference_Shi} and split learning\cite{lin2024split}, in both of which the AI model is divided into sub-models for deployment on both the sensor and server. In split inference, a sub-model operates on the sensor, extracting feature maps from raw sensory data, while the remaining model resides on the server, performing inference with a global view based on the feature maps uploaded by the sensor. In split learning, the devices train on-device sub-models locally in parallel, and upload activations to the server for on-server model training. In contrast to AI-on-sensor and AI-on-server, the split ISEA paradigm introduces a flexible computation allocation between the sensor and the server. This flexibility significantly enhances resilience in scenarios with diverse on-device and central computation resources~\cite{CS2018TWC,RN175}. Moreover, by keeping the raw data processing localized at the sensor and only transmitting abstracted feature maps to the server, split ISEA minimizes the exposure of sensitive sensory data, contributing to a more secure and privacy-aware system architecture. The optimization of the model splitting is the main concern in split ISEA. To minimize E2E latency, smart splitting strategies are proposed to reach a balance between local computation and data uploading subject to local computation/memory constraints. Furthermore, in the case of multi-tasks, efforts are made to jointly optimize model splitting, access control, and task scheduling among sensors.

\subsubsection{Cooperative ISEA}
ISEA equipped with a single sensor may encounter inherent limitations when perceiving complex environments~\cite{RN156,Cooper}. For instance, for a camera-based sensor, the information captured within captured images is inherently constrained by the camera's field of view (FoV). This limitation poses challenges for tasks like object detection and semantic segmentation, especially in scenes with occlusion. In the case of LiDAR-based sensing, point clouds exhibit data irregularities, characterized by an ``intense in proximity, sparse in the distance'' pattern, due to reflectivity reduction~\cite{PCinAir}. Cooperative ISEA emerges as a promising solution to overcome the constraints of FoV limitations and data irregularities by assigning a server to combine data or intermediate results generated by distributed cross-modal sensors for cooperative perception. The collaborative strategy often outperforms individual detection, thanks to the complementary nature of sensory outcomes. The collaboration within cooperative ISEA can occur at different levels: data level, feature level, and proposal level. This corresponds to the fusion of raw sensory data from various modalities and localizations through spatial alignment, the combination of feature maps computed from raw data in the feature space via concatenation or element-wise multiplication, and the integration of prediction results made by individual sensors utilizing their on-device models\cite{huang2022multi}.

\subsubsection{Hierarchical ISEA}
Hierarchical ISEA offers strong resilience for perception systems with heterogeneous capabilities of sensing, computation, and communication. Specifically, it can divide a global sensing task into several sub-tasks and then assign the sub-tasks of data collection and data analytics to system components according to their resources. Consider field monitoring with a sensor-edge-cloud architecture, where the expansive sensing can be decoupled into the detection of a set of subareas. Each edge node is responsible for a single subarea according to its physical location, performing local data analytics (e.g., feature extraction/fusion) from raw data or pre-processed results aggregated from sensors within their coverage. Then, the extracted information across different edge nodes is further aggregated to the central cloud to make globally informed decisions.

Hierarchical ISEA is also developed from the integration of ISEA with communication networks with a natural hierarchical architecture, for example, satellite-terrestrial edge computing~\cite{RN357}. Air-ground systems can realize high-resolution classification, perception, and environmental mapping. The complementarity of extraterrestrial and terrestrial sensing results in improved performance in modern sensing applications like autonomous driving and smart cities.

{ The five ISEA paradigms—split, AI-on-server, AI-on-sensor, hierarchical, and cooperative—are not mutually exclusive but represent distinct characteristic dimensions of ISEA deployment. For instance, the first three paradigms (split, AI-on-server, and AI-on-sensor) define model placement strategies with trade-offs in hardware, communication, and computational complexity, while cooperative ISEA addresses multi-device coordination challenges. Crucially, these paradigms can be combined adaptively: hierarchical structures may employ different placement strategies across layers, and cooperative systems can integrate hybrid architectures (e.g., split for latency-sensitive tasks and AI-on-sensor for privacy constraints). Such flexibility, however, necessitates advanced scheduling algorithms and communication optimization, underscoring the need for context-aware ISEA design.
\subsection{Lessons Learned}

\begin{itemize}
    \item The general architecture of ISEA includes sensors, edge brain, central brain and actuators. Sensors transform physical world information into digital signals. It is worth pointing out that sensors are not limited to standalone devices (e.g., cameras), but may coexist with other functions in shared hardware, such as RF sensing in ISAC systems. Edge brain encompasses all intelligent nodes in an edge network, which constantly communicate and collaborate with each other to accomplish ISEA tasks. The central brain, located in cloud data centers, focuses on time-insensitive operations such as global information aggregation and periodic model update distribution due to high access latency. The decisions made by edge brain are sent to actuators to interact with the physical environment. 
    \item Computation operations are integral to the execution of an ISEA task and can be deployed to various ISEA components based on device capabilities and collaboration mode. The characteristics of these computing operations vary substantially with the sensing data modality and AI backbone used\cite{RCNN_accl}. As these operations are distributed across different nodes, communication interface between ISEA components are necessitated, including vertical data transmission for offloading or data aggregation and horizontal data exchange for peer collaboration. 
    \item 
ISEA instances can be categorized into different paradigms depending on the placement of AI models onto nodes, network topology and collaboration protocols. It is importance to notice that these paradigms can be combined or switched between in a shared ISEA system depending on the task requirements.
\end{itemize}}

\begin{table*}[!htbp]

\centering
\caption{Summary of related works on digital air interface for ISEA }
\scriptsize

\begin{center}

\setlength{\tabcolsep}{0.8mm}{%
\begin{tabular}{|>{\centering\arraybackslash}m{0.09\textwidth}|c|m{0.28\textwidth}|m{0.15\textwidth}|m{0.4\textwidth}|}
\hline
\multicolumn{1}{|>{\centering\arraybackslash}m{0.09\textwidth}|}{\textbf{Approaches}}                       & \textbf{Ref.}      & \multicolumn{1}{>{\centering\arraybackslash}m{0.28\textwidth}|}{\textbf{Sensing Scenario}}                                              & \multicolumn{1}{>{\centering\arraybackslash}m{0.15\textwidth}|}{\textbf{Performance Metrics}}                    & \multicolumn{1}{>{\centering\arraybackslash}m{0.4\textwidth}|}{\textbf{Key Contribution}}                                                                                                                                                                     \\ \hline
\multirow{7}{*}{\centering JSSC} 
                 & \cite{Task-based_ADC_2021} & Sensing task-based ADC converters                                                                                                          & \multirow{2}{*}{\makecell[c]{Multivariate signals\\ classification accuracy}} & \multirow{2}{*}{\makecell[c]{Design task-specific ADCs extracting digital information\\ representations from multivariate continuous-time signals} }          \\ \cline{2-3}
                 & \cite{Task-based_ADC_MIMO_2020} & Sensing task-based ADC conversion for MIMO receivers                                                                                       &                                                               &                                                                                                                                                                                        \\ \cline{2-5} 
                                             & \cite{Task-oriented_NextG_2023} & Split ISEA for classification of sensed data                                                                         & Inference accuracy                          & An encoder-decoder pair trained for E2E accuracy without explicit signal reconstruction                                               \\ \cline{2-5} 
                                             & \cite{D2JSCC} & Transmission of sensed image to server                                                                       & Image reconstruction quality                            & Deep source coding jointly tuned with the channel coding rates for optimized E2E performance                                  \\ \cline{2-5} 
                                             & \cite{JSCC_Imp1} & Split ISEA for classification of sensed data                                                                  & Inference accuracy, reconstruction error                       & Use the importance of semantic features for downstream inference for bit and subcarrier allocation                                                                             \\ \cline{2-5} 
                                             & \cite{Task-oriented_SC_2023} & Explainable task-oriented communication                                                                                                    & Semantic channel capacity, reconstruction error                                          & Select task-related semantic features to transmit                                                                                                                                 \\ \cline{2-5} 
                                             & \cite{JSCC_Imp2} & Collaborative ISEA inference where sensors have different modalities & Sentiment score prediction error                                & Derivation of varied modality importance for unequal error protection                                        \\ \hline
\multirow{9}{*}{Access Control}              & \cite{RN193} & A UAV with image sensors transmits semantic triplets to users per their different queries                                                  & Triplet drop probability, semantic matching score             & Identification of personalized saliency to prioritize power allocation for key semantic triplets                                                                                       \\ \cline{2-5} 
                                             & \cite{liu2020who2com} & An agent with a degraded view selects a helper to receive the required information                                                  & Sensing accuracy, communication overhead                      & Proposes a handshake protocol where the requester sends a low-dimensional query to requester for semantic matching                                                                     \\ \cline{2-5} 
                                             & \cite{semdas} & A service requester aims to find task-relevant sensory data from data sources distributed in the network                                   & Missing rate, communication latency                           & Proposes sending queries for locating relevant data sources and joint semantic-channel matching                                                                                        \\ \cline{2-5} 
                                             & \cite{mass} & Multiple cooperative vehicles exchange sensory information via V2V networks                                                                & Total perception gain                                         & Design of a mobility-aware distributed sensor scheduling algorithm to maximize the expected perception gain                                                                            \\ \cline{2-5} 
                                             & \cite{RN131} & Multiple agents communicate repeatedly until a holistic view of the environment is formed                                                  & Object detection precision, communication overhead            & Exchanges spatial confidence maps among sensors to trigger feature transmission on critical areas                                                                                      \\ \cline{2-5} 
                                             & \cite{FLaloha} & FL where sensors located in a cell have asynchronous local updates                                                         & Aggregation error                                             & Proposes multichannel random access for data uploading with optimized access probability                                                                                               \\ \cline{2-5} 
                                             & \cite{10130012} & FL over massive MIMO systems with asynchronous local updates                                                               & Learning convergence rate                                     & Proposes a random access scheme utilizing an access class barring method to select the uploading devices                                                                               \\ \cline{2-5} 
                                             & \cite{jankowski2023adaptive} & Collaborative inference where the sensor wirelessly transmits intermediate features via JSCC for inference offloading                      & Inference accuracy, communication cost                        & Designs a transmission-decision neural network by jointly considering sample confidence and SNR levels                                                                                    \\ \cline{2-5} 
                                             & \cite{10133831} & Collaborative inference where the sensor can offload samples to the server or execute locally                                              & Average accuracy, inference latency                           & Design of hard-case discriminators and a collaborative scheduler adaptive to C$^2$ resource conditions                                                                  \\ \hline
\multirow{9}{*}{\makecell[c]{RRM}}   & \cite{RN205} & Connected vehicles scheduled to upload sensing data to multiple servers for model training and subsequent inference                        & Overall perception accuracy of all trained models             & Joint association, power and bandwidth allocation for maximizing the quality of training                                                                                               \\ \cline{2-5} 
                                             & \cite{RN206} & An ISAC system where one BS simultaneously senses and transmits data packet to multiple users                                                & Total power, network stability, detection performance         & Dynamic power and subcarrier allocation scheme for power minimization with queue awareness                                                                                             \\ \cline{2-5} 
                                             & \cite{wen2024multidevice} & A collaborative ISEA system where the BS aggregates data from multiple RF sensors                     & Sensing disciminant gain                                          & Joint allocation of communication time, quantization bit and transmission energy                                                                                                          \\ \cline{2-5} 
                                             & \cite{RN198} & A collaborative ISEA system where an access point (AP) aggregates data from multiple sensors and broadcasts the fused data                     & Total execution time                                          & Information extractor design and a task-oriented channel allocation mechanism                                                                                                          \\ \cline{2-5} 
                                             & \cite{Lau2022decentralized} & FL under hybrid data partitioning, i.e., with multiple groups of different-type sensors                                    & Model training performance, resource consumption              & Solving the data dependence issue and joint sensor scheduling, bandwidth allocation and quantization optimization                                                                      \\ \cline{2-5} 
                                             & \cite{RN184} & Autonomous driving where vehicles and the server are equipped with shallow and deep sub-models respectively                                & Inference accuracy, latency, energy consumption               & A multi-task-learning based module for joint offloading decision and resource allocation                                                                                               \\ \cline{2-5} 
                                             & \cite{RN175} & Inference in industrial IoT where models are deployed at consoles, edge servers and cloud with increasing complexity                 & Total delay and inference accuracy                            & Optimized task offloading decision and resource allocation given wireless network conditions                                                                                           \\ \cline{2-5} 
                                             & \cite{RN182} & An edge server supports the object detection of multiple vehicles                                                                          & Average precision, delay and energy                           & Dividing the task into real-time and delay-tolerant tasks with respective resource allocation optimization                                                                             \\ \cline{2-5} 
                                             & \cite{10445443} & A split inference system involving a UAV and a BS where the UAV can switch between uploading raw data or features                          & Total energy consumption                                      & Embedding the optimal power allocation into a tiny reinforcement learning algorithm for optimal mode selection                                                                         
                                             \\ \cline{2-5} 
                                             & \cite{RN183} & Split ISEA supporting inference request from multiple users                          & Task success probability                                      &  Joint optimization of compression ratio selection, user selection and resource allocation with task priorities                        \\ \cline{2-5} 
                                             & \cite{RN155} & An edge server supports inference requests from multiple sensors                                                                           & Inference throughput                                          & Optimal task scheduling and bandwidth allocation with batching and early exiting                                                                         \\                                                
                                             
                                             \cline{2-5} 
                                             & \cite{Model_Splitting_RRM_TWC2024} & Collaborative ISEA with model partitioning                                                                           & Weighted sum energy consumption                                          & Energy minimization by splitting point          selection and C$^2$ resource allocation                                                                     \\                                                
                                             
                                             \cline{2-5} 
                                             & \cite{lu2025rrm_veh} & Collaborative perception in vehicle networks                                                                          & E2E latency, perception probability                                          & Joint optimization of compression ratio, collaborator selection, offloading decision and computation resource allocation with Lyapunov-aided deep reinforcement learning                                                                                            \\ \hline
\end{tabular}%
}
\end{center}

\end{table*}

\section{Digital Air Interface for Integrated Sensing and Edge AI}\label{section-dai}

\begin{figure}[t]
    \centering
    \subfigure[]{\includegraphics[width=0.48\textwidth]{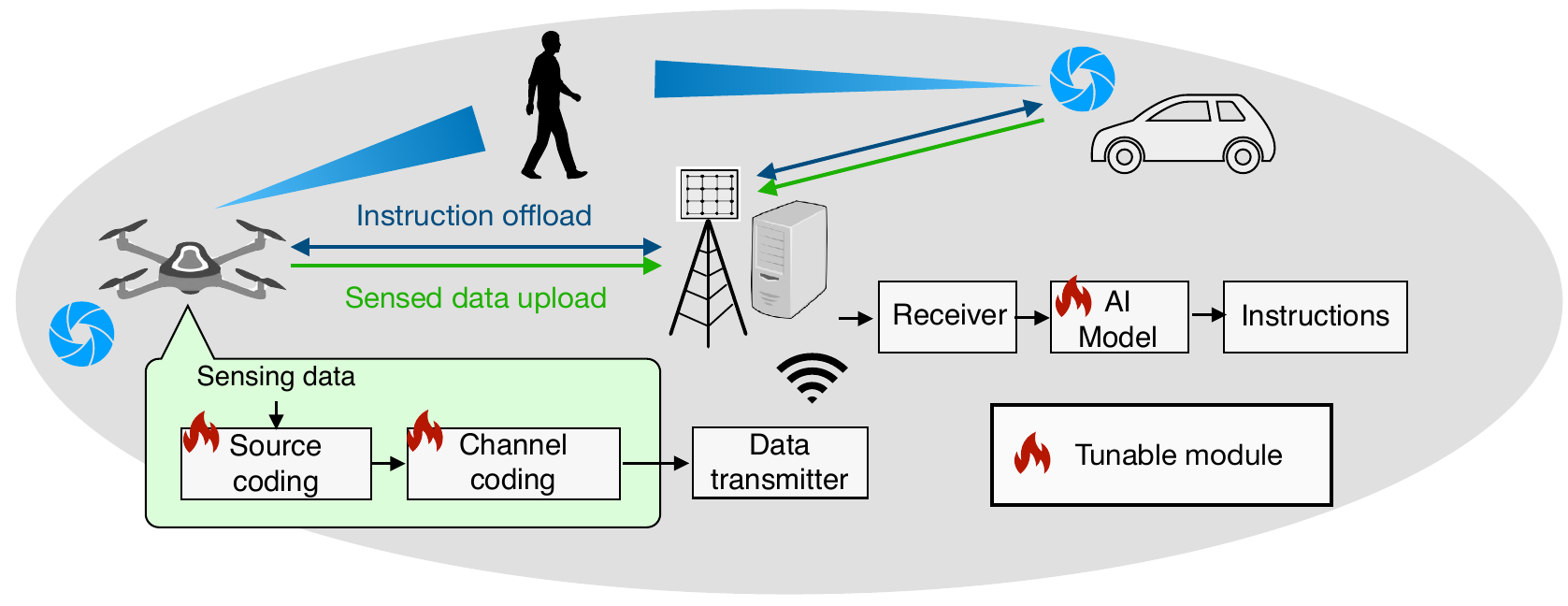}}
    \subfigure[]{\includegraphics[width=0.48\textwidth]{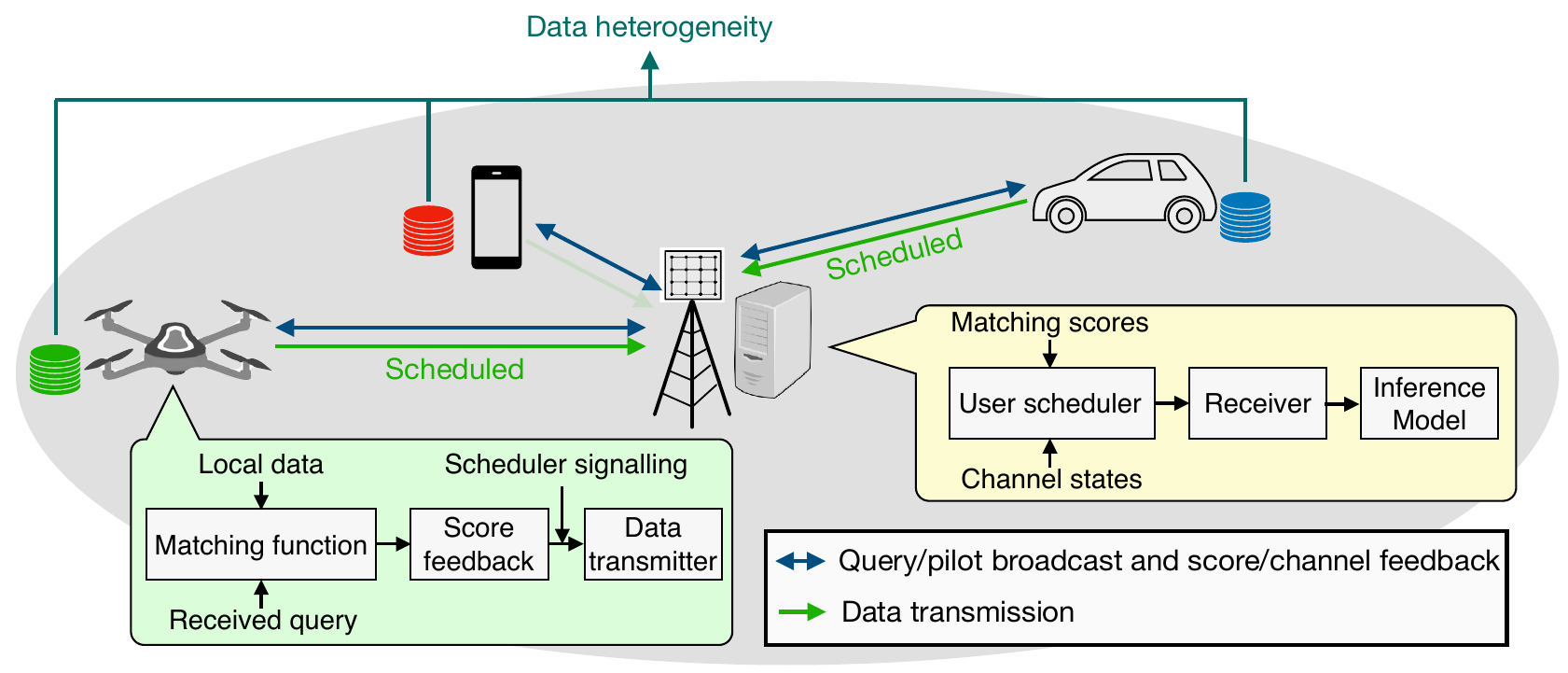}}
    \subfigure[]{\includegraphics[width=0.48\textwidth]{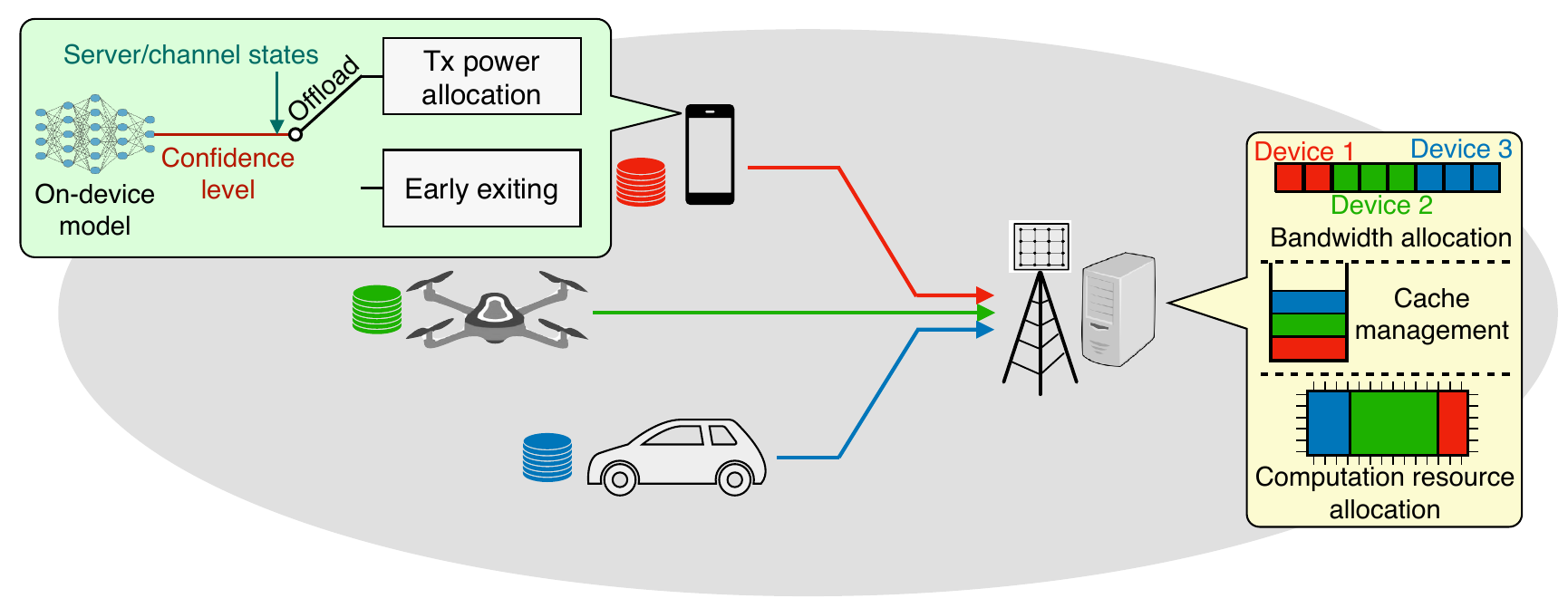}}
    \caption{
   Digital-air-interface techniques designed for ISEA can be categorized into three key aspects: (a) joint source-channel coding, (b) access control, and (c) radio resource allocation.}
    \label{fig:dai}
\end{figure}

{The operations of ISEA involve intense wireless horizontal and vertical data exchange, bottlenecking task performance metrics due to hostile channel fading and high dimensionality of sensory information. The existing digital air interface technologies, designed under rate-centric and best-effort principles, cannot guarantee the QoS in terms of latency and reliability desired by ISEA applications. In 6G, it is expected that the digital air interface will evolve in its physical and link layers oriented towards satisfactory task performance under potentially bursting request arrivals. In the ISEA context, the digital air interface design shall be designed with sensing awareness and oriented towards ISEA performance by considering task-related aspects such as sensing semantics, data importance, sensor importance, and computation capabilities. This section reviews research efforts on digital air interface dedicated to ISEA applications in the said direction from different aspects including JSCC, access control, and RRM, as illustrated in Fig.~\ref{fig:dai}.}

{
\subsection{Joint Source-Channel Coding}\label{subsection-dai-jscc}
JSCC in ISEA unites the separated coding blocks and optimizes them towards ISEA-oriented QoS requirements, e.g., inference accuracy at a split ISEA system where successive
feature delivery is involved~\cite{Task_oriented_6G_2023}. 
Many works have been done in this domain. 
The schemes in \cite{Task-based_ADC_2021} target the design of task-specific analog-to-digital converters (ADC) that perform sampling and quantization over a sensed continuous-time signal for efficient extracting of digital information desired by downstream signal classification instead of exact signal reconstruction.
Such specific learnable ADC design could be applied in MIMO communication channels where multi-variate signals are transmitted for identifying disparate sensing objects' echos with competitive accuracy performance but reduce the power consumption in traditional exact signal recovery \cite{Task-based_ADC_MIMO_2020}.  
DL-based JSCC enables E2E training targeting downstream ISEA tasks with varied performance metrics.
For example, one application scenario of JSCC is the low-latency transmission of sensed signals, e.g., images for on-server classification of the signal type, e.g., class of the observed object. In \cite{Task-oriented_NextG_2023}, aiming at remote classification of sensed signal, the authors apply a neural network encoder to map the signal source to symbols and a neural network receiver that directly predicts the label from received symbols without the redundancy of signal reconstruction. While demonstrating superior performance, the authors also point out its vulnerability to smart adversarial attacks. 
 In \cite{D2JSCC},  deep source coding is employed with an adaptive density model and jointly tuned with channel coding rates for optimizing the E2E performance metric derived via the analysis
of the Bayesian model in the JSCC system. 

Building on the E2E task metric, an effective and explainable JSCC technique is an explicit characterization of data importance for consideration in channel coding. In \cite{JSCC_Imp1}, considering split ISEA inference, the sensed data is first encoded into semantic features, whose importance to the downstream ISEA task is measured via the designed Semantic Task Relevance module and Inter-Semantics Relevance module. Then, the obtained importance is integrated into the optimization of bit and subcarrier allocation to maximize an objective considering both task performance improvement and distortion suppression. 
The work in~\cite{Task-oriented_SC_2023} applies the variational autoencoder to provide an explainable and transparent JSCC system that disentangles the extracted features and transmits only the sensing-relevant ones according to the derived semantic channel capacity bounds. Considering a collaborative ISEA scenario where multiple sensors have different modalities, \cite{JSCC_Imp2} utilized the framework of robust verification problem to derive the respective importance, i.e., the contribution to the inference task, of different modalities. Then, an unequal error protection scheme based on the derived importance is applied to protect important modalities.}

\subsection{Access Control }\label{subsection-dai-access}
Traditionally, access control involves protocols and resource allocation schemes to coordinate multiple devices under a shared wireless medium, the performance metrics of which include access latency, throughput, fairness, etc. In ISEA, while these metrics are still of practical value, the ultimate goal of access control design is to optimize downstream task performance by considering data semantics. It is worth pointing out that access control schemes can overlap with radio resource allocation, which is the theme of the next subsection. However, here we cover mechanism designs for connection establishments based on data or task importance, while research works where sensor scheduling is part of optimal resource allocation are reviewed in the next subsection. There are two mainstream approaches to multi-access techniques in ISEA, namely, active data source selection and random access. The former usually involves the requester using meta-information, e.g., low-dimensional queries and feedback, to identify and admit task-relevant sensors into the uplink transmissions. The scheduling criteria jointly examine the state of the channel and the semantic relevance of each sensor. The latter is an extension of conventional random access, where users contend for channel resources in a decentralized manner, involving the importance of data as a factor in the access protocol. In parallel, sample-wise offloading gates can be designed to determine whether a task should be offloaded to the server based on the confidence level of edge devices.
\subsubsection{Data Source Selection}
Data source selection serves as a common framework for scheduling the transmissions of data sources based on sensory semantics and communication metrics, e.g., channel conditions as coordinated by the data requester. A line of research adopts the query-and-feedback approach, where the information requester sends compact semantic queries to candidate providers, which feed back a matching score for scheduling relevant users. In\cite{RN193}, a UAV-based image acquisition scenario is considered where users retrieve images of interest from the UAV. A task-oriented access scheme is designed where the users match their query text with the low-dimensional semantic features encoded and transmitted by the UAV and only download original images that are sufficiently matched. The semantic encoder is designed with an awareness of users' subjective interest, and the radio resource allocation is optimized for reliable transmission of the extracted semantic information. Who2com\cite{liu2020who2com} considers a collaborative perception scenario where a sensor with degraded views seeks assistance from neighboring sensors for complementary observations. To enhance the communication efficiency, the degraded agent broadcasts a low-dimensional query to other sensors, which compute a matching score between the query and local observations via a learnable attention mechanism and feed back to the degraded agent. Based on the matching scores, the requester chooses one of the helping sensors with the most semantically relevant information to access the full observation. From a C$^2$ integrated perspective, semantic data sourcing is proposed in\cite{semdas}, which selects the data source based on joint semantic-and-channel matching, featuring a quality indicator comprising both semantic and channel factors. In the absence of feedback, the past uploaded data can be utilized to predict the value of sensors' current data. In this direction, a mobility-aware sensor scheduling scheme is developed in\cite{mass} for a V2V scenario where an ego vehicle determines and accesses information from an optimal sharer based on locality, channel states, and sensor qualities. The optimal sensor scheduling problem is formulated, for which a learning-while-scheduling approach is designed to predict the performance gain of accessing each sensor based on historical observations. Then, the ego vehicle accesses the data of the sharer bearing the maximum perception gain, compressed subject to the rate constraint. Besides, Where2comm\cite{RN131} targets a scenario where multiple sensors equipped with cameras and LiDARs communicate repeatedly until a holistic view of the environment is formed. The customized spatial confidence map indicates a sensor's knowledge level on a BEV coordinate of the perception region. By exchanging the binary confidence maps among agents, feature transmission is triggered only if one sensor has high confidence in a spatial region about which another sensor is uncertain. This procedure is repeated for multiple rounds until all agents share the same global view.  

\subsubsection{Random Access}
In ISEA applications with stringent latency requirements, even the centralized control complexity and frequent exchange of metadata and control signals become costly, calling for a revisit of random access techniques. The random access shall be designed for specific ISEA metrics such as convergence speed and inference accuracy. For FL over multiple uploading channels where sensors generate local updates intermittently, the authors of\cite{FLaloha} proposed to use multichannel ALOHA, a random access scheme instead of sequential polling, as the latter can lead to idle channels when the local update is not ready. The access probability for each device is determined based on the local updates and server feedback to minimize the aggregation error. A follow-up work on random access for FL over massive MIMO system is\cite{10130012}, where each device determines whether to access the channel to upload local model updates through an access class barring (ACB) check. The ACB factor is optimized to maximize the average number of devices involved in model aggregation, considering the different probabilities of availability of local updates across devices. In a cooperative ISEA inference scenario, \cite{AndersSemRA} proposes a query-based random access scheme where the requester broadcasts a semantic query for local matching, and each sensor attempts to transmit following the ALOHA protocol if the matching score exceeds a threshold. Compared with the query-and-feedback mechanism, this semantic-based random access scheme avoids the overhead of feeding back matching scores, which can be substantial if only a very small portion of sensors are relevant.
\subsubsection{Sample-wise Offloading Gates}
A typical scenario of ISEA is where sensors offload inference workloads to an edge server and download the results for subsequent actions. For efficient resource utilization, an on-sensor gate for offloading decisions can be applied to determine whether each sample needs to be offloaded to the server based on its predicted difficulty. 
In\cite{jankowski2023adaptive}, an early-exit classifier is added to the device sub-model, allowing devices to obtain the inference results locally without transmitting features to the server, where a lightweight transmission decision neural network (TD NN) is deployed at the device to determine whether the current sample need to be offloaded to the server for more accurate inference. The TD NN is trained to minimize a loss function by combining both the classification entropy and communication cost. The authors of\cite{10133831} design a device-cloud collaborative inference framework for object detection, exploiting the heterogeneous requirements for computation resources of tasks with different difficulties. A hard-case discriminator is proposed, which differentiates hard and simple cases based on the image content and decides whether each sample should be offloaded to the server or processed locally, considering the current resource-utilization status.

\subsection{Radio Resource Management}\label{subsection-dai-rrm}
RRM is a classic topic in wireless communications, which allocates power, spectrum and time among users for high throughput and fairness. RRM for ISEA shifts the conventional objective of rate maximization to task-oriented performance enhancements. To achieve this goal, the sensor connectivity and semantic relevance of sensors’ observations to the sensing task of interest shall be considered. It is important to note that the communication and sensing perspectives are coupled instead of orthogonal. For example, the geometrical positions of sensors impact both connectivity and semantics. In addition, the observations are correlated among sensors, as opposed to the conventional assumption of independent data streams.

\subsubsection{Power Control and Bandwidth Allocation}
The allocation of power and bandwidth resources to a given set of sensors shall consider the heterogeneous channels, data qualities, and computation capabilities of different sensors to meet the ISEA task performance requirement. In \cite{RN205}, an edge AI-assisted autonomous driving system is considered, where connected vehicles are scheduled to upload sensing data to multiple servers for perception model training. To maximize a defined metric, i.e., quality of training, the power and bandwidth allocation to different vehicles and time slots are jointly optimized. In \cite{RN206}, the authors consider ISAC settings where a BS simultaneously senses and transmits data packets, arriving in a Poisson process, to multiple users with a multi-functional OFDM signal. The subcarrier and power allocation problem is formulated to minimize the long-term average power consumption subject to sensing performance and queueing stability conditions and is solved via Lyapunov optimization. 
For multi-device cooperative RF sensing, the authors of \cite{wen2024multidevice} studies joint control of quantization bits, communication time allocation, and energy allocation to maximize the discriminant gain of sensing. In a collaborative ISEA system with a fusion center considered in\cite{RN198}, the inadequacy of uplink communication resources for uploading all sensing information necessitates information extraction, agent selection, and subchannel allocation according to channel states and importance of messages. The information extraction module is trained using the information bottleneck principle, the subchannel allocation is optimized to maximize the expected information gain, and message fusion is designed with a double-attention mechanism. 
The authors of \cite{Lau2022decentralized} first propose a primal-descent dual ascent training algorithm for FL under hybrid data partition, which involves several groups of sensors with different types, to resolve the data dependence issue. Then, the optimization problem of sensor scheduling, bandwidth allocation, and quantization is formulated to minimize the training residual term and resource consumption, which is solved in a decentralized manner. 

\subsubsection{Offloading Scheduling}
In split ISEA or AI-on-server ISEA, under limits on communication and computation resources, accommodating all sensors that require computation offloading can be infeasible. In such cases, the heterogeneity among requests can be leveraged to optimize the scheduling of task offloading for global objectives, e.g., utility and throughput. A two-tier edge AI-assisted autonomous driving system is outlined in~\cite{RN184}, which, similar to split inference, deploys the shallow layers of the perception model at the vehicles and the rest at servers. A multi-task learning model is trained to infer the optimized decisions for offloading decisions and resource allocation. In \cite{RN175}, the authors consider the inference phase in industrial IoT, where AI models with different computational costs and accuracy are deployed at three levels, i.e., consoles, edge servers, and the cloud. Based on the wireless network conditions, accuracy requirements, and computational resource utilization, the task offloading decision and resource allocation scheme are jointly optimized to minimize the long-term system cost, which includes processing latency and monetary loss due to wrong classification. The Lyapunov optimization technique is employed to solve the optimization problem via transformation, where both optimal and heuristic algorithms are proposed. For an ISEA system where the edge server supports the object detection of multiple vehicles, the authors in \cite{RN182} first divide the task into real-time and delay-tolerant tasks, where the former is executed with the edge server's GPU and the latter cached for future inference. A resource allocation problem is formulated for real-time tasks with the objectives of maximizing accuracy and minimizing latency. For delay-tolerant tasks, another problem is formulated of which the objectives include both accuracy and energy conservation. A distributed deep Q networks-based algorithm is proposed to solve both problems. 

\subsubsection{Joint Computation-Communication Resource Allocation}
In ISEA systems, computation and communication resources can both be bottlenecks of task performance, especially in view of time-varying channel conditions and limited on-sensor computation capability. This requires a joint controller coordinating communication and computation operations to ensure full resource utilization. In \cite{10445443}, a point-to-point split inference system involving a UAV and a BS is considered, where the UAV can switch between two modes, i.e., uploading raw data or extracted features, with different C$^2$ resource consumption. Such mode switching is jointly optimized with the transmit power allocation for multiple timeslots and computational speed to minimize the total energy consumption. The resultant optimization problem is solved by embedding the optimal power allocation into a tiny reinforcement learning algorithm for optimal mode selection. In light of the tradeoff between reducing communication cost and preventing semantic loss as regulated by the compression ratio, in \cite{RN183} a joint ratio selection and resource allocation problem is formulated for a multi-user semantic communication system aiming at supporting inference tasks for users. The solution approach is through alternative minimization that iterates between compression ratios and resource allocation. Further, a user selection algorithm is designed via branch and bound under communication resource constraints, considering that local tasks have different priority levels. Considering split ISEA where multi-user performs inference offloading with heterogeneous accuracy and latency requirements, the authors in \cite{RN155} propose integrated batching and early exiting to maximize the system inference throughput. Under a total bandwidth constraint and different channel states, the optimized allocation of computation and communication resources for inference throughput maximization is realized via a low-complexity sub-optimal algorithm and a tree search-based optimal algorithm. For collaborative ISEA, the splitting point between sensor and server sub-models can be jointly optimized with C$^2$ resources for minimized energy consumption subject to delay constraints\cite{Model_Splitting_RRM_TWC2024}. For collaborative vehicles, \cite{lu2025rrm_veh} jointly optimizes collaborator selection, compression ratio, offloading decision, and computation resource allocation to minimize the E2E delay subject to a perception probability requirement.
\section{AirComp-based Air Interface for Integrated Sensing and Edge AI}\label{section-aircomp}

\begin{figure*}[t]
\centering
\subfigure[]{\includegraphics[width=0.32\textwidth]{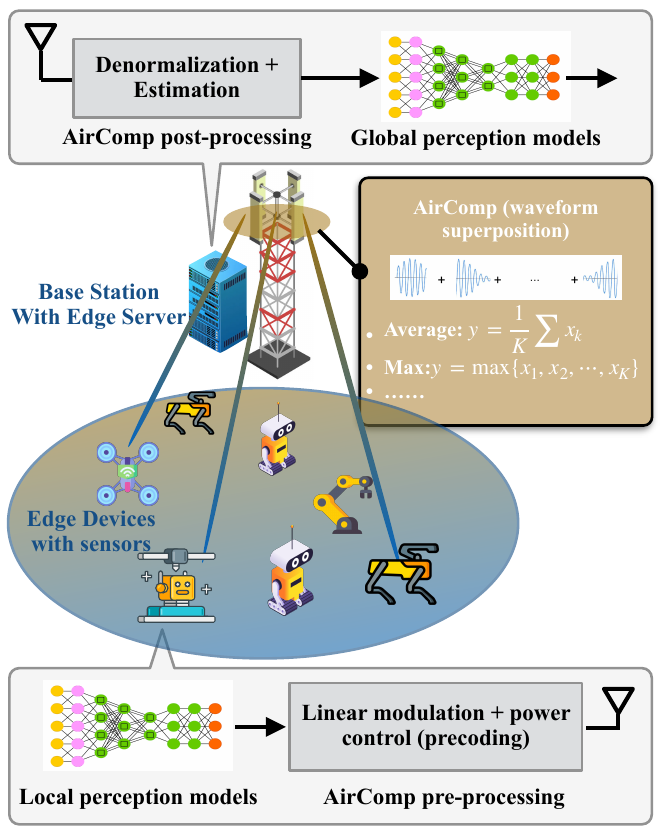}}
\subfigure[]{\includegraphics[width=0.65\textwidth]{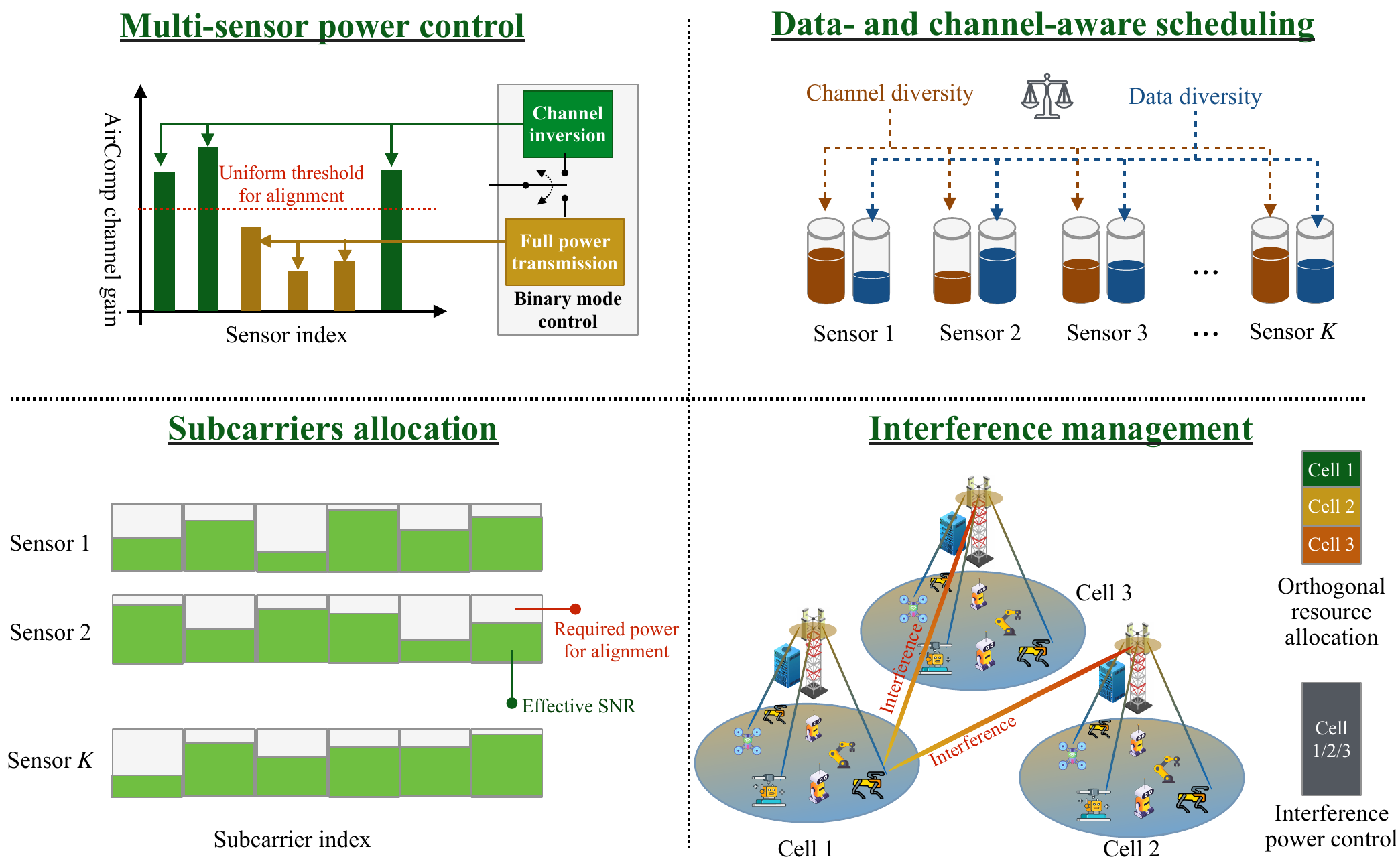}}
\label{fig:aircomp}
\caption{
From the AirComp-based air interface perspective, the paradigms and resource allocation strategies are specifically designed for ISEA. Subfigure (a) illustrates the general architecture, while subfigure (b) showcases various RRM schemes, including power control, sub-carrier assignment, sensor scheduling, and interference management.
}
\end{figure*}
While digital air interface benefits from modular design and compatibility with existing systems, analog communication schemes show potential in the most latency-stringent ISEA operations within limited resources as they promise ultra-low latency and high scalability. As a representative technology, AirComp leverages the waveform superposition property of wireless multi-access channels to realize
the desired aggregation function of distributed data by simultaneous transmissions. Its validness in ISEA results from the graceful performance degradation in AI-empowered sensing applications, i.e., model training/inference tasks are tolerant to moderate errors in data/feature
transmission with minuscule performance loss thanks to the inherent robustness of AI models. In this section, we review the advancements in developing AirComp-based techniques for ISEA, ranging from fundamental system design to RRM to practicality issues.

\begin{table*}[!ht]

\centering
\caption{Summary of Related Works on AirComp-based Air Interface for ISEA }
\scriptsize

\begin{center}
    
\setlength{\tabcolsep}{0.8mm}{%
\begin{tabular}{|>{\centering\arraybackslash}m{0.10\textwidth}|c|m{0.24\textwidth}|m{0.11\textwidth}|m{0.45\textwidth}|}
\hline
\multicolumn{1}{|>{\centering\arraybackslash}m{0.11\textwidth}|}{\textbf{Approaches}}                       & \textbf{Ref.}      & \multicolumn{1}{>{\centering\arraybackslash}m{0.24\textwidth}|}{\textbf{Sensing Scenario}}                                              & \multicolumn{1}{>{\centering\arraybackslash}m{0.10\textwidth}|}{\textbf{Performance Metrics}}                    & \multicolumn{1}{>{\centering\arraybackslash}m{0.45\textwidth}|}{\textbf{Key Contribution}}                                                                                                                                                                                                           \\ \hline
\multirow{9}{*}{\makecell[c]{AirComp-based\\Data Aggregation\\Design}} & \cite{GX2019IOTJ}  & Multi-modal sensors collect raw data and upload them to the server                                                                                                   & MSE                                            & Propose AirComp with transmit power control and receive beamforming in a MIMO system for fast sensory data aggregation                                                                                                \\ \cline{2-5} 
                                                       & \cite{Aggregation_gain_XChen}  & Sensory feature aggregation via AirComp for average-pooling                                                                 & Sensing (inference) uncertainty                & Propose a theoretical framework for quantifying the performance of multi-view sensing and figure out its relations to system parameters                                                                                                \\ \cline{2-5} 
                                                       & \cite{xiaowen2021interference}  & Sensory data aggregation using AirComp in a multi-cell network                                                                                                                       & {MSE}                           & {{Investigates optimal interference management policies to minimize the MSE in aggregated signals received at different APs.}}                                                                                          \\ \cline{2-5}
                                                       & \cite{Liu_TWC2020}  & Sensory data aggregation with peak-power constraints                                                                                                        &      MSE, peak power                                          &  Optimize the transmitting-receiving policy under the peak power constraints to minimize the aggregation MSE                                                                                                                                                                                                                                      \\ \cline{2-5} 
                                                       & \cite{Zhang_WCL2020}  & Sensory data aggregation with  sum-power constraints                                                                                                        & MSE, sum power                                 & Optimize the transmitting-receiving policy, under the sum power constraints, to minimize the MSE of aggregated signals at the receiver                                                                                                 \\ \cline{2-5} 
                                                       & \cite{WCL_AirComp_temporal}  & AirComp of spatial-and-temporal correlated signals                                                                                  & MSE                                            & Propose a low-complexity AirComp policy for achieving the minimum MSE                                                                                                                                                                  \\ \cline{2-5} 
                                                       & \cite{RIS_AirComp_TC} & RIS-assisted AirComp in single-cell IoT network                                                         & MSE                                            & Joint design of AirComp transceivers and RIS phase-shifts to minimize distortion                                                           \\ \cline{2-5} 
                                                       & \cite{RN358} & Sensory feature aggregation via AirComp for max-pooling                                                                 & Classification accuracy, MSE                  & A generalized AirComp framework for the max-function computation of features among sensors and its optimization                                                                   \\ \cline{2-5} 
                                                       & \cite{RN125} & BS and intelligent omni surfaces cooperate for joint multi-target sensing and multi-user communication                                                                    & Sensing SINR, data rate                        & Jointly optimize radar receive vector, transmit beamforming, and IOS coefficients to maximize the SINR of the multi-target under a data rate constraint. \\ \hline
\multirow{12}{*}{\makecell[c]{RRM}       }     & \cite{GX2020TWC} & FL data aggregation with AirComp over broadband channels                                                                                                                             & AirComp SNR, sensor truncation ratio           & Derives the tradeoffs between communication and learning metrics in truncated channel inversion                                                                                                                                        \\ \cline{2-5} 
                                                       & \cite{Xiaowen2020TWC} & Sensory data aggregation with AirComp over fading channels                                                                                                                           & MSE, average power                             & Optimal threshold-based power control for static channels and regularized channel inversion for time-varying channels                                                                                                                  \\ \cline{2-5} 
                                                       & \cite{Tao2021TWC} & Over-the-air FL data aggregation                                                                                                                                                     & MSE                                            & Optimal power control with heterogeneous gradient statistics                                                                                                                                                        \\ \cline{2-5} 
                                                       & \cite{Deniz2020TSP} & Over-the-air FL gradient aggregation                                                                                                                                                 & Convergence rate                               & Convergence analysis of AirComp-based gradient aggregation and comparison with digital transmission                                                                                                                                    \\ \cline{2-5} 
                                                       & \cite{RN187} & Over-the-air distributed PCA over noisy fading channels                                                                                                     & Convergence speed, power consumption           & Region-aware adaptive power control that exploits channel noise to escape saddle points                                                                                                                                                \\ \cline{2-5} 
                                                       & \cite{Shi2020TWC} & Over-the-air FL data aggregation with a multi-antenna edge server                                                                                                                    & Number of scheduled sensors                    & Jointly optimizes the receive beamforming and sensor scheduling to maximize the number of active sensors under error constraints                                                                                                       \\  
                                                       \cline{2-5}
                                                       & \cite{RN108} & Over-the-air FL gradient aggregation                                                                                                                                                 & Convergence rate                               & Dynamic and decentralized sensor scheduling considering CSI, local gradient and accumulated residual gradients                                                                                                                         \\ \cline{2-5} 
                                                       & \cite{10038617} & Over-the-air FL gradient aggregation                                                                                                                                                 & Convergence rate                               & Gradient and channel aware device scheduling that trades off between sensor quantity and quality with multi-objective optimization                                                                                                     \\ \cline{2-5}
                                                       & \cite{10341307}  & Over-the-air FL gradient aggregation                                                                                                                                                 & Convergence rate             & Probabilistic device scheduling using channel and gradient importance                                                                                                             \\ \cline{2-5} 
                                                       & \cite{airfusion} & Broadband over-the-air fusion of spatial features                                                                 & Minimum receive SNR                            & Optimal pairing of voxel-level subtasks and subcarrier to leverage feature sparsity and channel diversity                                                                                                                              \\ \cline{2-5} 
                                                       & \cite{cflit} & IT users and AirComp FL users sharing broadband channels                                                                                           & IT rate, FL convergence rate & An online subcarrier allocation algorithm to maximize the IT rate with FL convergence guarantees                                                                                                                     \\ \cline{2-5} 
                                                       & \cite{RN249} & Broadband AirComp for general data aggregation                                                                                                                                       & MSE                                            & Designs a broadband AirComp scheme where each sensors broadcast over multiple subcarriers to exploit diversity                                                                                                                         \\ \hline
                                                   \multirow{14}{*}{\makecell[c]{Practical\\AirComp}       }    & \cite{lan2020simultaneous} & AirComp in two cells with inter-cell interference                                                                                                                                    & AirComp degrees-of-freedom                     & A simultaneous signal-and-interference alignment scheme for interference management                                                                                                                                                    \\ \cline{2-5} 
                                                       & \cite{9770061} & Multi-cluster AirComp                                                                                                                                & MSE                                            & Distributed transceiver design to suppress interference                                                                                                                                                                          \\ \cline{2-5} 
                                                       & \cite{spectrumbreathing} & Over-the-air FL under interference                                                                                                                                                   & Convergence rate                               & Cascades gradient pruning and spread spectrum to suppress interference without bandwidth expansion with optimized tradeoff                                                                                                             \\ \cline{2-5}
                                                       & \cite{aircomp_byzantine} & Over-the-air FL under Byzantine attacks                                                                                                                                                   & Convergence rate                               & Implememt one-bit gradient quantization and majority vote to enhance robustness against Byzantine attackers                                                                                                             \\ \cline{2-5}
             & \cite{Robust_RS} & Wireless sensor network with function computation of  measurements                                                                         & Function estimation error                      & Propose an analog computation scheme that allows for an efficient estimate of linear and nonlinear functions                                                                                 \\ \cline{2-5} 
                                                       & \cite{Robust_RS1} & Sensory data aggregation using AirComp and RIS with imperfect CSI                                                                                           & MSE                                            & Propose a framework utilizing RIS to assist the AirComp-based data aggregation with imperfect CSI                                                                                                                                      \\ \cline{2-5} 
                                                       & \cite{Robust_RS2} & Sensory data aggregation using AirComp and RIS with imperfect CSI                                                   & MSE                                            & Optimized transceiver design for RIS-assisted AirComp with imperfect CSI and power constraints                                                                               \\ \cline{2-5} 
                                                       & \cite{Robust_Blind} & A wireless sensor network with multiple sensor nodes for data collection                                             & Function estimation error                      & Proposes a blind AirComp without requiring CSI access for low complexity and low latency data aggregation                                                                                                            \\ \cline{2-5} 
                                                       & \cite{GX2021TWC} & Over-the-air FL gradient aggregation                                                                                            & Convergence rate                               & Propose one-bit broadband digital AirComp for the aggregation high-dimensional stochastic gradients                                                                                           \\ \cline{2-5} 
                                                       & \cite{liu2024digital} & Sensory data aggregation using AirComp                                                                                                                      & MSE                                            & A framework for digital AirComp with digital modulation of data points and optimized bit-slicing for bit allocation between symbols                                                \\ \cline{2-5} 
                                                       & \cite{USRP2} & Distributing sampling and computing of spectrum sensing to multiple sensor nodes for cognitive radio                                        & Relative reconstruction MSE                    & Propose a cooperative wideband spectrum sensing scheme using AirComp for the summation of Fourier transform and establish a demo on the platform of USRP                                                                               \\ \cline{2-5} 
                                                       & \cite{USRP3} & Sensory data aggregation using AirComp                                                                               & MSE                                            & Real-time implementation of sensor AirComp achieved by using SDR nodes and extensive tests                       \\ \cline{2-5} 
                                                       & \cite{XSDR} & Over-the-air FL gradient aggregation                                                                                            & Training loss                                  & Propose a two-stage waveform pre-equalization technique and a protocol for over-the-air FL with hardware implementations                                              \\ \cline{2-5} 
                                                       & \cite{WSN}& Sensory data aggregation using AirComp for environment perception & Computation error                              & Implementation and evaluation of  AirComp over 15 hardware sensor nodes                                                                                           \\ \cline{2-5} 
                                                       & \cite{SDSDR} & Sensory data aggregation using AirComp                                                                               & Computation error                              & Establish a testbed using SDSDR devices to validate the AirComp-based sensory data aggregation                                                                                                                                         \\ \hline
\end{tabular}%
}
\end{center}

\end{table*}

\subsection{AirComp-based Data Aggregation Design} 
Data aggregation is crucial for ISEA systems, fusing observation, knowledge, and results from edge devices to achieve a comprehensive perception. AirComp accelerates this fusion process by integrating the fusion computation into communication, leveraging the waveform superposition property of wireless multi-access channels. The involved AirComp techniques, based on the desired functional form, can be classified into nomographic and non-nomographic function computation methods.
\subsubsection{Nomographic Data Aggregation} Nomographic functions are a set of multivariate functions that can be represented as a post-processed function over the sum of pre-processed functions of individual variables. Its instances commonly used in ISEA include weighted sum, arithmetic mean, geometric mean, Euclidean norm, and polynomials. The weighted sum, for example, is employed in FL for gradient aggregation\cite{GX2020TWC} and model fusion and multi-view sensing for feature pooling~\cite{RN358,XSDR}. This decomposable property lays the foundation for AirComped nomographic functions in ISEA. The pre-processed functions specified by ISEA tasks are offloaded to edge devices, of which the results are transmitted to the edge server simultaneously, forming a noisy sum of pre-processed functions through wave superposition. Finally, the desired nomographic function is reconstructed with channel distortions by feeding the received signal into its corresponding post-processing operation. To control the effects of channel noise, the parameters in the pre/post-processing algorithms will be tweaked according to system resources and functional properties, such as bandwidth, transmit power budget, channel fading, channel correlation, and signal correlation (see, e.g.,~\cite{Liu_TWC2020,Zhang_WCL2020,WCL_AirComp_temporal}).

    ISEA also involves the computation of high-dimensional nomographic functions, where the input and output of AirComp consist of high-dimensional vectors, such as vector-matrix multiplication leveraged in fully connected layers of neural networks. The input vector, consisting of the output features of the preceding layer, is multiplied by a trained matrix and passed through an activation function to obtain a new representation of features. In AirComp-based ISEA, the input vector is distributed over multiple devices and MIMO AirComp allows the desired vector-matrix multiplication achieved over the air~\cite{AirComp_SL_JSAC2023}. The core of MIMO AirComp is to create effective channel gains of space-frequency channels, through precoding and receiving methods, equivalent to the matrix elements of the matrix-vector multiplication specified by ISEA tasks. Furthermore, researchers also resort to reconfigurable intelligent surfaces (RIS) to directly manipulate the physical channels by programming the phase shifts of RIS~\cite{RN384}, which has the potential to improve the implementation of AirComp~\cite{RIS_AirComp_TC}. 
\subsubsection{Non-nomographic Data Aggregation} While the nomographic function computation dominates the computation tasks of ISEA, its non-nomographic counterpart, such as general non-linear operations, may also be used in sensory data aggregation.  One outstanding example is the max-pooling operation of multi-view sensing, which computes the element-wise maximum of features distributed across devices. Its integration into AirComp requires not only the aforementioned pre-processing and post-processing in nomographic functions but also a functional approximation due to its lack of decomposable property. To address this issue, the vector $p$-norm occurred in previous works for over-the-air approximation of the max function~\cite{RN358}. Such approximation inevitably results in a tradeoff between the approximation error and noise amplification.  

\subsubsection{End-to-End Signal Processing}
It should be emphasized that ISEA applications always require the integration of AirComp-based computation and subsequent complicated computation, such as distributed analytics and estimation. For example, in multi-view sensing with MVCNN architecture, the local features can be fused into a global feature map via average or max pooling. The obtained global feature is then taken as input to the following classifier for final classification. To optimize the E2E performance, AirComp processing algorithms should be designed using E2E metric optimization rather than minimizing the AirComp error. To address this issue, primary efforts have been made to characterize the E2E accuracy in terms of error metrics, such as discriminant gain, Mahalanobis distance, and MSE, using the theory of classification margin, information theory, and estimation~\cite{RN358,Aggregation_gain_XChen}. These studies lay a preliminary theoretical foundation for performance evaluation and algorithm design for AirComp-based ISEA. Moreover, the architecture of FL can be employed to propose a joint design of dynamic learning rate and AirComp transceiver~\cite{RN125}. This design enables the learning rate to adapt to the time-varying wireless environment, achieving an importance-aware distortion control that enhances the learning performance.

\subsection{Radio Resource Management}\label{subsection-aircomp-rrm}
While the above signal processing pipelines and performance analysis enable AirComp for ISEA, it still encounters several challenges in realistic settings, such as hostile channel states, inter-cell interference, heterogeneous computation capabilities, and data quality across devices. Combating these issues to optimize the E2E sensing performance requires task-oriented RRM throughout the AirComp process. The existing literature develops various schemes in aspects such as power control, sub-carrier assignment, sensor scheduling, and interference management schemes to optimize E2E performance metrics or their tractable surrogates, as illustrated in Fig.~7~(b) and reviewed in detail below.
\subsubsection{Power Control} Power control is critical for sensors subject to stringent battery-life and peak-power constraints. Classical AirComp power control aims at MSE minimization assuming independent data streams for either instantaneous\cite{Xiaowen2020TWC,Liu_TWC2020} or total power constraints\cite{Zhang_WCL2020}. However, in the ISEA context, these classical schemes become sub-optimal as some assumptions and metrics need to be revised for sensing. The first reason is the non-IID and non-ergodic distribution of sensory data as a natural result of sensor positions and mobility. For example, \cite{Tao2021TWC} takes the gradient statistics into consideration, which may vary across sensors and communication rounds. With the gradient mean and variance known, the optimal power control policy for MSE minimization, depending on channel gain, noise power and gradient statistics, is derived in closed form. A practical adaptive power control scheme is designed when the current gradient statistics are unknown via parameter estimation with historical statistics. 
As the second reason, although MSE is shown to be loosely relevant to the downstream task performance in ISEA, for example, learning convergence speed \cite{Deniz2020TSP} and inference accuracy\cite{RN358}, optimization for MSE is still inherently sub-optimal for ISEA tasks. Direct orientation towards the E2E metric yields more efficient power control scheme by, e.g., optimally trading off between sensor inclusion and aggregation reliability \cite{GX2020TWC}.
In \cite{ISCC_AI_2024}, an ISEA inference scenario is targeted where multiple devices equipped with RF sensing hardware participate in feature AirComp for inference at the fusion center. To enable E2E optimization, the minimum pairwise discriminant gain is analyzed and formalized as the optimization objective of joint power allocation, transmit precoding, and receive beamforming. The proposed scheme, leveraging successive convex approximation, allows resource allocation in terms of both channel states and feature importance. Interestingly, the AirComp noise, which classical power control aims to suppress, can be even beneficial for the downstream task.
In \cite{RN187}, the authors study over-the-air principal component analysis and discover that the perturbation to aggregated gradient caused by channel noise can assist the escaping from saddle points, rendering gradient MSE minimization inappropriate in saddle regions. For this reason, a region-adaptive power control scheme is then designed, which suppresses noise in non-stationary regions but reduces the power level within the saddle region to exploit the effect of noise. 

\subsubsection{Data- and Channel-Aware Sensor Scheduling}
From the communication perspective, it is well-known that the aggregation MSE of AirComp is limited by sensors suffering from deep fading, calling for sensor scheduling based on channel states. For example, the prevalent truncated channel inversion scheme, which assigns zero power to sensors in deep fading, can be viewed as a sensor scheduling mechanism based on a channel threshold \cite{GX2020TWC}. On the other hand, from the computing perspective, sensor scheduling should take into account the importance of heterogeneous data across sensors. Generic data-importance metrics include distance to supporting hyperplane, entropy, and gradient norm for FL, as well as discriminant gain and contribution estimators based on first-order expansion. Depending on the ISEA application scenario, data importance can be further complemented by sensing-relevant metadata such as location, view angle, and sensing coverage. Oriented towards AirComp-based ISEA performance, sensor scheduling shall integrate data and channel awareness. The relevant works are introduced below. 

Observing that the convergence performance of FL improves with the number of participating sensors, the authors in \cite{Shi2020TWC} have formulated the AirComp sensor scheduling problem as joint sensor selection and receive beamforming to maximize the number of scheduled sensors under an MSE requirement and power constraints, which is further transformed into a sparse and low-rank optimization. The authors then propose an algorithm based on difference-of-convex functions programming with near-optimal performance. In \cite{RN108}, the authors first propose a dynamic residual feedback mechanism to combat data distortion, where sensors, instead of discarding the local gradient when not scheduled for transmission, record historical gradients and transmit a weighted sum of the current and accumulated gradient when being scheduled. Then, a decentralized optimization framework based on the Lyapunov drift optimization methods is constructed to optimize the sensor scheduling and power control, adapting to both channel states and gradient importance. By also adopting the residual feedback mechanism, the authors in \cite{10038617} formulate the sensor scheduling problem as a multi-objective optimization problem with the tradeoff between scheduling more sensors and device quality maximization, where the latter is measured by the sum of a designed quality indicator comprising data qualities, channel gains, and energy consumptions, and then proposed a dynamic scheduling algorithm based on Lyapunov optimization. In \cite{10341307}, the authors study probabilistic device scheduling to resolve the bias issue in deterministic scheduling policies and propose an algorithm considering both channel conditions and gradient importance. For ISEA inference, \cite{airfusion} designed an over-the-air feature fusion protocol involving a sparsity pattern feedback phase to identify sensors with empty voxel feature vectors, which are thus excluded from AirComp. Subsequently, the optimal power control is derived in closed form. 

\subsubsection{Subcarrier Allocation}
To overcome the challenges of high-throughput fusion of high dimensional gradients/features, AirComp in ISEA is expected to operate on multiple subcarriers in a broadband system in the forms of multiplexing, where the aggregation of each dimension/task is assigned to one of the subcarriers~\cite{GX2020TWC}, diversity, where one symbol is simultaneously transmitted on several carriers, or a mixture of both. In view of the coexistence of data/task heterogeneity and different channel fades on different subcarriers, optimizing subcarrier allocation can significantly boost the AirComp performance. In \cite{cflit}, the authors consider the co-existence of information-transfer (IT) users and AirComp FL users in a broadband system sharing broadband channels, resulting in a subcarrier allocation problem between the two sets of users, which aims to maximize the IT rate subject to a guaranteed FL convergence rate. The convergence rate constraint is translated into a minimum number of subcarriers assigned for FL, based on which an online subcarrier allocation algorithm is designed. In \cite{airfusion}, an environment perception task is considered, which exhibits heterogeneous feature sparsity patterns across sensors. The designed AirFusion protocol uses a feedback process to enable awareness of sensor-wise sparsity on each voxel dimension and design a set of algorithms to optimally pair voxels with subcarriers such that the sparsity pattern matches the channel gain pattern. Exploiting broadband channels for diversity, \cite{RN249} considers an AirComp system where each sensor broadcasts its data simultaneously on multiple subcarriers, and the server estimates the aggregated data using the received signal on all subcarriers. An optimization problem to determine the subcarriers to transmit on and the invested power for each sensor is formulated and solved by alternating minimization.

\subsection{Practical AirComp}\label{subsection-aircomp-practical}
Despite many well-established theories and algorithms for AirComp, there is still a gap between theoretical assumptions and real-world conditions when deploying AirComp in practical ISEA systems. In this subsection, the existing solutions for addressing the implementation issues are reviewed in the aspects of robustness, compatibility, and prototype validation.
\subsubsection{Robust AirComp}
ISEA tasks, such as autonomous driving and robotics, are mission-critical, imposing stringent robustness requirements on ISEA algorithms in the face of non-ideal environments in practice. However, the common assumptions adopted by algorithm design, including perfect device synchronization and accurate CSI, could be inapplicable because of the unaffordable costs paid for feedback bandwidth restriction and large-scale networking.  For example, time-division duplexing (TDD) mode is mainly considered in existing AirComp architecture, where the signaling overhead during channel transmission can increase at a linear rate with respect to the number of devices, leading to high communication latency for massive-access networks. To this end, robust design under imperfect CSI has been extensively studied, with attempts made to achieve reliable functional computation under different scenarios, such as RIS-assisted AirComp~\cite{Robust_RS1,Robust_RS2} and single-input-single-output (SISO)/SIMO/MIMO AirComp~\cite{Robust_MIMO1,Robust_MIMO2,Robust_MIMO3}. By modeling channel estimation error in either stochastic (i.e., expectation) or deterministic (i.e., worst-case) models, robust AirComp transceivers are designed to combat channel distortions caused by erroneous channel feedback. Additionally, the authors in~\cite{Robust_Blind} have developed a CSI-free AirComp system, leveraging blind demixing to recover a sequence of signals without accessing any CSI, thereby achieving the desired function computation. 

On the other hand, moving devices, such as vehicles and UAVs functioning as sensors and data processors, play a principal role in ISEA systems. These devices exhibit a strong degree of mobility in contrast to mobile phones and fixed cameras in traditional networks, rendering their wireless connection potentially asynchronous. Specifically, using existing synchronization techniques in 4G/5G standards, the resulting time offset is inversely proportional to the leveraged bandwidth. This matter can be of crucial significance for AirComp-based ISEA since AirComp depends on synchronization among sensors to attain accurate wave superposition. To combat this issue, some research efforts have been made. As exemplified by~\cite{Robust_RS}, the random sequence technique is employed to transmit sensor data, thereby effectively countering misalignment in wave superposition. Signal processing methods can also be employed to control coarse synchronization. For instance, \cite{Robust_ML} uses a maximum-likelihood estimation approach to eliminate channel distortion caused by channel-gain variation and device asynchrony.

In addition, AirComp performance is prone to interference from jammers or neighboring cells, which may lead to sudden drops in the sensing task performance, i.e., AI outage. A set of RRM techniques has been developed for interference management. In\cite{lan2020simultaneous}, a scheme called simultaneous signal-and-interference alignment is proposed to manage inference in a two-cell system. It divides the MIMO channel space into an interference alignment space and signal alignment space, and by constraining the interference into the former, the number of interference-free dimensions for AirComp can be maximized. In \cite{xiaowen2021interference}, the centralized power control problem for a multi-cell AirComp network is tackled by profiling the Pareto front of multi-cell MSEs and solving a sequence of second-order cone programming problems, and a distributed algorithm designed by exploiting the interference temperature technique. For multicluster networks, \cite{9770061} developed a centralized branch-and-bound algorithm and a distributed algorithm for the transceiver optimization for transceiver design to minimize the aggregation MSE accounting for both signal misalignment error and interference. A novel technique, termed spectrum breathing, is developed in \cite{spectrumbreathing}, which features cascaded gradient pruning and spread spectrum to reduce interference under a given bandwidth budget. In addition, in \cite{aircomp_byzantine}, one-bit gradient quantization and majority-vote aggregation have been proposed in over-the-air FL systems to counter Byzantine attacks.

\subsubsection{Digital AirComp}
Traditional AirComp architecture is mainly based on uncoded analog transmission and waveform superposition of linearly modulated symbols. This indicates that analog AirComp can ensure average error performance but may fail to safeguard crucial sensory features for mission-critical ISEA tasks. Inspired by this, digital AirComp is put forward to protect data aggregation from channel adversity through nonlinear modulation and coding techniques. The approaches of integrating AirComp with existing digital techniques mainly consist of regarding AirComp as a special type of channel, encoding operations, or mapping between finite fields~\cite{liu2024digital}. Consequently, the designed computation/aggregation results can be retrieved by proposing signal processing techniques such as maximum-likelihood (ML) decoders or detectors~\cite{CodedAirComp1,CodedAirComp2,CodedAirComp3,CodedAirComp4,USRP1}. In these studies, nested lattice coding has the potential to achieve data aggregation with digital AirComp as its inherent linearity guarantees the decodability of coded data~\cite{CodedAirComp1,CodedAirComp2}. Additionally, the authors in \cite{USRP1} have developed a framework of digital AirComp with Low Density Parity Check Code (LDPC) coded data aggregation over non-orthogonal OFDM subcarriers, along with joint channel decoding and aggregation decoders customized for convolutional and LDPC codes. 

It is worth noting that digital AirComp presents a natural tradeoff between reliability and latency. At the expense of data rate reduction, reliable data aggregation is facilitated by a digital scheme featuring high-precision quantization and low-rate coding. This renders it feasible to flexibly adapt digital AirComp's coding, quantization, and modulation strategies to ISEA applications with diverse latency-reliability requirements. For instance,~\cite{GX2021TWC} proposes a one-bit AirComp scheme that quantizes the coefficients of stochastic gradient uploaded by each device into single bits. It is shown in~\cite{GX2021TWC} that such one-bit AirComp can achieve low latency for gradient aggregation while allowing for the resulting quantization errors satisfactory for FL tasks. The data importance also involves the latency-reliability tradeoff.~\cite{Layerwise_onebit} has demonstrated that optimizing the power control and compression ratio, based on the convergence effect of quantization and sparsification in different layers, can achieve the fastest convergence rate.

\subsubsection{Demo Development}
To validate the feasibility and effectiveness of AirComp in real-world scenarios, several prototypes have been developed in several platforms including Universal Software Radio Peripheral (USRP)~\cite{USRP1,USRP2,USRP3}, Xilinx Software-Defined Radio~\cite{XSDR}, Wireless Sensor Network~\cite{WSN}, and Self-developed Software Defined Radio (SDSDR)~\cite{SDSDR}. For instance, to demonstrate the concepts of AirComp-assisted FL, \cite{USRP1} and~\cite{SDSDR} have respectively leveraged the USRP and SDSDR platforms to verify the effectiveness of proposed transmission AirComp schemes. In~\cite{SDSDR}, the authors establish a testbed using SDSDR devices to validate a transmission scheme proposed in~\cite{Robust_RS}, where the testbed includes one antenna configured as the function computation and others emulating separate single-antenna sensor nodes. Additionally, the authors in~\cite{WSN} validate their proposed transmission scheme, which encodes values into Poisson-distributed burst sequences to perform mathematical functions via AirComp, using a wireless sensor network platform with 15 simple nodes and a central receiver.

\section{Advanced Signal Processing for Integrated Sensing and Edge AI}\label{section-advsp}
Signal processing is a crucial component in ISEA, serving, e.g., pre-processing of raw sensory datasets, noise removal, and compression before input to the air interface, as well as signal detection and multi-view multi-modal fusion at the receiver. While traditional analog and digital signal processing typically aims at the best reconstruction of the carried data, advanced signal processing techniques for ISEA shall aim at sensing task performance with potentially multiple objectives, such as sensing accuracy, E2E latency, and privacy.  This section reviews the research efforts on advanced signal processing techniques and frameworks, focusing on several emerging aspects during the evolution to ISEA within 6G. These aspects include {on-the-fly communication and computing} (FlyCom$^2$), multi-functional waveforms, and privacy-preserving techniques.

\begin{table*}[!ht]

\centering
\caption{Summary of Related Works on Advanced Signal Processing for ISEA }
\scriptsize

\begin{center}
\setlength{\tabcolsep}{0.8mm}{%
\begin{tabular}{|>{\centering\arraybackslash}m{0.10\textwidth}|c|m{0.21\textwidth}|m{0.12\textwidth}|m{0.48\textwidth}|}
\hline
\multicolumn{1}{|>{\centering\arraybackslash}m{0.10\textwidth}|}{\textbf{Approaches}}                       & \textbf{Ref.}      & \multicolumn{1}{>{\centering\arraybackslash}m{0.21\textwidth}|}{\textbf{Sensing Scenario}}                                              & \multicolumn{1}{>{\centering\arraybackslash}m{0.12\textwidth}|}{\textbf{Performance Metrics}}                    & \multicolumn{1}{>{\centering\arraybackslash}m{0.48\textwidth}|}{\textbf{Key Contribution}}                        \\ \hline
\multirow{5}{*}{\centering \makecell[c]{FlyCom$^2$}} & \cite{RN383} & Sensors capture raw data as tensors for fusion and PCA & Decomposition error                                             & Propose a framework of FlyCom$^2$ for DTD of local sensory datasets with streaming computation and progressive global aggregation                                         \\ \cline{2-5} 
                                                & \cite{Pyramid} 
                                                & Image applications requiring features extracted from raw images                                                                                             & Average recall, average precision                               & Exploit the inherent pyramidal hierarchy of deep convolutional networks to construct feature pyramids                                                                                                                                                                                               \\ \cline{2-5} 
                                                        & \cite{OcTr} & LiDAR-based 3D object detection                                                                                                                                       & Average precision                             & Propose an Octree-based Transformer to capture sufficient features from large scale 3D scenes                                                                                                                                                                          \\ \cline{2-5}
                                                        & \cite{Muscle} & LiDAR sensors collect point cloud streams transmitted and stored within limited data volume                                                          &                          Average precision                                &                                                                                                          A compression algorithm that exploits spatio-temporal relationships across multiple LiDAR sweeps to reduce the storage of LiDAR sensor data streams.                                                                                                                                                                                                   \\ \cline{2-5} 
                                                        & \cite{FlyCom_PC} & Fusing point clouds to learn a unified representation at a central server                                      & MSE                                                             & Propose a FlyCom2 framework to enable  PtCloud fusion with streaming  processing, progressive data uploading integrated with AirComp, and the progressive output of a global PtCloud representation.                                                                                    \\ \hline
\multirow{14}{*}{\makecell[c]{Multi-Functional\\Waveforms}}  & \cite{RN90} & RF-Sensing shares spectrum and hardware with communication in millimeter wave band                                                                                     & Spectrum efficiency, MSE                                        & Overview radar-communication coexistence and dual-functional radar-communication (DFRC) systems and propose a transceiver architecture and frame structure for a DFRC BS                                                                                   \\ \cline{2-5} 
                                                        & \cite{MVW_Chen2021} & Simultaneous communication and radar-sensing with unified spectrum and transceivers                                                  & Bit error rate, average error, MSE                              & Propose a code-division orthogonal frequency-division multiplex (CD-OFDM) scheme with interference cancellation methods for joint communication and sensing                                                                                                                                         \\ \cline{2-5} 
                                                        & \cite{Aboulnasr2016} & A joint radar-communication system using a dual-functional waveform                                                                                                   & Bit error rate                                                  & Exploit orthogonal waveforms for embedding a sequence of information bits during each radar pulse                                                                                                                                                                                                   \\ \cline{2-5} 
                                                        & \cite{Tedesso2018} & A joint radar-communication system using a dual-functional waveform                                                                                                   & Symbol error rate, ambiguity function & Propose to use BPSK and QPSK signals for both bit modulation and pseudo-random coded radar                                                                                                                                                                                                          \\ \cline{2-5} 
                                                        & \cite{KaiWu2021} & A FH-MIMO radar-based dual-function radar communication                                                                                           & MSE                                                             & Design a waveform and a estimation method for FH-MIMO receivers to estimate the frequency hopping sequence and channel parameters                                                                                                                                                                   \\ \cline{2-5} 
                                                        & \cite{LFM-MSK2011} & A radar system with dual-functional LFM signals                                                            & Ambiguity function                                              & proposes to generate a sensing-communication compatible waveform by modulating LFM signal by minimum shift keying.                                                                                                                                                                                  \\ \cline{2-5} 
                                                        & \cite{LFM-CPM2019} & A joint radar-communication system using a dual-functional waveform                                                                                                  & Bit error rate                                                  & Exploit the combination of LFM signals and continuous phase modulation signals for simultaneous data transmission and target detection                                                                                                                                          \\ \cline{2-5} 
                                                        & \cite{RN378} & An uplink system with a full-duplex BS illuminating an imaging scene while receiving data from a user                                    & Degrees of freedom                                              & Propose a unified signal space analysis framework based on the degrees of freedom metric to characterize the trade-offs between sensing (imaging) and communication                                                                                                                                 \\ \cline{2-5} 
                                                        & \cite{RN390} & ISAC with a unified dual-functional waveform for simultaneous communication and sensing                                                          & Data rate, Cramér-Rao bound                                     & Analyze the fundamental sensing-communication tradeoff under a Guassian channel based on the tool of Cramér-Rao bound (CRB)-Rate region                                                                                                                                                             \\ \cline{2-5} 
                                                        & \cite{FLiu2018TSPdual} & An MIMO-radar system which communicates with downlink users and detects radar targets simultaneously                               & Sum data rate, radar detection probability, MSE                 & Design a dual-functional waveform for MIMO radar, controlling the tradeoff between radar and communications performance                                                                                                                                                                              \\ \cline{2-5} 
                                                        & \cite{FLiu2018TWC} & An MIMO-radar system which communicates with downlink users and detects radar targets simultaneously                               & SNR, peak-sidelobe-ratio, MSE                                   & Propose a unified waveform  omited by transmit array incoporating radar probing beampartterns as well as communication signals                                                                                                                                                                      \\ \cline{2-5} 
                                                        & \cite{Damith2021} & MIMO radar with full-duplex hybrid beamforming for both sensing and communication                                                                                     & Transmitter radar power                                         & Optimize the transmit beamforming with concurrent multiple beams for communication and sensing and mitigate cross-interference therein                                                                                                                                                              \\ \cline{2-5} 
                                                        & \cite{Dokhanchi2021JCS} & Multi-target localization from a signal data stream                                                                          & Signal-to-clutter-plus-noise ratio                 & Design a precoder to maximize multicasting rate while maximizing the radar SCNR                                                                                                                                                                                                                     \\ \cline{2-5} 
                                                        & \cite{multi_obj_isac2024} & Multi-user and multi-target ISAC                                                                          & Sum rate, sensing error                 & Utilize constructive interference to combat multi-user interference and design ISAC waveforms to achieve the optimal Pareto front                                                                                                                                                                                                     \\ \cline{2-5} 
                                                        & \cite{RN99} & Simultaneous communication with an OFDM receiver and  target parameters detection               & Cramér-Rao bound, data rate                                     & Optimize subcarrier powers in a time-frequency region of interest to realize a performance trade-off between radar and communications                                                                                                                                                               \\ \hline
\multirow{10}{*}{\makecell[c]{Privacy-Preserving\\Signal Processing\\for ISEA} }              & \cite{PrivacyCode} & Multiple servers cooperate for linear inference requests from sensors                                                                                                                                  & Latency                                                              & A coding scheme based on Shamir's secret sharing to ensure that the original data cannot be recovered from the colluding servers                                                                                                                                                                                                                                                  \\ \cline{2-5} 
                                                        & \cite{EdgeSanitizer} & Split ISEA inference                                                                                                                                                 & DP, data utility                                                              &  Preserve privacy by a data minimization model and adaptive noise                                                                                                                                                                                                                                                                                                   \\ \cline{2-5} 
                                                        & \cite{Roulette} & Split ISEA inference                                                                                                                                               & DP, accuracy                                                              & Preserve privacy by a feature encryptor trained without exposing ground-true labels to the server                                                                                                                                                                                                           \\ \cline{2-5} 

                                                        & \cite{PrivacyForFree} & Over-the-air FL data aggregation                                                                                                                                           & DP, convergence rate                                               & Adaptive power allocation for orthogonal and AirComp FL with privacy guarantee and conditions for privacy without performance loss \\ \cline{2-5} 
                                                        & \cite{Deniz2022ISIT} & Over-the-air collaborative inference                                                                                                                                       & DP, accuracy                                        & Over-the-air ensemble inference with model privacy guarantee                                                                                                            \\ \cline{2-5} 
                                                        & \cite{AirFLDP1} & Over-the-air FL gradient aggregation                                                                                                                                           & DP, convergence rate                                                              & Adaptive noise injection per-device to achieve a desired DP level                                                                                                                                                                                                                                               \\ \cline{2-5} 
                                                        & \cite{AirInfDP1}  & Over-the-air collaborative inference                                                                                                                                      & DP, accuracy                                                              & Feature-aware device selection followed by noise injection to preserve privacy                                                                                                                           \\ \cline{2-5} 
                                                        & \cite{AirFLDP2} & Over-the-air FL gradient aggregation                                                                                                                         & DP, convergence rate                                                              & Random data dimension reduction for privacy amplification and power control scheme robust to CSI attacks                                                                                                                                                                                     \\ \hline
\end{tabular}%
}
\end{center}

\end{table*}

\subsection{On-the-Fly Computing and Communication}\label{subsection-advsp-flycom}
\begin{figure*}
    \centering
    \includegraphics[width=0.75\textwidth]{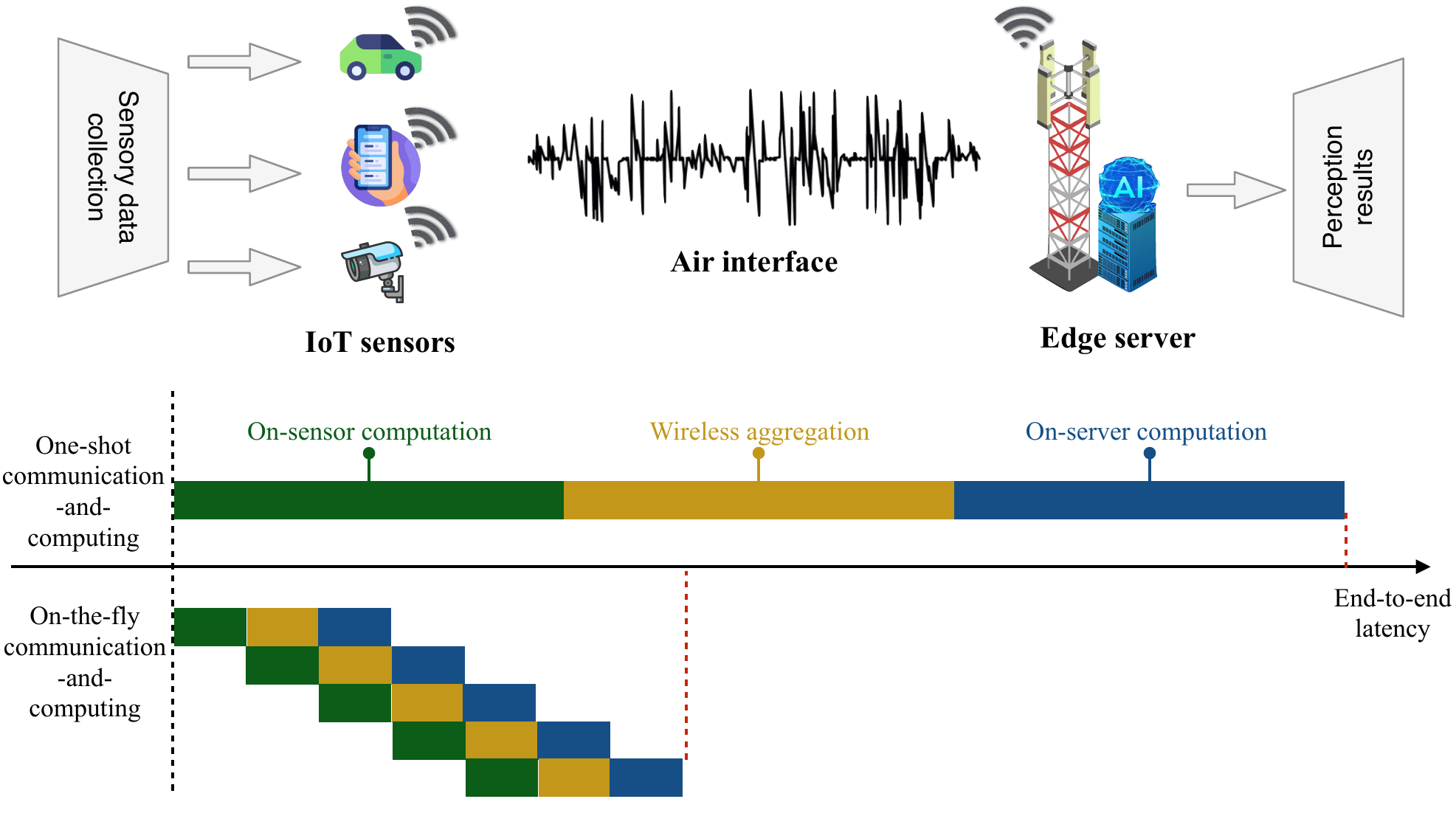}
    \caption{Compared to one-shot communication and computation, on-the-fly techniques decompose the sequential processing of high-dimensional data into streaming operations with low-dimensional communication and computation, thereby significantly reducing the E2E latency of ISEA.
}
    \label{fig: flycom}
\end{figure*}
The existing architectures of ISEA primarily rely on a framework that separates high-dimensional computing and communication processes. For instance, in split ISEA, intermediate features are extracted from extensive on-device data using AI models deployed on edge devices. These features are then transmitted to a central server for global detection. However, this sequential approach to computation and communication presents two main drawbacks regarding E2E latency and computation constraints. First, the E2E latency in ISEA, composed of computing and communication latency, may not meet the stringent latency requirements of time-sensitive applications. Second, this architecture imposes heavy computation requirements, including memory and complexity, on edge devices. In ISEA, sensory data collected by sensors can consist of several million data points, such as LiDAR-based environmental perception. The large volume of data makes it challenging for lightweight local sensors to perform feature extraction and other data analytics operations. Additionally, executing complex and energy-consuming computation tasks at sensors in a large-scale sensor network is undesirable due to considerations of network lifetime and energy efficiency.

The above challenges stemming from high-dimensional computation and communication can be addressed by the innovative framework of FlyCom$^2$~\cite{RN383, DEFT}. 
As illustrated in Fig.~\ref{fig: flycom},
the main idea behind this framework is to break down the sequential processing of high-dimensional data into streaming operations of low-dimensional communication-and-computing. The concept of FlyCom$^2$ was initially introduced in~\cite{RN383}, where a FlyCom framework was developed for distributed tensor decomposition (DTD) in distributed data analytics. The proposed FlyCom$^2$ offers several solutions to the aforementioned issues. First, it utilizes a low-complexity random sketching technique to reduce the computational complexity of on-device processing. This approach allows for efficient computation on edge devices. Second, the E2E performance of an ISEA system incorporating FlyCom$^2$ continuously improves as the framework accumulates low-dimensional local sketches. These sketches are leveraged by the server to infer patterns in the distributed dataset. Consequently, FlyCom$^2$ exhibits graceful degradation due to link disruptions or packet losses. Third, FlyCom$^2$ enables parallel streaming communication and computation, which halves latency compared to the sequential operations of traditional one-shot algorithms.

Despite the potential to improve efficiency and reduce on-device computation constraints, the development of FlyCom$^2$ in ISEA is still nascent. In essence, FlyCom$^2$ represents leveraging incremental operations of parallel computing and communication to improve system performance continuously. This implies that in particular ISEA tasks, the involved data processing algorithms and air interface techniques shall be jointly designed based on task-specific requirements and properties. Several examples are introduced below.
\begin{itemize}
    \item \textbf{FlyCom$^2$ with large language model (LLM) fine-tuning:} With human-like perception and understanding capabilities, LLMs are expected to be vital for supporting intelligent sensing tasks. In this case, parameter-efficient fine-tuning is always required to customize LLMs to particular sensing requirements, where a popular technique is to formulate a prompt ensemble~\cite{PromptEnsemble}. It has been shown in \cite{DEFT} that such a prompt ensemble, which comprises a set of lightweight prompts, can be boosted progressively via incremental prompt transfer and combination. This inspires the proposal of on-the-fly boosting (FlyBoosting) to progressively broaden the range of solvable problems of a prompt ensemble, thereby realizing communication-and-computation efficient multi-device device-edge cooperative fine-tuning (DEFT). Among others, to further enhance efficiency, goal-oriented communication techniques, such as importance-aware scheduling and resource allocation, can be designed based on the effectiveness scores of individual prompts and communication conditions of participants (e.g., sensors and servers). 
    
    \item \textbf{FlyCom$^2$ with hierarchical feature extraction:} There are various neural network architectures and AI models that facilitate coarse-to-fine pattern recognition and semantic image synthesis by using hierarchical feature extraction. Examples of such models include MUSH and Pyramid~\cite{Pyramid}. These models enable incremental computation, making them suitable for developing a FlyCom$^2$ framework for ISEA.
    
    \item \textbf{FlyCom$^2$ with tree-structured data processing:} Sparse data processing commonly utilizes tree-structured representations to encode the data space in a compact form. Examples of such representations include octrees and $k$-dimensional (KD) trees, which are widely employed to describe point clouds in a 3D space~\cite{OcTr,Muscle,FlyCom_PC}. Analyzing data over these tree structures involves incremental operations such as tree search, data compression, and feature extraction.

    In typical ISEA scenarios, efficient collaborative sensing is essential. In this context, by leveraging FlyCom$^2$'s ability to decompose sequential high-dimensional data processing into streaming operations of low-dimensional communication and computation, the incremental processing of tree-structured data can be efficiently performed in a distributed manner, facilitating collaborative sensing and analysis among multiple sensors.
    
    \item \textbf{FlyCom$^2$ with online algorithms:} Indeed, there is a set of online algorithms specifically designed to handle the continuous arrival of data points. They feature immediate decision-making based on the currently available data without requiring access to the entire dataset, thereby supporting real-time scenarios. For example, the random sketching technique used in~\cite{RN383} was originally proposed to learn principal components from the continuous streams of data vectors~\cite{Sketching}. The integration of online algorithms with the FlyCom$^2$ framework allows for ultra-low-latency processing and transmission of real-time sensory data, making it well-suited for dynamic ISEA scenarios such as healthcare, autonomous driving, and AR/VR.
\end{itemize}

\subsection{ISEA with Multi-Functional Waveforms }\label{subsection-advsp-mulfuncwaveform}

From a physical layer perspective, advancing ISEA critically relies on the foundational design of waveforms. Multi-functional waveforms are sophisticated signal designs that integrate multiple functions, such as sensing, communication, localization, and radar, into a unified structure, leveraging advanced signal processing to optimize performance.
Previous attempts in this field focused on extending the functionalities of existing systems. For example, communication symbols were modulated onto traditional radar waveforms, such as space-time coding over orthogonal radar waveforms, code shift keying using optimized coded radar signals, and embedding communication symbols in frequency-hopping multi-input multi-output (FH-MIMO) radar systems~\cite{Aboulnasr2016, Tedesso2018, KaiWu2021}. 
Some early works in this field explored the performance of sensing using signals generated through linear frequency modulation (LFM) and continuous phase modulation~\cite{LFM-MSK2011, LFM-CPM2019}. These schemes aimed to utilize existing architectures to meet the limited demands of additional functionalities.
More recent research focuses on the novel design of multi-functional waveforms for simultaneous monostatic sensing and communication (SMSC). In SMSC, a transceiver array emits dual-functional signals to illuminate the scene and gather environmental information through echo signals. At the same time, forward signals are reflected to a communication receiver for information delivery. The advantage of SMSC is that it allows the uncertainty of communication messages to be incorporated into the transmitted waveforms without compromising the coherent processing of monostatic sensing. This eliminates the need for cooperation overhead~\cite{RN378}. However, the distinct requirements of sensing and communication pose challenges in the design of multi-function waveforms, such as the Cramér-Rao bound (CRB) and data rates, leading to a tradeoff in performance~\cite{RN378,RN98}. To address these tradeoffs, research has focused on beamforming techniques, where beam patterns are formed in a joint radar-communication MIMO system based on the directions of sensing targets and communication users to mitigate mutual interference~\cite{FLiu2018TSPdual,FLiu2018TWC,Damith2021,Dokhanchi2021JCS}. In multi-user and multi-target scenarios, multi-user interference is a critical challenge, which is addressed in \cite{multi_obj_isac2024} by introducing constructive interference. Additionally, time-frequency waveforms have been revisited to optimize subcarrier powers in a specific time-frequency region of interest, achieving a favorable performance tradeoff between radar and communications~\cite{RN99}.

Integrating advanced AI algorithms with multi-functional waveforms, ISEA is able to enhance both sensing and communications in a variety of ways. First, AI models can be leveraged to extract valuable information from sensory data gathered by multi-functional waveforms (e.g., intelligent micro-Doppler signature detection and radio map construction). Second, AI-assisted sensing and communication frameworks offer diverse E2E performance metrics—including accuracy, IoU, semantic data rate, as well as traditional measures like CRB, ambiguity functions, and throughput—to holistically guide waveform design and system development. Finally, certain AI algorithms, such as reinforcement learning and deep unfolding, can effectively address the complex tradeoff between sensing and communication, a challenge that is often intractable using conventional analytical optimization techniques.

{
\subsection{Privacy-Preserving Signal Processing in ISEA}
\label{subsection-advsp-privacy}
Privacy issues arise with the frequency of sensory data exchange among different entities in ISEA. The sensing data security and privacy in networks received early attention when the embedded sensing is implanted into personal devices, e.g., mobile phones \cite{Privacy_Personal_Sensing_2009}, and architectural and protocol designs were discussed in different sensing scenarios such as smart cities\cite{Security_Smart_City_2017}, IoTs\cite{Privacy_Survey_2017}, and mobile crowdsensing systems\cite{Privacy_Mobile_Crowdsensing_2020}. In ISEA, a major challenge is to design privacy-preserving signal processing schemes to desensitize sensory data such that the sensing task performance requirement can be achieved with minimum privacy leakage. It is notable that compared with uploading raw data, the split ISEA paradigm, where sensors need to upload extracted features to the edge brain for inference, is inherently advantageous in privacy-preserving as the features are generally more abstract than raw data. Nevertheless, these features are still prone to multiple types of malicious attacks, e.g., model inversion attacks attempting to recover the original data\cite{model_inversion_survey}.

To alleviate the privacy issue, multiple privacy-preserving techniques have been developed for digital air interface-based split ISEA. \cite{PrivacyCode} considers split ISEA with multiple honest-but-curious edge servers to cooperate to complete linear inference requests from multiple sensors. To achieve privacy preservation, the data is divided into pieces for distributed computation at multiple servers, and a coding scheme based on Shamir's secret sharing is applied to ensure that the original data cannot be recovered from the collision of a number of servers. Apart from traditional signal processing, e.g., coding, one popular approach is to use deep encoders to transform the sensory data into a less sensitive representation and then, optionally, inject additional distortion. For example, EdgeSanitizer, proposed in\cite{EdgeSanitizer}, tackles the privacy issue in split inference by applying a data minimization model to the raw features to obtain less sensitive features, followed by injecting adaptive Laplacian noise. More noises are injected to feature dimensions less relevant to the inference performance. Theoretical analysis shows that this scheme can satisfy a given level of differential privacy (DP) while guaranteeing data utility. In another framework named Roulette\cite{Roulette}, a similar procedure is proposed where the sensor model is both a feature extractor and an encryptor with the uploaded data further obscured by a noise injector. In addition, the authors design a novel split-learning procedure for training the sensor model without exposing either the input data or the ground truth label to the server.  

Intriguingly, data aggregation by AirComp is inherently more privacy-preserving than by orthogonal access, as the superposition of signals from multiple sensors, distorted with channel noise, conceals individual private information\cite{De-RPOTA}. Theoretically, it is proved in\cite{PrivacyForFree} that AirComp-based FL can achieve a certain DP level ``for free'', i.e., without affecting the learning performance, thereby outperforming orthogonal access. For split ISEA inference, it is also shown in\cite{Deniz2022ISIT} that AirComp-based ensemble inference achieves much better accuracy compared with orthogonal access while guaranteeing model DP. Nevertheless, when the desired privacy level is beyond what can be achieved by channel noise and signal superposition, additional signal processing methods shall be introduced to AirComp-based data aggregation. In\cite{AirFLDP1}, the authors propose a DP-guaranteeing AirComp scheme where each sensor individually controls the noise added to its gradient to achieve a desired DP level. The level of individually injected noise is derived analytically considering gradient randomness. In a cooperative ISEA scenario where multiple sensors perform feature AirComp for inference, it is proposed in\cite{AirInfDP1} to first select devices with potentially high contributions to the accuracy and then inject Gaussian noises to uploaded features. In addition to noise perturbation, it is shown in\cite{AirFLDP2} that random data dimension reduction with resemblance to sparsification can further amplify privacy. }

\section{ISEA with Other Trending Advancements in 6G}
Apart from the techniques mentioned above, the ISEA performance can be boosted by other trending advancements in 6G. In this section, we provide a thorough discussion of these opportunities and a review of relevant works, including the integration of ISEA with advanced physical layer techniques, satellite networks, and advanced networking.
\label{section-6gadv}
\subsection{ISEA with Advanced Physical-Layer Techniques}
\label{subsection-6gadv-phy}
\subsubsection{Extreme-Large-Scale MIMO}
Extreme-large-scale MIMO (XL-MIMO) refers to the use of a massive number of antennas, typically in the range of hundreds or thousands, in a wireless communication system to enable spatial multiplexing and beamforming techniques to enhance the system capacity and coverage. Specifically, XL-MIMO leverages spatial multiplexing to simultaneously serve multiple users with the same spectrum and time resources. This improves spectral efficiency and enables higher data rates. Moreover, with many antennas, XL-MIMO can effectively mitigate interference by spatially separating users and applying advanced interference cancellation techniques. The ways of realizing XL-MIMO include the employment of large-scale arrays, intelligent surfaces, and distributed MIMO~\cite{RN369}.  XL-MIMO also reveals a property of non-stationary power distribution and channel statistics~\cite{RN369,RN375}. Specifically, the extreme-large aperture of XL-MIMO renders the approximation of plane wave propagation in traditional MIMO systems invalid, leading to multi-antenna signal processing in the near field~\cite{RN377}. On the other hand, different sub-arrays of XL-MIMO can experience different channel fading, e.g. line-of-sight (LOS) and non-LoS, which causes channel non-stationarity~\cite{RN375}.

To boost the area of ISEA, XL-MIMO can be utilized in the following ways:
\begin{itemize}
    \item \textbf{Enhanced sensing capabilities:} XL-MIMO can support sensing purposes, such as target detection, localization, and tracking, through its massive antenna arrays and radio wave signal processing. For instance, using dense arrays with high-frequency signals, e.g. millimeter waves, the phase and amplitude information recorded over a two-dimensional aperture to reconstruct a focused image of a target~\cite{3D_imaging}. Using the spatial resolution provided by many antennas, XL-MIMO also improves the accuracy and reliability of sensing applications. The angular resolution achieved by dense array systems, such as MIMO radar, has been shown to increase linearly with increasing effective array aperture~\cite{3D_imaging,MIMORadar2009}. In this vein, a key challenge could be addressing the near-field beam squint effect, the beams of different frequencies focusing at different locations, that is caused by the vast array and the huge bandwidth~\cite{XL_NF_localization}.
    
    \item \textbf{Massive and cell-free sensor access:} In the scenario of densely placed sensors, traditional MIMO cannot ensure orthogonality as sensors become indifferent in the angular domain. However, XL-MIMO can realize beamforming in both direction and distance in the near field, thereby allowing simultaneous orthogonal access by a massive number of sensors\cite{NearFieldMag}. Moreover, the strong association between the XL-MIMO channel and sensor location can be leveraged for sensor localization and semantic-and-channel aware sensor scheduling.  Furthermore, it is possible to leverage XL-MIMO with its powerful beamforming techniques to realize cell-free communication. In a cell-free communication system, users are no longer associated with a specific cell or BS. Instead, they can connect to multiple APs simultaneously, forming a dynamic and flexible network configuration. The APs work cooperatively to serve the users by jointly transmitting and receiving signals. This allows for real-time analysis and decision-making at the edge for ISEA, reducing the latency and communication overhead associated with transmitting raw sensor data to a centralized processing unit.
\end{itemize}

Finally, XL-MIMO can be jointly optimized with ISEA algorithms and protocols in terms of resource allocation, beamforming, and signal processing. Specifically, by analyzing the sensing and communication requirements of different users and applications, XL-MIMO can intelligently allocate antennas, power, and bandwidth to maximize the overall ISEA performance.

\subsubsection{Terahertz Techniques}

{
THz techniques combine massive bandwidth with sub-millimeter wavelengths, enabling both ultra-high-speed data transmission and high-resolution sensing in a single band. By unifying these capabilities within an ISEA framework, networks can simultaneously “sense” and “connect” their environments. Several works provide insights into these aspects. For instance, \cite{THz_ML_2022} highlights THz’s strong potential for fine-grained environmental scanning and imaging, demonstrating how advanced machine learning algorithms can extract meaningful insights (e.g., object detection, material properties) in real-time. This, in turn, enables rapid decision-making (e.g., dynamic beam adjustments) within ISEA.
By integrating camera data with THz signals, \cite{Kim_JSAC_2024} shows that the two modalities outperform the use of either modality alone: 1) cameras provide rich contextual and positional information for THz beam selection, alignment, and movement prediction; 2) THz links offer ultra-broadband backhaul for camera feeds while providing additional environmental-sensing capabilities in scenarios where vision alone might fail. This bidirectional relationship exemplifies the broader vision of ISEA, which leverages multiple sensory modalities.
Meanwhile, \cite{kim2024vomtc} incorporates object detection (e.g., people, vehicles, obstacles) directly into the communication process to proactively mitigate link blockages or signal degradation, thereby reflecting ISEA’s objective of distributing AI-driven contextual information to the network’s edges for more resilient and efficient wireless systems.}

\subsubsection{Reconfigurable Intelligent Surfaces}
RIS are innovative technologies that enhance wireless communication by intelligently controlling the propagation environment \cite{IRS_Comm_Tutorial, WC_RIS_2019, RIS_Comm_2021}. They can also improve the accuracy and reliability of wireless sensing by manipulating electromagnetic waves to focus on specific targets or areas. Specifically, integrating RIS into ISEA extends beyond the limitations of conventional wireless sensing by (1) creating additional LoS links for sensing, (2) providing controllable propagation paths, and (3) enabling adaptive RIS beam scanning \cite{IRS_Wireless_Sensing_2024, Sensing_IRS_2022}.

Building on the integration of RIS for sensing, several research opportunities emerge:
\begin{itemize}
\item  \textbf{Optimization of RIS placement and configuration:}
   The deployment of RIS in sensing systems significantly affects both sensing accuracy and communication efficiency. For example, placing the RIS at a higher altitude might establish a clearer LoS path, while a lower altitude can reduce path loss with the sensing target. Determining the optimal RIS placement requires accounting for environmental characteristics and dynamically adjusting sensing coverage to reduce blind spots.

\item  \textbf{Joint beamforming techniques:}
   Reflection/refraction control of the RIS must be jointly designed with conventional transmit beamforming to maximize sensing performance, rather than simply focusing on maximizing SNR as in typical communication scenarios. For instance, \cite{IRS_Beamforming_2019} demonstrates how adaptive RIS beamforming can maximize the received signal power at RIS sensors—a crucial metric for sensing. Moreover, via controlled refraction/reflection of sensing signals, RIS allow receiving and distinguishing echo signals from multiple targets\cite{multi_func_ris_sens2025}. In general, beamforming and RIS control strategies should be explored to balance communication efficiency and sensing accuracy, as RIS allows manipulation of both communication and sensing signals\cite{ni2024ris_sensing}.

\item  \textbf{Robust sensing algorithm design:}
   In an RIS-enabled sensing paradigm, sensors collect signals reflected from the target for subsequent analysis. This is particularly challenging when multiple targets must be distinguished from overlapping signals originating from the same direction, making the estimation problem more complex than in conventional MIMO communications.

\item  \textbf{Distributed RIS sensing:}
   Within the ISEA framework, multiple RIS can be deployed cooperatively, each acting as a local estimator. The distributed information among these RIS units is then shared to enhance sensing accuracy. As a result, efficient and accurate low-complexity distributed sensing paradigms and algorithms are essential.

   \item  \textbf{Energy efficiency and security:} 
   Employing RIS can lower the overall energy consumption of wireless networks by minimizing high-power transmissions and optimizing signal paths for sensing. Nonetheless, effective power control is necessary to mitigate the added costs of operating RIS. Meanwhile, incorporating RIS into the network also calls for enhanced security mechanisms, leveraging intelligent signal manipulation to safeguard against potential threats.

\end{itemize}

\subsection{Space-Ground Empowered ISEA }\label{subsection-6gadv-global}
\begin{figure}[t]
    \centering
    \includegraphics[width=0.85\linewidth]{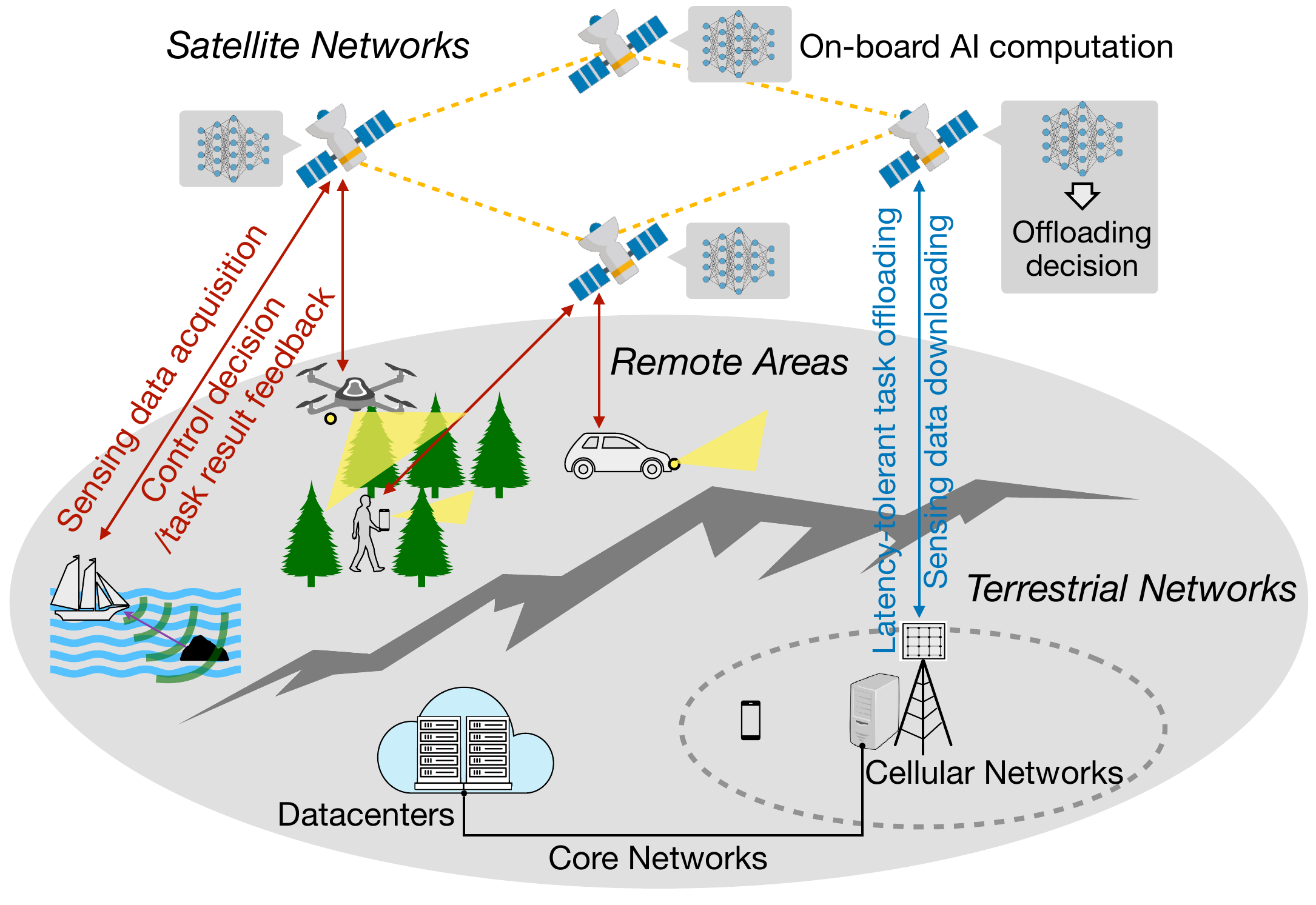}
    \caption{
   ISEA techniques can be enhanced by the space-ground integrated network through the utilization of ubiquitous satellite sensing and communication coverage.}
    \label{fig: glob}
\end{figure}
ISEA can be implemented in space and ocean networks to achieve uninterrupted global connectivity as envisioned in 6G, which is shown in Fig.~\ref{fig: glob}. These networks have distinct characteristics compared with terrestrial networks in aspects of sensing modalities, mobility, channel propagation, computing architecture, topology, and QoS. The existence of extra-terrestrial and maritime terminals leads to new sensing modalities such as orbital images and under-water sonars, while these non-terrestrial network nodes, e.g., low Earth orbit (LEO) satellites, typically have scarce computing resources compared with ground-based nodes such as BS. Also, the communication links in non-terrestrial network nodes usually have limited bandwidth, high latency, and short contact windows due to mobility. Not only are these constraints challenging, but tasks carried out in non-terrestrial networks, such as forest fire alarming and rescuing, are usually mission-critical and demand high reliability. It is thus motivated to design ISEA protocols and algorithms dedicated to those networks to address the aforementioned challenges.

In conventional bent-pipe architectures for satellite-centric remote sensing, satellites simply relay commands from the ground and transmit raw data via low-rate links, resulting in communication bottlenecks. Similar to the ISEA paradigm, a C$^2$ framework integrated addresses this challenge by employing onboard computation for data filtering and semantic matching, thereby enabling more efficient and scalable downlink communications.
A few representative works are introduced as follows. A typical satellite-terrestrial collaborative object detection scenario is considered in \cite{RN385}, where the satellite captures remote sensing images to be processed and transmitted to the ground station for AI-based object detection. A task-oriented communication scheme is developed to combat the low-rate issue of satellite-terrestrial links. The satellite first identifies between the region of interest and background with a lightweight on-board model, downsamples the latter, and transmits each processed image block with priorities determined by their contribution to the detection task.
Driven by a similar motivation of reducing downlink data volume, the authors of \cite{RN373} propose an orbital edge computing (OEC) system that places energy-efficient GPUs on nanosatellites to enable onboard computation. Then, the system employs a technique called intelligent early discard, which filters the captured image using the onboard model, transmitting only useful images to the ground station and discarding obscured ones. Further, to tackle the computation resource limits, an image frame can be divided into multiple tiles processed in parallel on multiple satellites.  In \cite{RN376}, the satellites carry out a hyperspectral imaging task, collecting electromagnetic spectrum in hundreds of frequency bands to be transmitted and processed at the ground station for pixel-level classification. An onboard band-selection system is proposed to reduce the downlink communication cost, which selects only a subset of bands to transmit based on the image content, network conditions, and power budgets. The adaptive band selection problem, which aims to maximize the utility, is composed of accuracy and costs and is solved via reinforcement learning under energy and latency constraints.  

{Aligning with the ISEA framework, another line of research focuses on harnessing ground-based devices for sensing, with satellites serving as the overarching edge servers.}
Specifically, 
\cite{RN387} considers the virtual network function placement problem in a satellite edge computing system where multiple satellites cooperate to endow devices collecting sensory data with computational capabilities. The said problem is formulated to maximize the sum payoffs of scheduled devices subject to network resource constraints, for which a decentralized algorithm is developed by a potential game approach. 
In \cite{RN371}, multiple field robots are controlled by an integrated satellite-UAV network with one UAV equipped with sensors, computation resources, and communication modules and one satellite connected to the UAV. The UAV senses the system state and analyzes the sensing data jointly with the remote center through a satellite connection to generate control commands that are transmitted to the robots via orthogonal channels. Under latency constraints, the power allocation among channels is optimized to maximize the control performance measured in linear quadratic regulator cost and associated with the downlink data rate. 
Moreover, a flexible space-ground ISEA framework, termed fluid inference, is proposed by \cite{chen2024Eng}. It adaptively splits AI workloads across satellites, edge servers, and end devices, leveraging real-time resource constraints to slash data traffic and improve responsiveness. By orchestrating partial or full inference traverse satellite networks, the accuracy-latency tradeoff is optimized under limited resources in computation and communication.

\subsection{ISEA with Advanced Networking}\label{subsection-6gadv-netw}
Though most ISEA applications can be realized within an edge network, certain ISEA operations still require edge-cloud cooperation or collaboration between multiple edge networks, such as hierarchical FL, model downloading, and offloading of heavy computational loads. Traditional protocols under the best-effort principle cannot guarantee low latency and high reliability per request by ISEA applications. This requires advanced techniques that involve task-oriented coordination of the transportation and networking layers in transport and core networks. To this end, ultra-low latency networks such as time-sensitive networking (TSN) and deterministic networking (DetNet) can be integrated into the ISEA architecture. TSN employs time-division multiplexing, assigning dedicated timeslots to prioritized time-sensitive flows such that they are isolated from best-effort flows and achieve predictable queueing delays through centralized management. Through these mechanisms, time-sensitive traffic enjoys bounded maximum latency and extremely low packet loss rates. With similar designing objectives, DetNet focuses on the network and higher layers, enabling deterministic latency through a set of protocols, including flow synchronization, path setup, traffic engineering, etc. 

Some early attempts at integrating ISEA and networking techniques are introduced as follows. In \cite{industedge}, an edge-cloud collaboration platform called IndustEdge is proposed to provide low-latency and reliable access to intelligence for smart industry applications, e.g., fault detection. It adopts the TSN protocol for the link layer and further provides an extensible edge-cloud orchestration component for flexible microservice deployments onto edge or cloud nodes. Also, IndustEdge provides an algorithm library that includes various tasks (e.g., classification and clustering) and supports different collaboration modes, all with transparent life-cycle management. Another work \cite{9815186} considers a collaborative multi-access edge computing system where systems in different domains can be interconnected for computation offloading. The proposed task deterministic networks (TDN) aims to achieve E2E deterministic communication for multi-access edge computing systems involving heterogeneous access and interconnections between multiple edge computing systems via transportation networks. To this end, TDN features cross-domain collaboration mechanisms to integrate TSN and DetNet protocols in different layers and designs a seamless multi-controller working scheme coordinating edge access and global deterministic forwarding in aspects of traffic scheduling and resource allocation.

\section{Research Opportunities and Open Issues}\label{section:outlook}

\subsection{ISEA Meets Foundation Models}\label{subsection-6gadv-foundmodels}

\begin{figure}
    \centering
    \includegraphics[width=0.99\columnwidth]{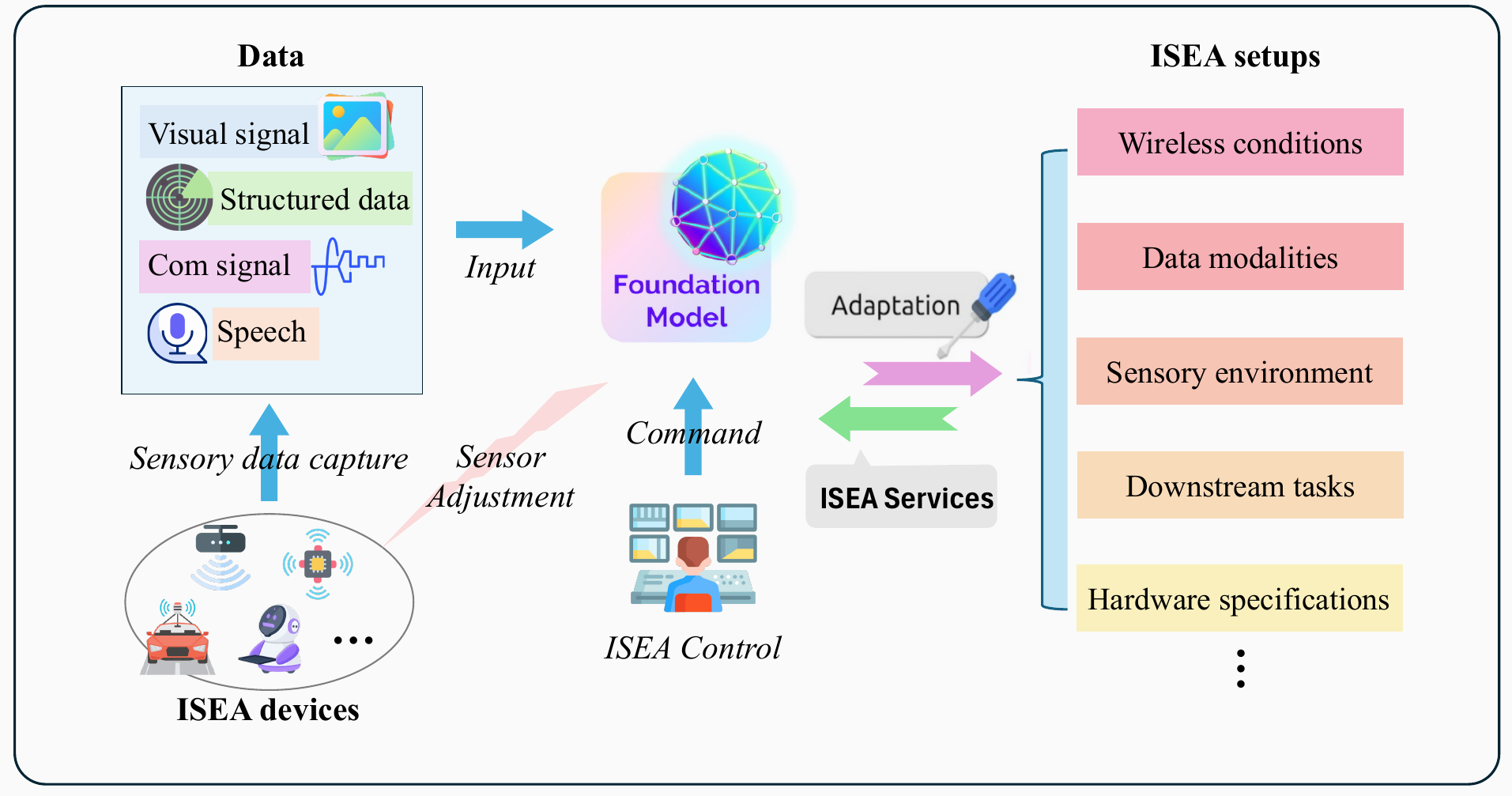}
    \caption{
    The workflow of foundation model-assisted ISEA. 
   Foundation models can assist ISEA by undergoing pre-training and fine-tuning tailored to ISEA-specific settings and sensory data.  
   }
    \label{fig: FoMo}
\end{figure}

{With the boom of large-scale foundation models, ISEA is believed to be further accelerated and improved by the generalized intelligence pre-trained models. As illustrated in Fig.~\ref{fig: FoMo}, foundation models can serve as an intelligent agent that understands sensing data and network conditions to make real-time configuration adaptations for accomplishing human commands. We discuss the foundation model-assisted ISEA from two perspectives: model adaptation and deployment, respectively.

The adaptation of foundation models that are pretrained for general purposes to ISEA can be roughly categorized into two levels. The first level is the direct adaptation of a large language model to sensing tasks, including instruction reasoning, planning, and simulation, which is an efficient and low-cost approach. To harness LLMs for sensing without fine-tuning, detailed textual descriptions of the sensing environment are required to prompt LLMs to make sensing decisions and issue control commands. For example, a prompt can detail vehicle states, including coordinates, speed, past trajectories, current velocity, and acceleration in LLM-assisted autonomous driving, accompanied by a chain of thoughts to help the LLM decompose the procedure of understanding current states and take action. This approach leverages LLMs’ ability to understand complex and novel scenarios with zero-shot or few-shots of training data.

Another level of adaptation is to fine-tune foundation models which, in addition to textual prompts, take the sensing data directly as inputs. Take camera-based visual sensing as an example. The architecture of vision foundation models can be applied, which are typically built on a transformer encoder architecture and pre-trained in image classification, image-text pairing, or image captioning tasks. Its adaptation to specific ISEA tasks can be achieved by downstream task-aware fine-tuning. For instance, one popular vision model, called SAM \cite{SAM}, has been employed in different perception scenarios, e.g., 3D object detection by projecting LiDAR point cloud to BEV images, generating prompts for each grid in the images to detect masks for foreground objects. The zero-shot transfer capability of SAM is leveraged to generate segmentation masks and 2D boxes. However, the empirical study shows that the vision foundation model is not capable of handling the sparse and noisy points \cite{SAM3D} in practical ISEA scenarios. Moreover, current vision foundation models still require handcraft adaptations and designs to accommodate the properties of visual information, such as the sparsity and temporality of the point cloud, for reliable and accurate edge sensing. Facing the time-varying sensing environments, data distribution and task requirements in ISEA, an important future direction is to automate the process of real-world foundation model adaptations.
The ultimate goal of foundation model adaptation is to incorporate different modalities, such as text, audio, and images, for improved robustness and accuracy in ISEA sensing tasks. Typically, the feature extractor for 2D images and 3D point clouds varies significantly due to the heterogeneity between sparse points and dense pixels. The direct fusion of different types of data sources overlooks the field of view outside the intersected area \cite{MSeg3D}.  The crafted design of a uniform multi-modal data extractor and fusion module is a main challenge in the multi-modal foundation model training for diverse ISEA tasks. 

Considering the deployment of foundation models in the wireless edge, a few precursor works have investigated the placement of foundation models and the caching of the query and response for low-latency accessing of AI-Generated Content (AIGC), which inspires the implementation of foundation models for ISEA infrastructure.
A hierarchical network architecture is designed in \cite{Joint_FM_Caching_and_Inference, EdgeGPT} to utilize the computation of devices, edge servers, and the central cloud to pass on the on-device inference task.
The inference relay is terminated once it hits one cached response, hence reducing the execution time of inference tasks. 
In the ISEA system, the multi-agent cooperation of computing and communication can be orchestrated to access the foundation model-empowered sensing with low latency. 
The work in \cite{EdgeGPT} considers the decomposition of a user request for cooperative edge inference, where the GPT on the cloud does the task planning and distributes the subtasks to edge clusters to accomplish the different parts of a user request.
A collaborative distributed diffusion-based AIGC framework is proposed in \cite{Collaborative_DF_AIGC} to partition a diffusion process for integrated central and edge inference to provide a better personalized user experience.
Specifically, the requests from multiple devices can share a few denoising steps in the diffusion inference stage to reduce the computational overhead and improve resource utilization efficiency. 
The considered task decomposition, planning, and distributed execution for ISEA is another interesting direction for foundation model implementations in ISEA systems.
Moreover, the edge cooperative fine-tuning of pre-trained foundation models to align the ISEA performance with task preference is another interesting direction~\cite{DEFT}, which is closely related to MEC and distributed learning systems.
Despite these pioneer works, timely and resource-efficient access to the foundation model service in the ISEA system still requires a further systematic framework and refinements to enhance the user experience under limited communication and computation capabilities, especially in view of the network congestion caused by massive requests to the central cloud. Special characteristics and novel acceleration techniques of foundation model computation shall be considered and integrated natively into ISEA design, such as key-value (KV) cache, speculative decoding, sparse attention, etc. }

{
\subsection{Convergence of ISEA and ISAC}

ISEA and ISAC have largely been two independent lines of research due to their different technical motivations. While ISEA mainly focuses on optimized communication and computation for AI-empowered sensing, ISAC aims at coexistence and joint design of RF sensing and communication within shared hardware and spectrum, particularly focusing on multi-functional waveforms. However, as two key functions coexisting in 6G, ISEA and ISAC are mutually complementary in multiple aspects, e.g., sensing modalities and AI backbones, while also potentially contending for spectrum and communication resources. Therefore, the convergence of ISEA and ISAC is a promising future direction, aimed at deep integration and cooperation of both for improved task performance. In what follows, we will first discuss how ISEA and ISAC interact with and assist each other, and then list several research challenges within a converged framework of ISEA and ISAC. 

First, consider assisting ISAC with ISEA. The first aspect is the use of ISEA sensory information for optimized ISAC operations. Sensing modalities in ISEA, usually with high resolution and rich semantic information, can enable situational awareness for ISAC by, e.g., scene identification, link blockage prediction, and user localization\cite{ISEA_ISAC2, ISEA_ISAC4}. This improves the performance of several critical ISAC techniques, such as beamforming and power allocation, while reducing the pilot overhead\cite{ISEA_ISAC1}. The gain will be particularly significant for high-mobility scenarios where ISEA-assisted predictive control reduces the negative effects of constant environmental changes. The second aspect is direct augmentation of ISAC sensing data with the ISEA counterpart. With properly designed AI models that translate information between domains, the high-resolution ISEA sensing data can enhance the quality of ISAC sensing data by localization refinement, time-domain interpolation, vision-aided generative reconstruction, etc. Then, spectrum resources for sensing can be saved to improve the communication performance without sacrificing sensing performance. Further, ISEA can provide AI inference for ISAC signal processing \cite{ISEA_ISAC3} and efficient communication of ISAC sensing data between network nodes to realize efficient and intelligent network-as-a-sensor. Conversely, ISAC can assist ISEA in several aspects. The sensing modalities of ISAC, i.e., RF sensing, can complement typical ISEA sensing, as it generally has a larger sensing range and higher robustness to ambient lighting conditions. While light-based sensing can suffer from occlusion, ISAC can realize non-LoS localization\cite{ISAC_ISEA5}. Compared to LiDARs, ISAC is a more power-efficient method to obtain depth information which is critical for environment perception with visual features. In addition, ISAC sensing data can be used for ISEA model training \cite{ISEA_ISAC_6} and estimation of several ISEA sensing parameters such as camera pose, relative speed, etc\cite{ISAC_Camera_Pose}.

Several research directions are warranted by the convergence of ISEA and ISAC. First, the scheme for integrating sensing data from different sources with different modalities shall be developed, which combats several practical issues such as time synchronization, different coverage, relative location measure error, etc. Second, despite mutual assistance, ISEA and ISAC operations inevitably use the same spectrum and hardware, leading to resource allocation problems. Specifically, the radio resources shall be optimally allocated for ISAC sensing waveforms, user traffic and ISEA traffic for sensing feature transmission, and the computing resources allocated for processing of different sensing modalities. Some fundamental tradeoffs remain to be analyzed and optimized. For example, in the sensing-communication tradeoff, the allocation of more resources for sensing improves sensing resolution but requires a higher sensory feature compression ratio that degrades transmission reliability, necessitating optimal control for maximizing the sensing performance.

\subsection{Ultra-Low-Latency ISEA}

\begin{figure}
    \centering
    \includegraphics[width=0.65\linewidth]{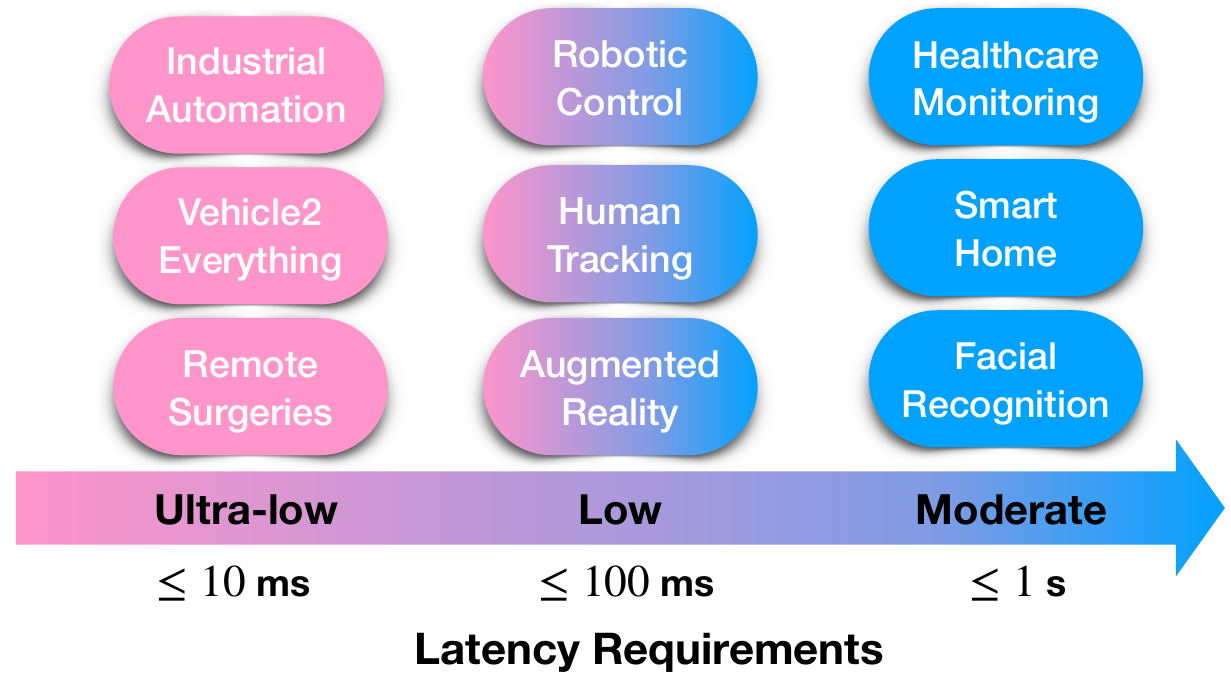}
    \caption{
   The latency requirements of ISEA tasks can generally be categorized into three levels: ultra-low-latency, low-latency, and moderate latency.
   }
    \label{fig: ultra-lola}
\end{figure}
The growing demand for real-time intelligent decision-making has been seen in mission-critical applications such as autonomous driving, remote surgeries, and industrial automation, as shown in Fig. \ref{fig: ultra-lola}.
This necessitates {ultra-low-latency} (ultra-LoLa) ISEA where the sensing task must be completed within strict latency constraints while simultaneously guaranteeing sensing performance.
The E2E latency of ISEA typically consists of sequential processes, including data sensing, on-sensor computation, and feature transmission, where communication latency must be under several milliseconds to meet the requirements of ultra-LoLa tasks.
Prevalent digital systems, e.g., LTE, aim at building errorless bit pipes that are agnostic of the task characteristics and latency deadlines. In view of their insufficiency in achieving ultra-low latency, we explore research opportunities to achieve ultra-LoLa ISEA with two suitable digital transmission schemes: {short packet transmission} (SPT) and uncoded transmission.

URLLC has been defined as one of the primary 5G missions, aiming to achieve latencies below one millisecond and packet-error rates of $10^{-5}$ or lower, thereby enabling mission-critical applications in 5G networks \cite{Petar-IEEEProc-2016}. A central challenge for URLLC lies in balancing tradeoffs among data rate, latency, and reliability—an issue long investigated by information theorists \cite{Verdu-TIT_2010}.
Guided by these theoretical insights, one practical approach adopted in 5G engineering for implementing URLLC is SPT \cite{popovski2019wireless}. The strict deadlines imposed by short packets (or blocklengths) lead to a non-negligible transmission error probability, which depends on the specific operating regime \cite{Petar-IEEEProc-2016}. At the cost of lower data rates, SPT restricts URLLC deployments to relatively low-rate transmissions (e.g., sending commands to remote robots or uploading sensing data such as humidity, temperature, or pollution) \cite{Nallanathan-TWC-2020,zhao2023joint}.
Nevertheless, researchers have explored diverse methods to mitigate these limitations by designing customized SPT techniques. These include non-coherent transmission \cite{Liva-TCOM-2019}, optimal frame structures \cite{Petar-TCOM-2017}, power control \cite{Quek-TCOM-2018}, wireless power transfer \cite{Schmeink-JSAC-2018}, and multi-access schemes \cite{Nallanathan-TWC-2020}.

However, existing SPT approaches targeting URLLC fall short of meeting the ultra-LoLa ISEA requirements. This shortcoming arises because ISEA applications are typically data-intensive and often require the transmission of high-dimensional features, which conflicts with SPT’s limited payload sizes and leads to communication bottlenecks. The problem is further exacerbated for tasks subject to strict latency requirements. Moreover, the E2E performance of sensing tasks cannot be accurately measured through decoding error probability alone, as that metric is more suited to assessing communication reliability.
To address these challenges, the first framework that enables ultra-LoLa ISEA leverages the interplay between SPT and multi-view sensing—improving accuracy while adhering to stringent latency constraints \cite{ZW2024ultra-LoLa}. Within such a framework, a fundamental tradeoff is identified and optimized between communication reliability and the number of sensing views, both of which are governed by the packet length.
This integration of SPT and ISEA opens numerous opportunities for future work. One key direction is the simultaneous improvement of accuracy and reduction of latency through a joint optimization of packet length and coding rate, which requires further exploration into the effects of feature dimensionality and quantization on packet selection~\cite{zeng2025ultra-lola}. From an energy-efficiency standpoint, designing power control techniques that accommodate both fading channels and feature importance is also a promising avenue. Additionally, extending the ultra-LoLa inference paradigm to incorporate AirComp constitutes another compelling research direction.

The aforementioned SPT is task-agnostic and prioritizes link reliability at the expense of data rate.
Unlike SPT, another technique for achieving ultra-LoLa ISEA is to enable uncoded feature transmission, thereby avoiding the need for channel code transmission \cite{QS2024tocm}.
This method leverages the inherent robustness of classifiers to absorb bit errors caused by channel perturbations, maintaining the desired sensing accuracy under latency constraints.
Retransmission and multiview are investigated to enhance classification robustness, with their benefits validated in scenarios where the classification margin is insufficient to counteract channel distortions. 
Uncoded feature transmission opens up numbers of directions for realizing ultra-LoLa ISEA:
 1) adaptive modulation schemes for ISEA systems by considering channel conditions and classification margins; 2) importance-aware power control over quantized bits for enhancing the sensing performance.
}

{
\subsection{Practical Challenges in ISEA}
\label{sec:practical-challenges}

This subsection discusses challenges in practical implementation of ISEA from the perspectives of scalability, energy resource constraints, heterogeneous and dynamic networks and controlling overhead, highlighting key issues and potential research directions.

\subsubsection{Scalability}
The scalability of ISEA systems is a critical challenge due to the massive number of sensors, edge devices, and AI models that must operate cohesively in 6G networks. As envisioned, 6G will support billions of interconnected devices, generating vast amounts of multi-modal sensory data (e.g., from RGB cameras, LiDARs, and RF sensors) that require real-time processing and transmission. This scale introduces several issues:
\begin{itemize}
    \item \textbf{Cooperation of numerous devices:} As more nodes in edge networks are equipped with AI and sensing capabilities, the number of sensors participating in cooperative ISEA is envisioned to continuously grow. For example, in cooperative autonomous driving scenarios, the number of agents in popular datasets have increased from $2$ in F-Cooper\cite{fcooper} to $6$ in OPV2V\cite{xu2022opencood} to $12$ in V2X-Sim\cite{v2xsim}. While such a trend enables more holistic sensing, it also increases communication overhead for multiple access and complicates scheduling and resource management. Moreover, coordinating distributed learning (e.g., federated learning) across thousands of devices introduces synchronization and convergence issues.
    \item \textbf{Data volume and processing overload}: Developments in resolutions and ranges of sensing technologies leads to higher dimensionality of sensory data. For example,  point clouds from LiDARs now have data rates up to several Gbps \cite{autolidar}. Sensors and edge brain with limited computational capabilities may struggle to process these data streams in real time. Also, the communication rate of data exchange grows proportionally, leading to a communication bottleneck.
    \item \textbf{Model scalability}: State-of-the-art multi-modal foundation models consistently grows in the numbers of parameters and required computation operations, known as the scaling law. Deploying such models requires significant memory and computational resources, which may exceed the capabilities of resource-constrained edge devices and edge servers. 

    \textbf{Research directions}: AirComp can be leveraged to aggregate sensory data efficiently \cite{RN358}, as its latency does not grow with the number of devices. However, with numerous devices, several challenges requires research efforts, such as device scheduling and synchronization. Neural network-based data down-sampling can be leverage to cope with the high data rate, and its interplay with wireless communication warrants future research. Model compression techniques, including quantization and pruning, can reduce the footprint of AI models, enabling their deployment on resource-limited devices\cite{10262344}. Efficient inference techniques, such as speculative decoding and expert parallelism, can accelerate inference on large-scale models. Finally, hierarchical edge-cloud architectures can offload computationally intensive tasks to higher-tier nodes, balancing scalability and performance.
\end{itemize}

\subsubsection{Energy Constraints}
Energy efficiency is an important concern for ISEA, given the resource-constrained nature of edge devices and sensors. The integration of sensing, AI inference, and communication in 6G networks demands significant energy, particularly for real-time, mission-critical applications.
\begin{itemize}
    \item \textbf{Sensing energy costs}: High-resolution sensors like LiDARs and cameras consume substantial power for data acquisition. For example, a typical LiDAR module can consume from 10 to 30 watts\cite{lidarspecs}, posing challenges for battery-powered devices like drones or IoT sensors.
    \item \textbf{AI inference and training}: Edge AI operations, including inference and distributed learning, are energy-intensive. Deep learning models, even when optimized, require significant computational resources for forward and backward passes, especially for real-time tasks like robotic control or semantic segmentation. Federated learning, while reducing data transmission, still incurs high energy costs for local model updates.
    \item \textbf{Communication energy costs}: Transmitting high-dimensional sensory data or model updates over wireless channels consumes considerable energy, particularly in mmWave or THz bands, which require high transmit power to overcome path loss. The frequent data exchanges in ISEA for cooperative sensing or computation offloading further exacerbate energy demands.

    \textbf{Research directions}: Energy-efficient designs are critical for ISEA. Selective sensor wake-up based on query or environmental information can reduce the long-term average sensing power consumption, thereby extending the network lifetime. Energy-efficient JSCC can minimize transmission energy by allocating power across different messages \cite{jssc_shortpacket} or adopt lightweight models to reduce the computation energy\cite{lightweight_jscc}.  Dynamic power control for AirComp can optimize energy use during data aggregation \cite{Xiaowen2020TWC}. Additionally, energy harvesting technologies, such as RF energy harvesting or solar-powered sensors, can extend the operational lifetime of devices. Finally, task offloading frameworks that balance local and edge computation based on energy availability can enhance overall efficiency.
\end{itemize}

\subsubsection{Heterogeneous and Dynamic Networks}
The heterogeneity of 6G networks, encompassing diverse devices, sensing modalities, and communication protocols, poses significant interoperability challenges for ISEA. Seamless integration and cooperation among these components are essential for achieving optimal E2E task performance.
\begin{itemize}
    \item \textbf{Diverse sensing modalities}: ISEA integrates multiple sensing modalities (e.g., RF, LiDAR, and RGB-D cameras), each with distinct data formats, sampling rates, and processing requirements. Harmonizing these modalities for joint processing or fusion is challenging, especially when devices operate on different temporal and spatial scales. For instance, aligning RF sensing data and video streams with different streaming rates and updating frequencies requires synchronization mechanisms.
    \item \textbf{Heterogeneous devices and protocols}: 6G networks will include a mix of legacy and next-generation devices, ranging from low-power IoT sensors to high-end autonomous vehicles. These devices may use different communication protocols (e.g., NB-IoT, LTE-M, 5G NR) and AI model formats, complicating cooperative tasks such as distributed inference or sensor data sharing. Moreover, hierarchical ISEA requires joint control and optimization across edge, transport and core networks.
    \item \textbf{Network dynamics}: The dynamic nature of 6G networks, with high device mobility and varying channel conditions, affects sensing performance. For example, in vehicular networks, rapidly changing topologies due to high-speed vehicle movement can disrupt cooperative sensing or model synchronization, leading to degraded task performance.

    \textbf{Research directions}: To enhance interoperability, standardized data formats and protocols for multi-modal sensory data exchange are needed. Semantic communication can be used to reconcile different sensing modalities under a unified framework. Adaptive modulation and coding schemes tailored to sensing tasks can accommodate diverse device capabilities. Additionally, LLM-empowered network agents can dynamically manage resources and protocols to maintain seamless cooperation under different task requirements and network status\cite{wirelessllm}. 
\end{itemize}

\subsubsection{Controlling Overhead}
Although the joint ISEA design enhances task performance, it also introduces additional overhead for signaling and computation operations in control algorithms. These overhead becomes particularly significant in ultra-low-latency or massive access applications.
    \begin{itemize}
        \item \textbf{Signaling overhead:} The time-varying nature of sensing context requires frequent transmission of control decisions and feedback information between sensors and edge nodes. Meanwhile, the raised dynamic resource mapping between communication, AI and sensing introduces higher multiple access control overhead compared to 5G’s static scheduling\cite{10500734}. In addition, while the unified waveform design in ISAC reduces redundant signal processing delays, it requires additional reference signals for cross-functional channel estimation (e.g., simultaneous CSI and radar sensing parameter extraction).
        
        \item \textbf{Computation overhead:} Real-time cross-domain optimization demands more baseband processing due to 1) the coupled sensing-communication beamforming (e.g., solving high-dimensional Pareto fronts); 2) AI task QoS monitoring (e.g., perceptual quality metrics for generative AI). This places heavy computing burden on 6G network controllers.

        \textbf{Research directions}: Digital twins of the environment can reduce the information feedback to the minimum amount by incrementally updating the digital twin at the server. For example, the server can estimate the wireless channel using ray-tracing in the digital twin, which only requires low-dimensional object position feedback but not CSI feedback. The computation overhead of control algorithms can be simplified by approximated optimizations and decentralized computation. Moreover, advanced baseband hardware, e.g., in-memory computing\cite{zeng2024inmemory}, can increase processing speed by up to 1000$\times$ compared with existing semiconductor devices, making it a promising solution for ultra-low-latency applications.
    \end{itemize}
}
\section{Concluding Remarks}\label{section-conclusions}
Driven by the rapid evolvement of AI and sensing technologies, a future digital era is emerging with ubiquitous perception sensing on 6G infrastructure. A key to realizing this vision is the integrated design of communication, computing, and sensing to support sensing data analytics at the wireless edge, raising a task-oriented paradigm, ISEA. In this paper, we have presented a comprehensive overview of ISEA and a detailed survey of relevant techniques. In particular, we establish the ISEA framework by providing concrete definitions of its key design principles and architectures. Under the unified principle of processing and transmitting sensory information for E2E metrics, we review enabling techniques for ISEA from various aspects and discuss several promising future research directions. We hope that this survey can be a useful reference for researchers entering this emerging field.

\bibliographystyle{IEEEtran}
\bibliography{references}
\begin{IEEEbiography}
[{\includegraphics[width=1in,height=1.25in,clip,keepaspectratio]{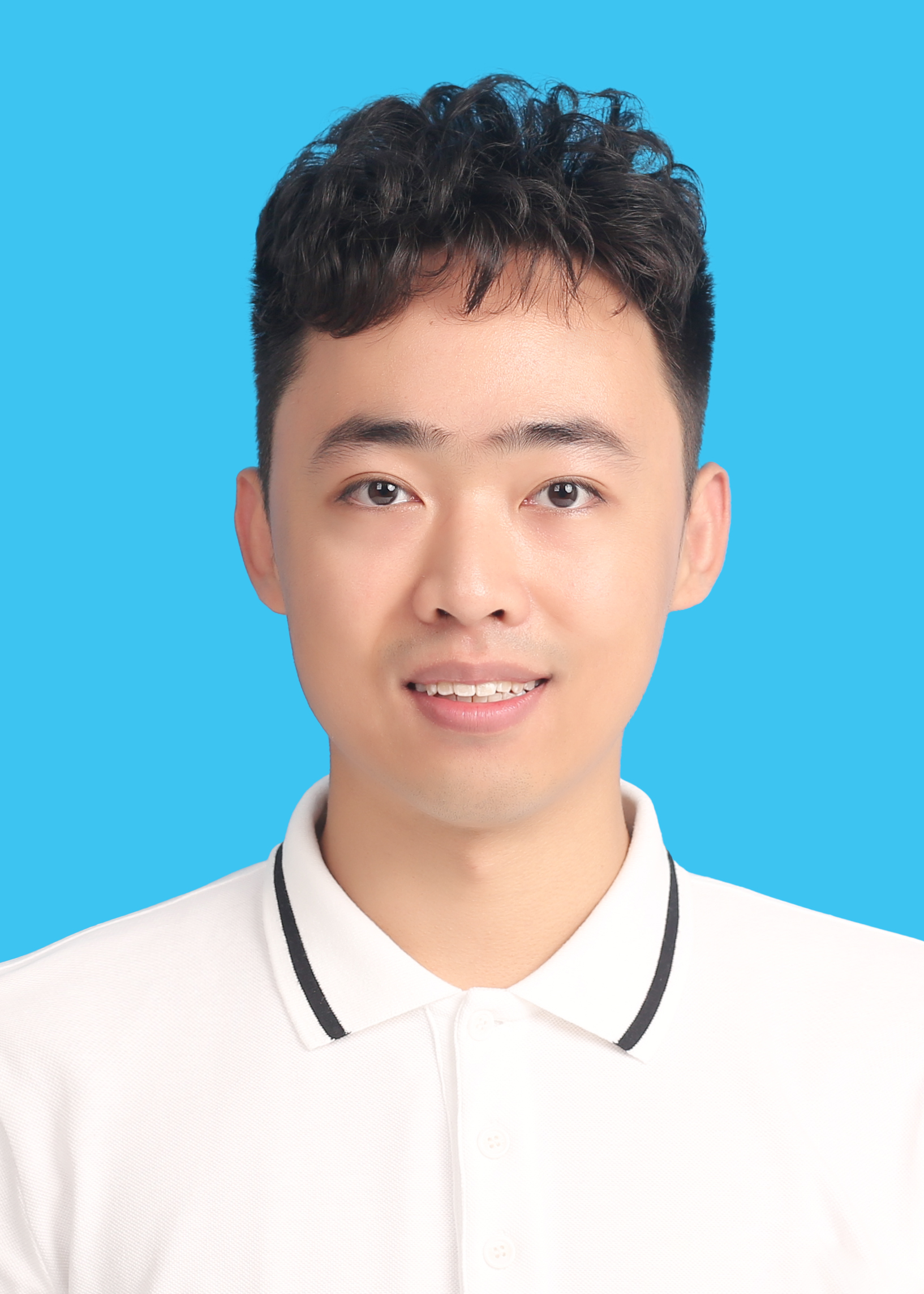}}]{Zhiyan Liu} (Graduate Student Member, IEEE) received the B.Eng. degree from the Dept. of Electronic Engineering, Tsinghua University, Beijing, in 2021. He is currently working towards the Ph.D. degree with Dept. of Electrical and Electronic Engineering, The University of Hong Kong (HKU), Hong Kong. His recent research interests include edge intelligence and distributed sensing in 6G wireless networks. He was a recipient of Hong Kong Ph.D. Fellowship.
\end{IEEEbiography}

\begin{IEEEbiography}
[{\includegraphics[width=1in,height=1.25in,clip,keepaspectratio]{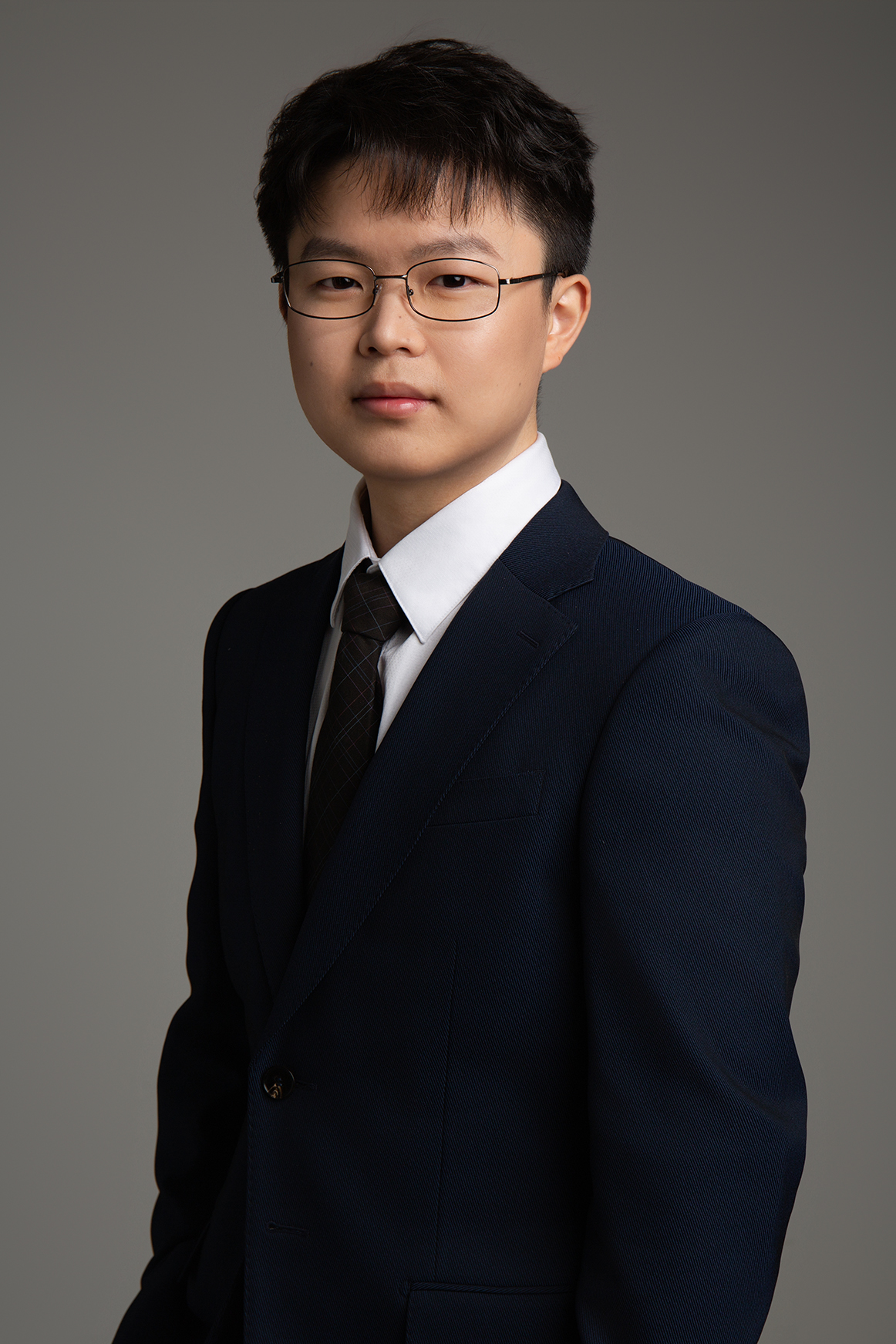}}]{Xu Chen} received the B.Eng. and M.Eng. degrees from Harbin Institute of Technology (HIT), Harbin, China, in 2020, and the Ph.D. degree from The University of Hong Kong, Hong Kong, in 2024. He is currently a senior standard engineer with Xiaomi Communications Co., Ltd.. He is a recipient of Hong Kong PhD Fellowship (HKPF). His research interests include wireless communications and AI for wireless.
\end{IEEEbiography}

\begin{IEEEbiography}
[{\includegraphics[width=1in,height=1.25in,clip,keepaspectratio]{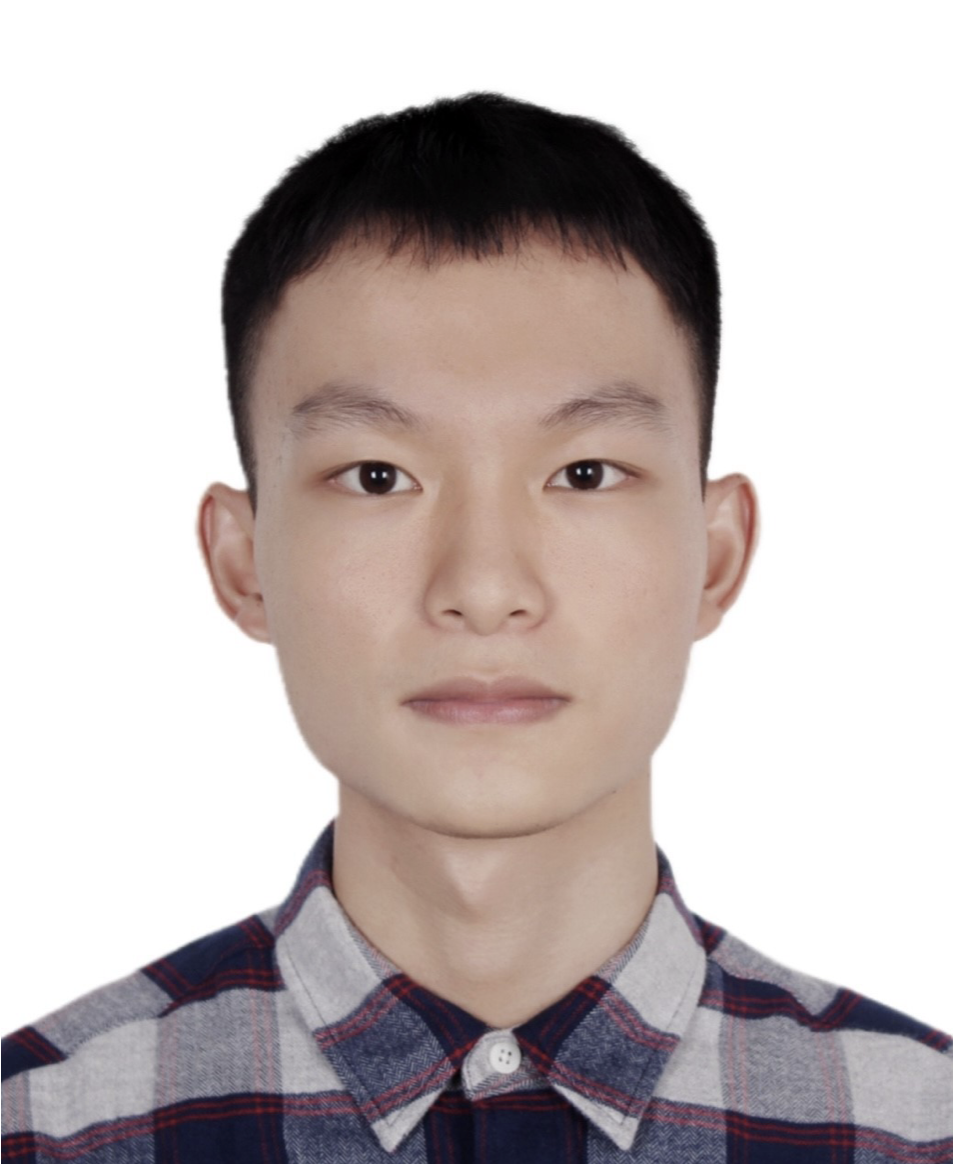}}]{Hai Wu} received the B.Eng. degree from Southern University of Science and Technology, Shenzhen, in 2020, and the Ph.D. degree from The University of Hong Kong, Hong Kong, in 2024. His recent research interests include deep learning and efficient deployment of artificial intelligence.
\end{IEEEbiography}

\begin{IEEEbiography}
[{\includegraphics[width=1in,height=1.25in,clip,keepaspectratio]{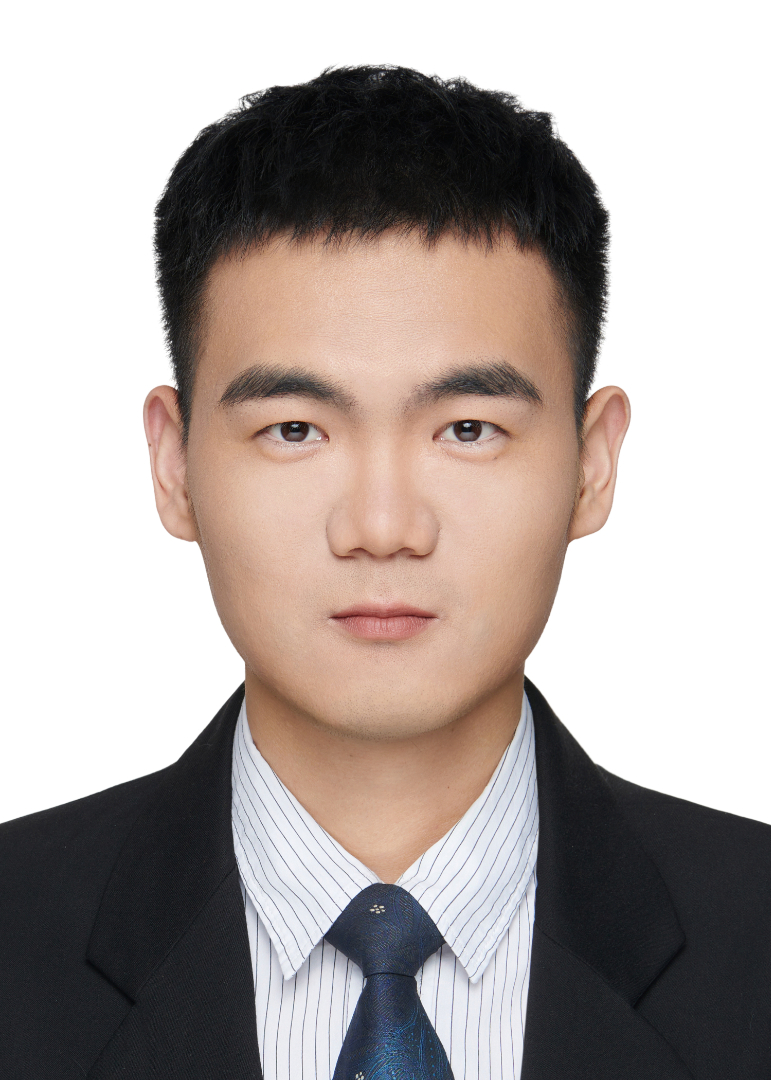}}]{Zhanwei Wang} (Graduate Student Member, IEEE) received the B.Eng. degree in Information Engineering and the M.Eng. degree in Information and Communication Engineering from Xidian University, Xi’an, China, in 2018 and 2021, respectively. He is currently pursuing the Ph.D. degree in the Department of Electrical and Electronic Engineering at The University of Hong Kong. His research interests include wireless communications, edge intelligence, distributed sensing, and atomic receiver.
\end{IEEEbiography}

\begin{IEEEbiography}
[{\includegraphics[width=1in,height=1.25in,clip,keepaspectratio]{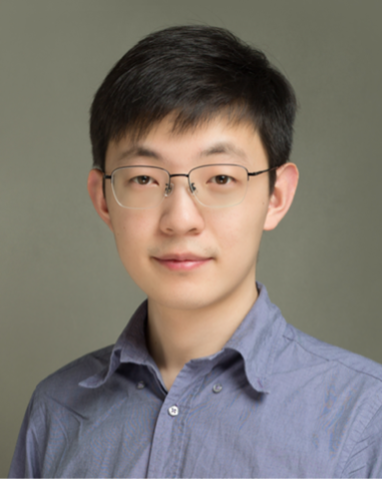}}]{Xianhao Chen}(Member, IEEE) received the B.Eng. degree in electronic information from Southwest Jiaotong University in 2017, and the Ph.D. degree in electrical and computer engineering from the University of Florida in 2022. He is currently an assistant professor at the Department of Electrical and Electronic Engineering, the University of Hong Kong, where he directs the Wireless Information \& Intelligence (WILL) Lab. He serves as a TPC member of several international conferences and an Associate Editor of ACM Computing Surveys. He received the Early Career Award from the Research Grants Council (RGC) of Hong Kong in 2024, the ECE Graduate Excellence Award for research from the University of Florida in 2022, and the ICCC Best Paper Award in 2023. His research interests include wireless networking, edge intelligence, and machine learning.
\end{IEEEbiography}

\begin{IEEEbiography}
[{\includegraphics[width=1in,height=1.25in,clip,keepaspectratio]{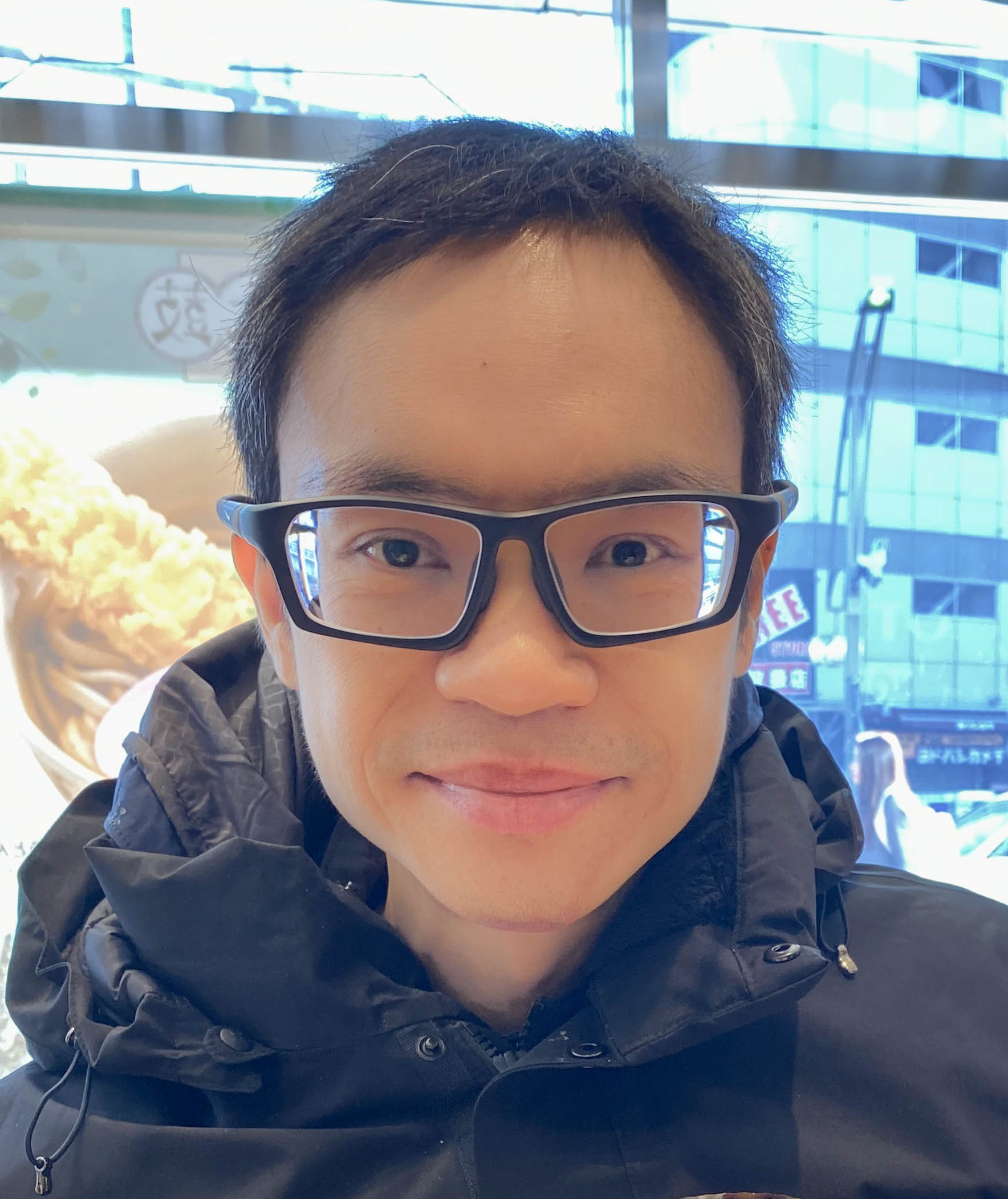}}]{Dusit Niyato} (M'09-SM'15-F'17) is a professor in the College of Computing and Data Science, at Nanyang Technological University, Singapore. He received B.Eng. from King Mongkuts Institute of Technology Ladkrabang (KMITL), Thailand and Ph.D. in Electrical and Computer Engineering from the University of Manitoba, Canada. His research interests are in the areas of mobile generative AI, edge intelligence, quantum computing and networking, and incentive mechanism design.
\end{IEEEbiography}

\begin{IEEEbiography}
[{\includegraphics[width=1in,height=1.25in,clip,keepaspectratio]{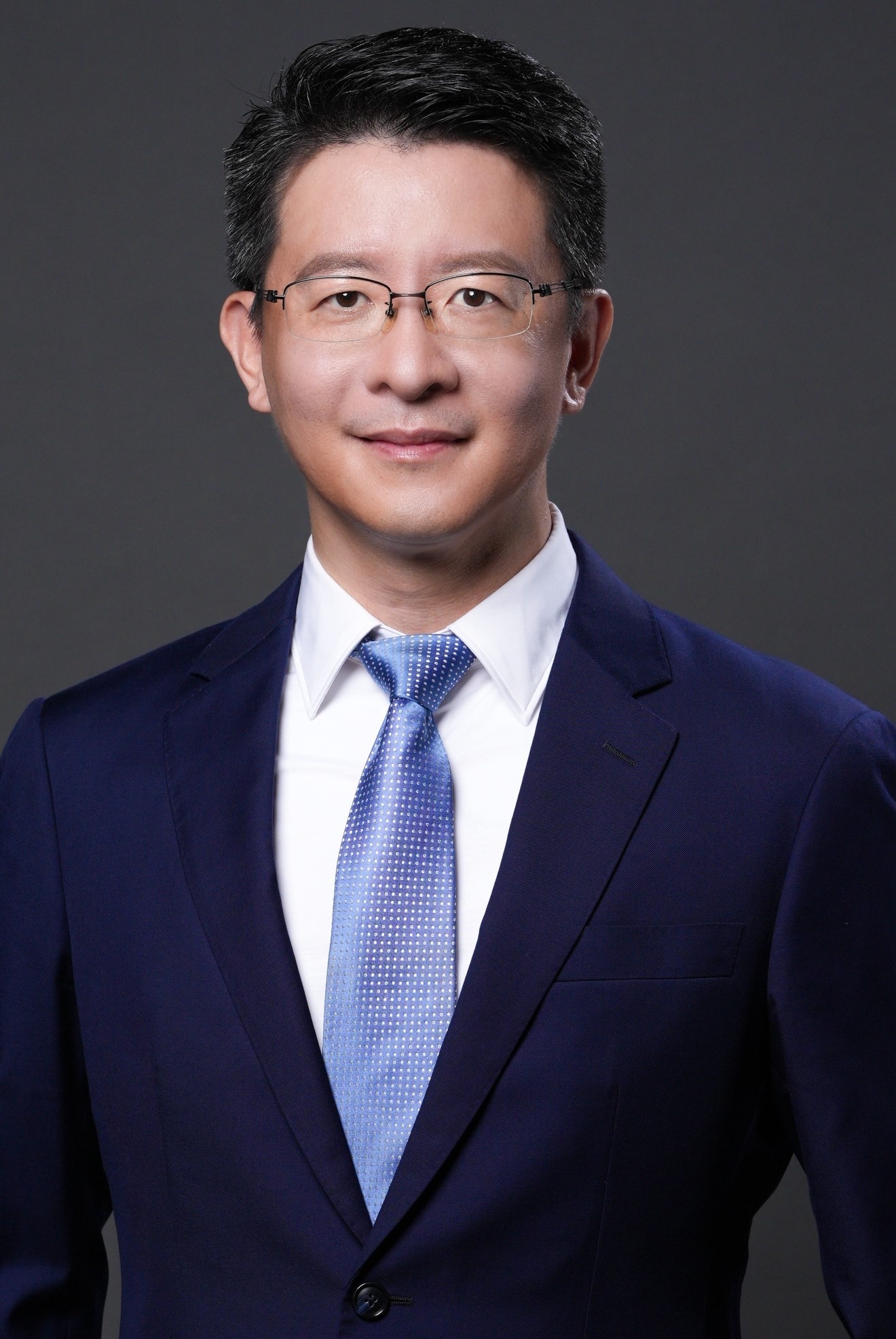}}]{Kaibin Huang} (Fellow, IEEE) received the B.Eng. and M.Eng. degrees from the National University of Singapore and the Ph.D. degree from The University of Texas at Austin, all in electrical engineering. He is the Philip K H Wong Wilson K L Wong Professor in Electrical Engineering and the Department Head at the Dept. of Electrical and Electronic Engineering, The University of Hong Kong (HKU), Hong Kong. His work was recognized with seven Best Paper awards from the IEEE Communication Society. He is a member of the Engineering Panel of Hong Kong Research Grants Council (RGC) and a RGC Research Fellow (2021 Class). He has served on the editorial boards of five major journals in the area of wireless communications and co-edited ten journal special issues. He has been active in organizing international conferences such as the 2014, 2017, and 2023 editions of IEEE Globecom, a flagship conference in communication. He has been named as a Highly Cited Researcher by Clarivate in the last six years (2019-2024) and an AI 2000 Most Influential Scholar (Top 30 in Internet of Things) in 2023-2024. He was an IEEE Distinguished Lecturer. He is a Fellow of the IEEE and the U.S. National Academy of Inventors.

\end{IEEEbiography}

\end{document}